# WHITE PAPER
## ON
# NUCLEAR ASTROPHYSICS
## AND
# LOW ENERGY
# NUCLEAR PHYSICS

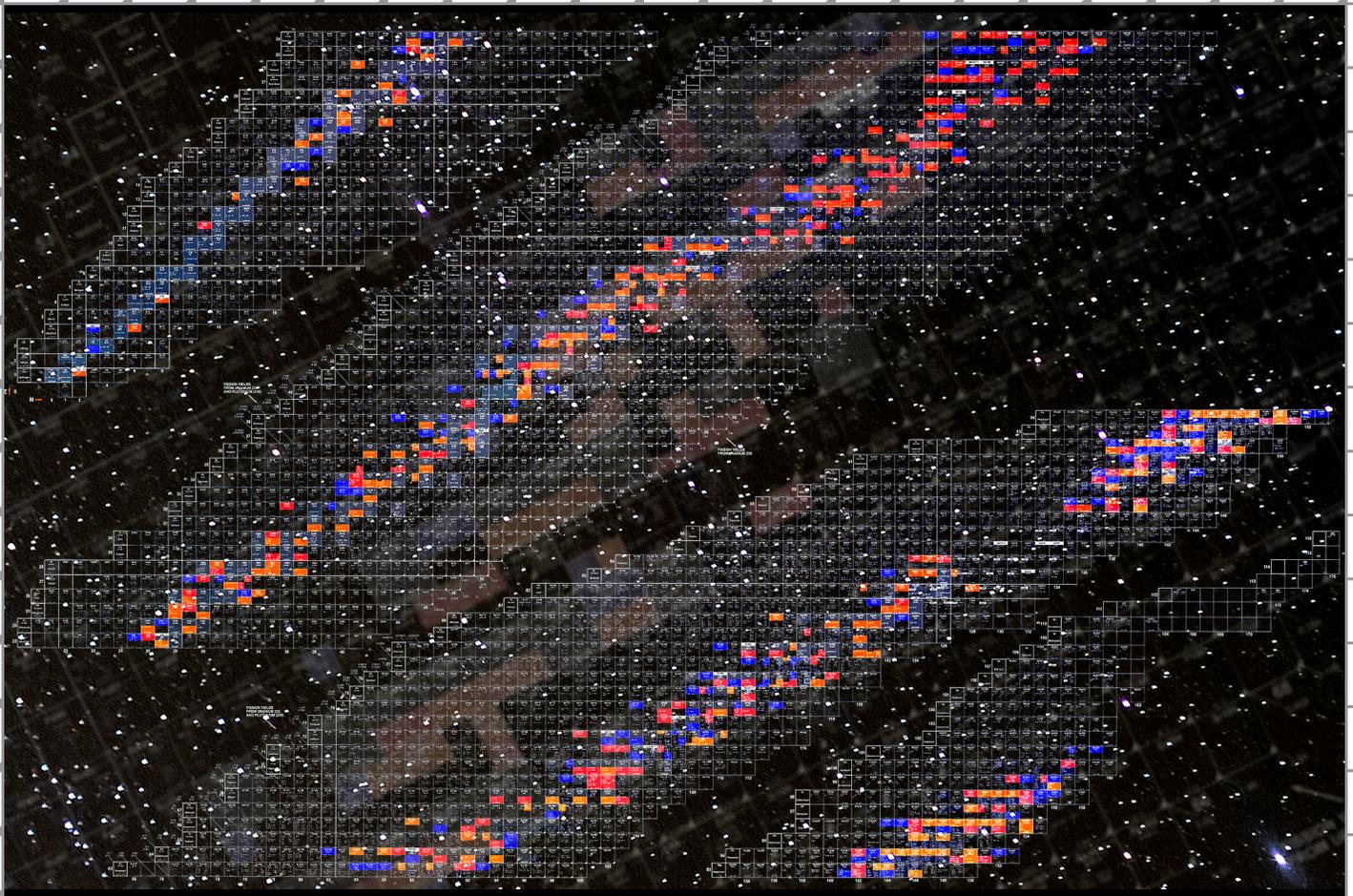

# PART ONE
# NUCLEAR ASTROPHYSICS

# JANUARY 31 2015

# WHITE PAPER ON NUCLEAR ASTROPHYSICS AND LOW ENERGY NUCLEAR PHYSICS

## PART 1: NUCLEAR ASTROPHYSICS

FEBRUARY 2016

Edited by:                    Hendrik Schatz and Michael Wiescher

Layout and design:            Erin O'Donnell, NSCL, Michigan State University

Individual sections have been edited by the section conveners:

Almudena Arcones, Dan Bardayan, Lee Bernstein, Jeffrey Blackmon, Edward Brown, Carl Brune, Art Champagne, Alessandro Chieffi, Aaron Couture, Roland Diehl, Jutta Escher, Brian Fields, Carla Froehlich, Falk Herwig, Raphael Hix, Christian Iliadis, Bill Lynch, Gail McLaughlin, Bronson Messer, Bradley Meyer, Filomena Nunes, Brian O'Shea, Madappa Prakash, Boris Pritychenko, Sanjay Reddy, Ernst Rehm, Grisha Rogachev, Bob Ruthledge, Michael Smith, Andrew Steiner, Tod Strohmayer, Frank Timmes, Remco Zegers, Mike Zingale

# ABSTRACT

This white paper informs the nuclear astrophysics community and funding agencies about the scientific directions and priorities of the field and provides input from this community for the 2015 Nuclear Science Long Range Plan. It summarizes the outcome of the nuclear astrophysics town meeting that was held on August 21-23, 2014 in College Station at the campus of Texas A&M University in preparation of the NSAC Nuclear Science Long Range Plan. It also reflects the outcome of an earlier town meeting of the nuclear astrophysics community organized by the Joint Institute for Nuclear Astrophysics (JINA) on October 9-10, 2012 Detroit, Michigan, with the purpose of developing a vision for nuclear astrophysics in light of the recent NRC decadal surveys in nuclear physics (NP2010) and astronomy (ASTRO2010). The white paper is furthermore informed by the town meeting of the Association of Research at University Nuclear Accelerators (ARUNA) that took place at the University of Notre Dame on June 12-13, 2014. In summary we find that nuclear astrophysics is a modern and vibrant field addressing fundamental science questions at the intersection of nuclear physics and astrophysics. These questions relate to the origin of the elements, the nuclear engines that drive life and death of stars, and the properties of dense matter. A broad range of nuclear accelerator facilities, astronomical observatories, theory efforts, and computational capabilities are needed. With the developments outlined in this white paper, answers to long standing key questions are well within reach in the coming decade.

# Contents

# 1 EXECUTIVE SUMMARY NUCLEAR ASTROPHYSICS

This white paper summarizes the outcome of the nuclear astrophysics town meeting that was held on August 21-23, 2014 in College Station at the campus of Texas A&M University in preparation of the NSAC Nuclear Science Long Range Plan. The meeting was organized jointly with a meeting of the low energy nuclear physics community and had a total of 270 on-site participants with 60 on-line participants. The nuclear astrophysics town meeting featured eight plenary talks on issues at the interface of nuclear structure, weak interaction and nuclear astrophysics and was organized in eight working group sessions to identify the most compelling research questions and the most urgently needed scientific equipment for the field.

This white paper reflects also the outcome of an earlier town meeting of the nuclear astrophysics community organized by the Joint Institute for Nuclear Astrophysics (JINA) on October 9-10, 2012 at the airport in Detroit, Michigan, that included 150 scientists with the purpose of developing a vision for nuclear astrophysics in light of the recent NRC decadal surveys in nuclear physics (NP2010) and astronomy (ASTRO2010). The white paper is furthermore informed by the town meeting of the Association of Research at University Nuclear Accelerators (ARUNA) that took place at the University of Notre Dame on June 12-13, 2014. The ARUNA meeting hosted 59 participants and the results will be presented in an independent white paper to NSAC.

The present white paper informs the nuclear astrophysics community and funding agencies about the scientific directions and priorities of the field and provides input from this community for the 2015 Nuclear Science Long Range Plan. Open questions in nuclear astrophysics cut across most areas of nuclear science. In particular nuclear structure, nuclear reactions, and neutrino physics play a critical role. While we summarize the important developments in all areas that are needed for progress in nuclear astrophysics (including astrophysics and astronomy), many of these developments are described in more detail in the respective white papers of the low energy nuclear physics and the fundamental symmetries and neutrino communities.

Nuclear astrophysics is a modern and vibrant field addressing fundamental science questions at the intersection of nuclear physics and astrophysics. Broadly these questions can be grouped into three themes:

- The origin of elements in our universe from the Big Bang to the present time. This theme addresses the build-up of light and heavy elements through a broad variety of nuclear processes in a multitude of stellar environments

- The nuclear engines for the life and death of stars from the first stars to our sun. This theme deals with the understanding of the critical nuclear reaction sequences, dense matter properties, and neutrino processes that drive the different phases of quiescent and exploding stellar burning scenarios

- The composition and state of matter in the crust and core of neutron stars. This theme investigates the fate of matter at extreme density conditions and the underlying physics of nuclear matter.

New questions emerge frequently in nuclear astrophysics driven by surprises in observations, experiments, or theory. Recent examples include the nature of the first stars, the mechanism of type Ia supernovae, the physics of super-bursts, the origin sites of intermediate mass elements, or the impact of neutrino or other weak interaction processes on astrophysics environments. In some areas, progress enables the field to move beyond qualitative questions. Examples include the Big Bang or the Sun, where the thrust has turned towards precision measurements of nuclear reaction rates that when combined with neutrino physics enable the use of these scenarios to answer questions related to new elementary particles, the nature of the universe, and the properties of neutrinos.

With technological advances in nuclear physics, astronomy, and computational physics, and the developments outlined in the ASTR2010 and NP2010 decadal reviews, the field is in an unprecedented position to answer many of the open questions in the next decade.

With the Facility for Rare Isotope beams (FRIB), most of the rare isotopes produced in stellar explosions and neutron stars finally become available for a broad range of laboratory studies.

Upgrades of stable beam facilities, many housed at university laboratories, will enable unprecedented precision measurements of stellar reaction rates with novel direct and indirect techniques. This effort should be complemented by the development of a high-intensity stable beam accelerator facility located deep underground to provide substantially reduced background conditions for the direct measurement of the extremely low cross sections at stellar energies.

While the US nuclear astrophysics community should maintain their leadership in nuclear astrophysics experiments with stable and radioactive ion beams major efforts in building new high flux neutron facilities for studying nuclear reactions critical for neutron capture and photodisintegration processes are underway in Europe. The US can complement that by maintaining experimental opportunities at facilities such as LANSCE at Los Alamos, HIγS at TUNL Laboratory, and Jefferson-Lab. New efforts have emerged at fusion physics facilities such as OMEGA and NIF that for the first time allow a direct study of the impact of the stellar plasma effects that are crucial for low temperature burning in stars and the ignition of fusion-driven bursts in the core of white dwarfs or the crust of neutron stars.

Progress in the theory of nuclear structure, weak interactions, and nuclear reactions, is in part driven by increasing computational capabilities, while generating new opportunities for advances in nuclear astrophysics. Extrapolations of theoretical predictions of nuclear properties to astrophysical regimes in energy, neutron or proton richness, and density are becoming much more reliable, and uncertainties can be better quantified, as phenomenological models are increasingly replaced with more fundamental approaches for broad ranges of nuclei and nuclear matter.

New observations often open up new areas of research in nuclear astrophysics. Large scale surveys of metal poor stars and high resolution spectroscopic follow-up with the largest available telescopes is providing a fossil record of chemical evolution that reaches back to the very first supernova explosions, soon after the Big Bang, opening links to cosmology and galaxy formation questions. New observational initiatives, such as PAN-STARR and LSST, will open up time domain astronomy with the prospect of the discovery of rare explosive astrophysical scenarios as yet unseen. And with advanced LIGO beginning operation, there is a realistic chance to detect gravitational waves from merging neutron stars for the first time, providing a direct link to fundamental nuclear astrophysics questions related to neutron stars and high density matter.

In the following we summarize the findings and recommendations of the nuclear astrophysics community to enable the exploitation of these opportunities towards transformational advances in nuclear astrophysics:

1. FRIB's unprecedented intense beams of fast, stopped, and reaccelerated rare isotopes offer game changing opportunities for nuclear astrophysics, in particular in the areas of explosive nucleosynthesis and neutron stars.

   - We strongly support the timely completion of the Facility for Rare Isotope Beams (FRIB) and the implementation of the full science program as the highest priority for the nuclear astrophysics community.

   - To operate a broad nuclear astrophysics program we strongly recommend the development and implementation of critical equipment such as SECAR, GRETA, and the HRS for nuclear astrophysics measurements.

2. To address the compelling questions in nuclear astrophysics and to operate an effective and competitive nuclear astrophysics program a broad range of nuclear probes, techniques, and theory is essential. This requires effective utilization of the available nuclear physics facilities, in particular university-based laboratories, and strong theory support.

   - We recommend to appropriately support operations and planned upgrades at ATLAS, NSCL, and university-based laboratories as well as the utilization of these and other facilities for enabling measurements with the broad range of beams required to achieve the science goals in nuclear astrophysics. It is essential that strong support for research groups is provided.

   - We recommend strengthening support for nuclear theory and the founding of an FRIB theory center that addresses the needs of a broader nuclear astrophysics community. In addition we recommend focused multi-institutional research collaborations in theory and simulation to take advantage of new opportunities created by increased computing capabilities and large data science.

3. High intensity underground accelerator measurements have emerged as a critical tool for directly studying reactions in stellar burning that govern stellar evolution and provide the seeds for explosive nucleosynthesis.

   ▪ We recommend the construction and operation of a high intensity underground accelerator facility for the study of stable beam reactions near quiescent stellar burning conditions.

4. Interdisciplinary centers are important for advances in nuclear astrophysics as they overcome field boundaries between nuclear physics and astronomy, and bring together the diverse experiment, theory, and observation communities that comprise the field of nuclear astrophysics. Data compilation, dissemination, and distribution are essential components for such interdisciplinary efforts.

   ▪ We recommend the continued support for the operation of the Joint Institute for Nuclear Astrophysics as Physics Frontiers Center and other field bridging initiatives.

   ▪ We recommend continued robust support of the operation of data centers and other data compilation efforts of importance for nuclear astrophysics.

5. Education and innovation are key components of any vision of the future of the field of nuclear science.

   ▪ We fully endorse the recommendations of the Education and Innovation White Paper.

In nuclear astrophysics, nuclear science, astronomy, and astrophysics are closely intertwined. The identification of future directions and priorities can therefore not be performed in isolation within a subfield. While the recommendations in this white paper serve as input of the nuclear astrophysics community (including astrophysicists and astronomers) into the 2015 Nuclear Science Long Range Plan, the broad look at the entire field of nuclear astrophysics that was required to arrive at these recommendations, resulted in the additional identification of needs that must be addressed by the astronomy community to enable nuclear astrophysics to achieve its scientific goals. While many of these needs align with the priorities of the astronomy community as outlined in the ASTRO2010 decadal survey, there are a number of additional observational capabilities that will be important for nuclear astrophysics. These include UV spectroscopy, which is essential for determining stellar abundances for a number of elements but where there is currently no future space capability beyond HST planned, X-ray spectroscopy with much increased sensitivity and timing resolution (NICER, ASTRO-H, and LOFT will be important, but capabilities beyond these instruments will eventually be required), and MeV gamma-ray spectroscopy capabilities. Continued operation of the Green Bank Radio Telescope is also important for observations related to neutron stars and the nature of dense matter and its equation of state.

This white paper is divided into two main sections - a discussion of the open scientific questions together with strategies to address them, and a discussion of the 'tools' that are needed to pursue these strategies. The science discussion is organized in topical sections encompassing the major questions of the field concerning the origin of the elements, stars and stellar evolution, core collapse supernovae, compact object binary mergers, explosions of white dwarfs, neutron stars, the Big Bang, and Galactic chemical evolution. Each topical section consists of an introduction for non-experts, a list of open questions, a discussion of the context of these questions in light of previous work, and the major strategies that need to be pursued to address these questions. The 'tools' section discusses accelerator facilities, theoretical developments, computational needs, astronomical observatories, tools related to managing and disseminating data and codes, and the role of centers. Science and 'tools' sections are closely connected through numerous cross references.

# 2 SCIENTIFIC CHALLENGES IN NUCLEAR ASTROPHYSICS

## 2.1 What Is The Origin Of The Elements?

### 2.1.1 Introduction For Non-Experts

The origin of the elements is one of the fundamental questions in science. How did the universe evolve from a place made of hydrogen and helium, with minute traces of lithium, to a world with the incredible chemical diversity of 84 elements that are the building blocks of planets and life? Our understanding of the answers to this basic question is incomplete in major aspects. What we do know is that a wide variety of nuclear reaction sequences in stars, stellar explosions, and, possibly, collisions of neutron stars, build up the elements step by step. Some of these reactions involve the fusion of stable nuclei over millions of years, others use extremely unstable nuclei as stepping stones to build up new elements within seconds. For some of these reaction sequences, experiments have provided data on reaction rates that allow the prediction of what elements they may have created. For most, however, knowledge is still very limited. Some processes have only very recently been discovered, and some may still await discovery. Even for some of the long identified processes, the astrophysical sites have yet to be identified with certainty. Thus, knowledge in nuclear physics and in astrophysics must be extrapolated through applications of theoretical models, and thus involves considerable uncertainties. Experiments in nuclear physics and astronomical observations are key to progress here. The answers to all these questions will not only inform us about the origin of the basic building blocks of nature, but will also address questions about the formation of galaxies, about the formation of stars and planets, about the interiors of stars and stellar explosions, and about the properties of neutrinos.

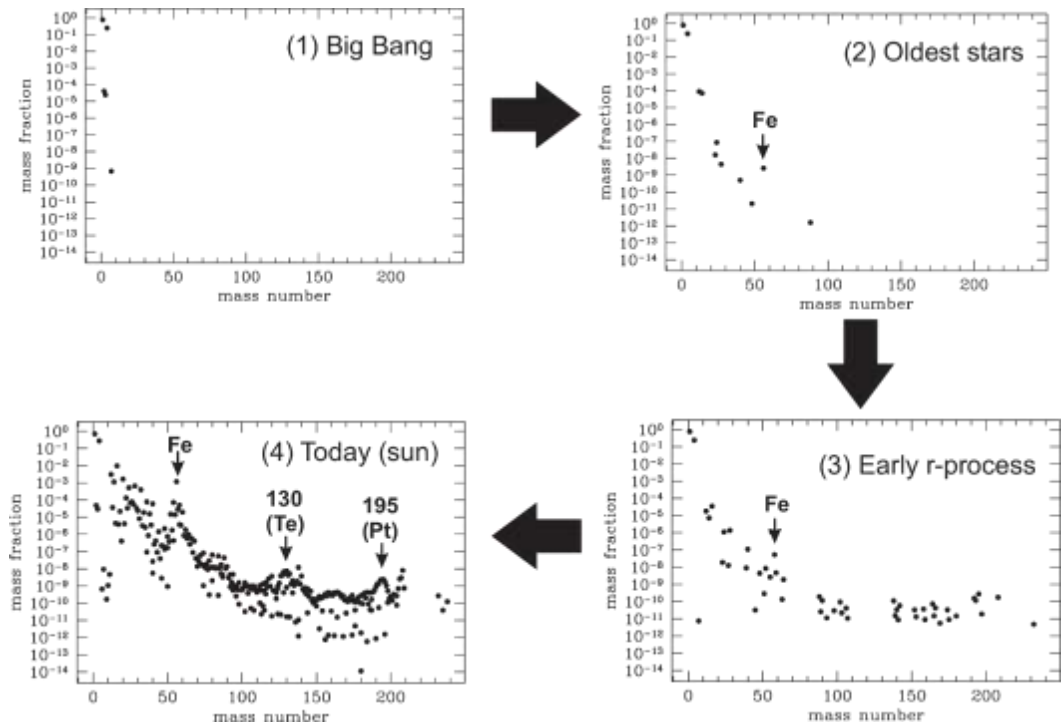

Figure 1: How did nature create the elements found today in the solar system (panel 4) from the few elements present at the time of the Big Bang (panel 1)? This is one of the key open questions nuclear astrophysics seeks to answer. Observations that for example show an emerging abundance pattern in old stars (panel 2) or r-process contributions from early explosive nucleosynthesis (panel 3) are now filling in the gaps and reveal a step by step history of the evolution of the elements. The future goal of nuclear astrophysics is to identify and understand the underlying nuclear processes and their corresponding astrophysical environments.

### 2.1.2  Current open questions

- What was the nature of the first stars, what are their nucleosynthetic signatures, and can we find these signatures today?

- What are the rates of the key nuclear reactions in stars that define the sequence of stellar evolution and characterize the patterns of stellar life?

- How can observations of solar and supernova neutrinos be exploited to deepen our understanding of stellar burning mechanisms?

- What defines the relative abundances of carbon and oxygen in our universe, which, in turn, determines the stages of stellar evolution, defines the seed for stellar explosions, and provides the base for the origin of life on Earth?

- Where are the 54 elements beyond iron created, that are traditionally attributed to a rapid neutron capture process (r-process)?

- Why is the r-process so robust, producing similar abundance patterns event by event?

- What is the contribution of neutrino-driven winds in core collapse supernovae to nucleosynthesis? And what role do neutrino properties play?

- What is the origin of the unexpectedly high abundance of the neutron deficient stable isotopes of molybdenum and ruthenium that are traditionally attributed to a p-process?

- How can we use element abundance observations in stars and presolar grains to validate complex stellar models?

- What is the quantitative contribution of different types of stellar sites to the origin of the elements (initial mass function), how does this evolve with time or galaxy type, and what is the effect of many stars being binaries?

- What are the ranges of elemental and isotopic variability for stars hosting exoplanets?

## 2.1.3  Context

Nuclear astrophysics has come a long way in explaining the origin of the elements since Margaret Burbidge, Geoffrey Burbidge, Willy Fowler, and Fred Hoyle in 1957 provided the first comprehensive theory of the origin of elements in stars. According to current understanding light elements up to slightly beyond iron are formed in a series of stellar evolution phases that are defined by the local fuel conditions in the stellar interior. The stellar main sequence and the red-giant phase of stellar evolution are maintained by proton (hydrogen) and alpha (helium) capture reactions, respectively. Later phases of stellar life are characterized by a complex network of fusion, photodisintegration, and capture reactions that set the conditions for core collapse as the final phase of stellar life. Traditionally, the pattern of abundances of heavier isotopes beyond iron found in the solar system has been interpreted as pointing to four distinct origin processes: a main slow neutron capture process (s-process) known to occur in lower mass, thermally pulsing red giant stars (TP-AGB stars), a weak s-process known to occur in massive red-giant stars, a rapid neutron capture process (r-process), and a photodissociation-driven process (p-process) producing the rare neutron-deficient isotopes of some elements. While freshly produced s-process species have been observed in red giant stars, the sites of the r- and p-process are not known with certainty. However, a p-process occurs naturally in models of core-collapse supernovae, and in some models of thermonuclear supernovae. On the other hand, many possible models for the r-process have been proposed – possibilities include various sites in core-collapse supernovae and merging neutron stars – but all these sites have difficulties explaining the entirety of observational data.

On the nuclear side, decades of laboratory efforts have succeeded in directly measuring many of the relevant neutron capture cross sections for the s-processes. Stellar s-process models can therefore be compared rather reliably to the observed abundances of s-process isotopes in the solar system, in stars, and in meteoritic grains thought to originate from the condensed ejecta of ancient red giant stars elsewhere in the Galaxy. Critical issues still remain with neutron capture reactions on long-lived radioactive nuclei that determine the branching points of the s-process. These are of particular interest since a detailed analysis of isotopic abundances of nuclei associated with branching points in meteoritic inclusions can be used as a thermometer and pycnometer for determining the internal conditions of the actual s-process sites. Also of interests are neutron capture rates associated with the presumed s-process endpoint in the Pb and Bi range, which serve as a tool for distinguishing between r-process and s-process scenarios for the production of Pb isotopes.

Despite impressive developments in accelerator and detector technology, however, experimental determination of the rates in stellar fusion, the p-process, and the r-process still remain largely elusive - in the case of stellar fusion because of the small cross sections that determine the stellar evolution time-scales, in the case of the r- and p-processes in fast explosive environments because the nuclei involved are unstable and difficult to produce in laboratories.

In addition to these laboratory challenges, with advances in observations and stellar modeling a more complex picture of the origin of the elements is emerging. Observations of the composition of old, chemically primitive stars in the halo of the Galaxy (so called metal poor stars) provide snapshots of the compositional evolution of the Galaxy forming a "fossil record" of chemical evolution (see Fig. 1). These observations indicate that elements above Ge, maybe up to Te, previously attributed to the r-process are instead produced by multiple processes of unknown nature with distinct compositional signatures. The observations also indicate that stellar nucleosynthesis products have changed over time, as the initial metal content of a star can dramatically alter its evolution and nucleosynthetic output. At the same time advanced stellar models have led to predictions of hitherto unknown nucleosynthesis processes, including extended reaction sequences in the first stars, proton-rich neutrino-driven winds in core-collapse supernovae, the so called νp-process, and an intermediate neutron capture process (i-process) in low metallicity stars. The jury is still out as to what extent these processes occur, and whether they may explain some of the unanticipated observational signatures. Nuclear data are urgently needed to predict their characteristic abundance patterns so that authoritative comparisons to observations can be made.

### 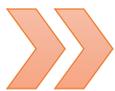 2.1.4 Origin of the Elements Strategic Thrust 1: The Nuclear Physics of Element Synthesis and Model Validation

For the understanding of the origin of the elements, knowledge of the underlying nuclear reactions is of fundamental importance (see Fig. 2). Only with reliable nuclear physics can one reliably predict the abundance signatures of various nucleosynthesis processes and unravel their contributions to the elements found in nature. And only with reliable nuclear physics can nucleosynthesis models be validated against observations. Once a nucleosynthesis process is identified, detailed observations of produced abundances in connection with reliable nuclear physics open the door to validate stellar models and constrain conditions inside stars and stellar explosions that are otherwise not accessible. For example, thanks to decades of careful experimental work, the s-process is now used as a sensitive probe of mixing processes in stellar interiors. The goal for the coming decade is that other nucleosynthesis processes can come to be used in a similar fashion.

However, experimental information on the element producing nuclear reactions in nature is surprisingly sparse. Reactions among stable nuclei occur in stars at relatively low densities and temperatures on timescales of millions or billions of years. Measuring these very slow reactions is a huge experimental challenge, and has only been achieved in rare cases of fusion or capture reactions in the pp-chains between low Z nuclei. Expanding the scope of experiments to hydrogen and helium induced reactions on higher Z-nuclei requires the development of new facilities with high intensity beams and increased background reduction capabilities above and under the ground (see section 2.2). On the other hand, when conditions produce faster reactions, as in stellar explosions, the nuclei involved are unstable, because additional reactions on these unstable nuclei can occur before the nuclei decay. Measurements are then equally challenging, as it is extremely difficult to produce sufficiently intense beams of unstable nuclei to study these reactions. Again, measurements have only succeeded in a very small number of cases. In the coming decade we will be able to address these challenges with the advent of a new generation of radioactive beam facilities (see section 3.2).

The critical reactions for stellar nucleosynthesis include the reactions influencing stellar evolution (see section 2.2), including the $3\alpha\to{}^{12}$ and ${}^{12}C(\alpha,\gamma){}^{16}O$ reactions that alter nucleosynthesis throughout the evolution of a star since they determine the carbon/oxygen ratio at the end of helium burning, a seed for multiple subsequent burning events in the later phases of stellar evolution. Also important are fusion reactions such as ${}^{12}C+{}^{12}C$ and ${}^{12}C+{}^{16}O$ since they define the seed for the subsequent phases of stellar evolution towards supernovae (see section 2.3) as well as the seed conditions on nova explosions on the surface of white dwarfs (see section 2.5). In addition, there are a number of reactions that are not critical for energy production, but nevertheless have a strong impact on nucleosynthesis. These include neutron-producing reactions, such as ${}^{13}C(\alpha,n){}^{16}O$ and ${}^{22}Ne(\alpha,n){}^{25}Mg$. These reactions determine strength and extent of the weak s-process, and, because s-process nuclei serve as seeds for the p-process, the nucleosynthetic outcome of the p-process. Of similar importance are the neutron capture rates on abundant nuclides that absorb neutrons, so called neutron poisons. During advanced burning stages and during explosive nuclear burning triggered by the shock wave passing through the star when it explodes as a supernova, proton, neutron, and $\alpha$ induced reactions on heavier stable and unstable nuclei become important.

Masses, $\beta$-decay properties, and neutron capture rates on hundreds of unstable nuclei are critical for modeling various r-processes and the i-process. In the case of the r-process, the nuclei are very far from stability (see Fig. 4) and many have not yet been produced in laboratories to date. Nevertheless, progress has been made. A wide range of mass measurements for increasingly unstable nuclei have been successfully carried out using time-of-flight and Penning trap techniques. $\beta$-decay measurements now reach beyond the N=50 shell in the Ga–Ge region covering the beginning of the r-process, and similar measurements at RIKEN are now verging on the r-process waiting points in the Rb–Zr region. FRIB will be essential in expanding the reach of r-process experiments to cover a significant portion of the r-process path (see section 3.2). Neutrino interactions play an important role in the r-process and can also produce some rare isotopes in the so called $\nu$-process.

For the recently discovered i-process, a neutron capture process with time scales intermediate to the s- and r-process, the critical nuclei are close to stability. However, accurate neutron capture rates are needed, which are very difficult to determine experimentally for unstable nuclei. Techniques to carry out such measurements, such as the surrogate approach using (d,p) and other transfer reactions, are critical. Pioneering measurements have been carried out, for example in the $^{132}$Sn region. Promising progress has also been made in utilizing inverse photodissociation or Coulomb breakup processes as in the case of $^{60}$Fe, but all these techniques need to be developed further through experimental and theoretical work. $\beta$-decay, proton capture, (p,$\alpha$), and (n,p) reactions on unstable neutron-deficient nuclei need to be understood for models of the $\nu p$-process as well as nucleosynthesis in nova explosions.

p-process models require reliable ($\gamma$,n), ($\gamma$,p), and ($\gamma$,$\alpha$) reactions on hundreds of stable and unstable neutron-deficient nuclei. The need for experimental data is underlined by findings of large discrepancies between statistical model predictions and measurements of reactions that involve $\alpha$-particles. Measurements can be performed with $\gamma$-beams (see section 3.6) or, taking advantage of quasi-virtual photons, via Coulomb breakup. However, in many cases, a measurement of the inverse particle induced capture reaction, and the application of time-reversal invariance, is preferable and is currently a standard tool for p-process studies. Currently the community worldwide is developing techniques to measure the relevant capture reactions using radioactive targets or beams. The ReA3 facility at the NSCL and later at FRIB is ideal for such measurements at astrophysical energies.

Nuclear theory is critical to complement experimental information (see section 3.6). Even with new facilities expected to fill in much of the missing information in the coming decade or two, theory is needed to reliably predict properties of nuclei beyond experimental reach, and to determine corrections to the measured nuclear data due to the extreme stellar environments. Nuclear theory is also essential when using indirect experimental approaches to determine neutron induced reactions on unstable nuclei, which are a particular challenge as both, target and projectile, are unstable. Of particular importance is reaction theory, to extrapolate experimental reaction data into yet unexplored regions of stable beam reactions or to translate nuclear structure data into cross section predictions for processes far off stability. Renewed effort is required also for the application of statistical model reaction theory which is critical for the analysis and conversion of Coulomb dissociation and transfer reaction data into reaction rates.

Finally it will be important to carefully assess and characterize the uncertainties of experimental and theoretical data. Recently, the community has begun to develop techniques to describe and use uncertainties of astrophysical reaction rates within nuclear astrophysics models (see section 3.10).

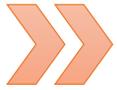

### 2.1.5 Origin of the Elements Strategic Thrust 2: Advancing models of individual nucleosynthesis processes

The understanding of individual nucleosynthesis processes in a variety of scenarios will directly benefit from advances in models of the various nuclear-driven astrophysical environment such as stars (see section 2.2), core-collapse supernovae (see section 2.3), and thermonuclear explosions (see section 2.5). However, progress in the quest for the origin of the elements requires complementary modeling efforts discussed in this section. There are many reasons for this. For some nucleosynthesis processes, the conditions needed to create new elements in accordance with observations, are not always produced naturally in current stellar models. In other instances, state of the art models do not extend to the regions, or phases, where nucleosynthesis occurs. In addition, computational limitations often prevent the use of state of the art stellar models for anything but very crude nucleosynthesis estimates. Finally, there are a number of processes where a stellar site has not been identified and site independent parameter studies are needed to complement attempts to adapt specific stellar models. Nucleosynthesis research therefore needs specific model approaches tailored to reliably predict element synthesis based on complete sets of nuclear data (see section 3.7).

**Massive Stars:** Our current understanding of the nucleosynthesis contribution of massive stars and core-collapse supernovae is based on 1D explosions induced by a parameterized piston or parameterized thermal energy deposition. Work in the past decade has highlighted the shortcomings of this approach. 3D simulations for the last phases of stellar evolutions have demonstrated the likely existence of deep convective dredge up and mixing processes that disperse the "onion shell" structure of a late star, modifying the conditions for important shell-burning nucleosynthesis reactions and for nucleosynthesis in the supernova shock front. 3D simulations of stellar explosions indicate the development of high velocity nickel "bullets" and other observed features that one dimensional simulations fail to match. Simulations of neutrino-powered explosions, using spectral neutrino transport, result in nucleosynthesis products qualitatively different in composition from either the parameterized bomb/piston nucleosynthesis models or older models using gray neutrino transport. These models have shown the importance of neutrino captures in the supernova ejecta, which significantly alter nucleosynthesis predictions and result in better agreement with observations. Physics beyond neutrino interactions, such as acoustic oscillations or magnetic field interactions may also play a role and need to be explored.

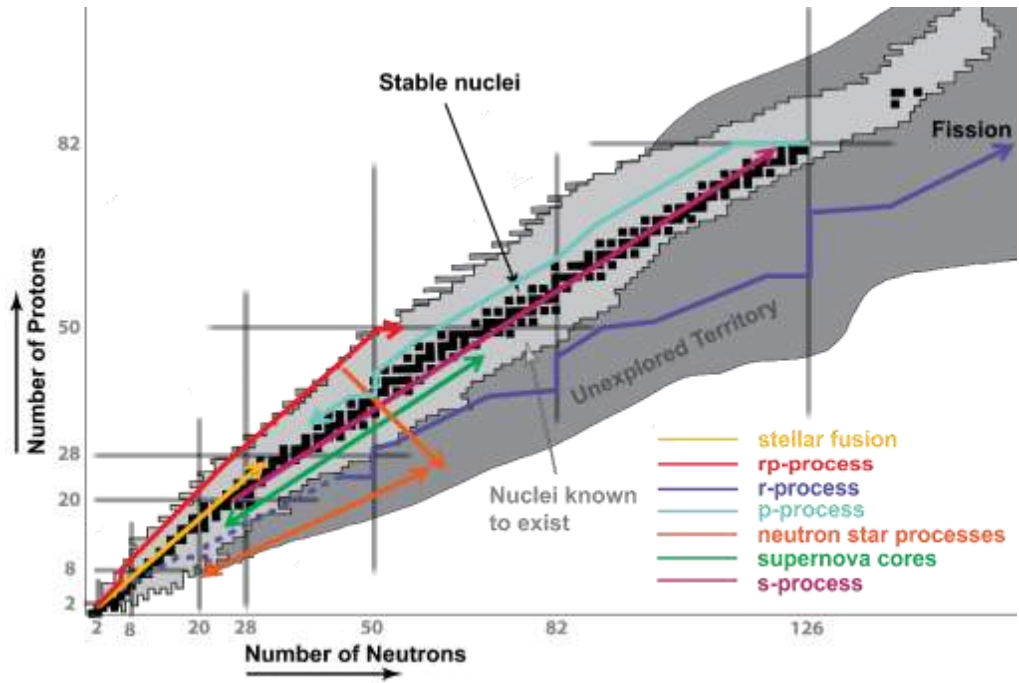

Figure 2: Schematic outline of the various nuclear reaction sequences in astrophysical environments (colored lines) on the chart of nuclides. Stable isotopes are marked as black squares. A broad range of nuclei are produced in astrophysical environments. The FRIB radioactive beam facility will provide access to the unstable nuclei that participate in many astrophysical processes, most of which have never been observed in a laboratory. Stable, gamma, and neutron beam facilities are needed to measure reactions with stable nuclei in stellar burning and the s-process. Note that many of these processes such as the νp-process, supernova core processes, and neutron star processes have only been identified in the last decade and are not well understood. The recently discovered i-process operates parallel to the s-process a few mass units towards the neutron rich side and is not yet included in this figure. (Figure from Frank Timmes)

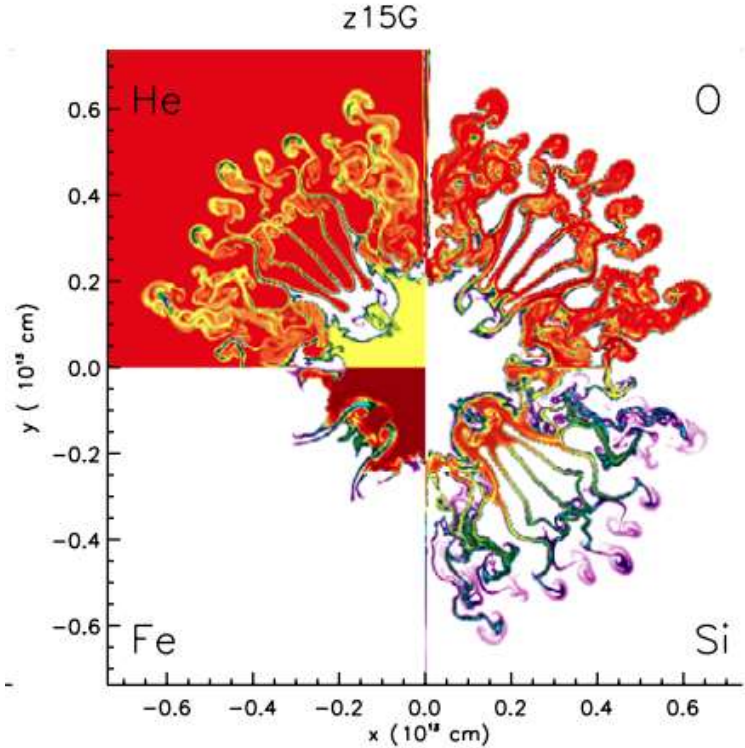

Figure 3: Model calculation of element formation (helium, oxygen, silicon, and iron) in a supernova of a first generation zero metallicity star. Multi-dimensional models can now predict not only the elements produced but also their expected distribution in the ejecta for a broad range of stars all the way to the first generation. (Figure from C. C. Joggerst et al. 2010 ApJ 709 11)

Given the importance of these findings, it is essential that we follow 3D, first-principles core collapse supernova models that employ spectral neutrino transport and other essential supernova physics through not just the explosion phase, but until the supernova shock breaks out from the surface of the star, and further until the supernova remnant forms (see section 3.8). Only from such extended models can we fully understand the impact of the core-collapse supernova engine on the isotopic composition and velocity distribution of the ejecta (see Fig. 3). As a second step, we must use these first-principles models, which will be limited in number because of their extreme computational cost, to calibrate simpler, parameterized models as a replacement for the bomb/piston models. These new parameterized models, which have yet to be identified, must be computationally frugal, to enable explorations in a wide parameter space of stellar masses, metallicity and progenitor physics, yet capture the essential impact that the neutrino-heated, convectively active central engine has on the nucleosynthesis.

**Neutrino-driven winds:** Reliable models of neutrino transport in core collapse supernovae are critical to the understanding of neutrino driven winds. These winds are expected to occur in the wake of the outgoing shock wave driving the supernova explosion as accretion onto the newly-formed proto-neutron star comes to an end. The large neutrino flux emerging from the proto-neutron star drives ejection of very hot material. Recent research shows that even small changes in the neutrino physics can alter nucleosynthesis drastically and make the difference between proton- and neutron-rich winds, the former being candidates for explaining some p-process abundances, the latter for a weak r-process. Either way, neutrino-driven winds are the prime candidates for producing heavy elements in the germanium–tellurium element range.

**r-process:** The r-process is thought to be responsible for the origin of about half of the heavy elements beyond germanium, and is the sole production site for uranium and thorium. The major challenge for the field has been to understand the nuclear physics of the extremely neutron rich exotic nuclei involved in the process (see Fig. 4), and to come up with a credible astrophysical scenario where the necessary extreme conditions (free neutron densities of the order of grams per $cm^3$ and more) occur frequently enough to explain the rather gradual heavy-element enrichment of the Galaxy observed in the abundance signatures of very metal poor stars.

There are indications from stellar abundance observations that the lighter heavy elements from strontium to maybe tellurium are produced in several different sites. There is now an opportunity to understand the origin of these elements: First, observations of ultra metal-poor stars are improving and their numbers are increasing. Second, the conditions necessary to produce these elements are less extreme than for producing heavy r-process nuclei. Nucleosynthesis studies based on current simulations show that these lighter heavy elements can be synthesized in neutrino-driven winds and in fast rotating stars. Third, in both cases the nuclear reactions and nuclei involved are not very far from stability, and most of them can be constrained in the coming years by experiments and theoretical models. It is thus crucial to identify the key nuclei and reactions that need to be measured. A recent effort has been successful in identifying the most critical nuclei with respect to mass measurements, decay studies, and neutron capture measurements. Fourth, chemical evolution models together with stellar abundance observations will reveal the relative contribution of the astrophysical sites (see section 2.8).

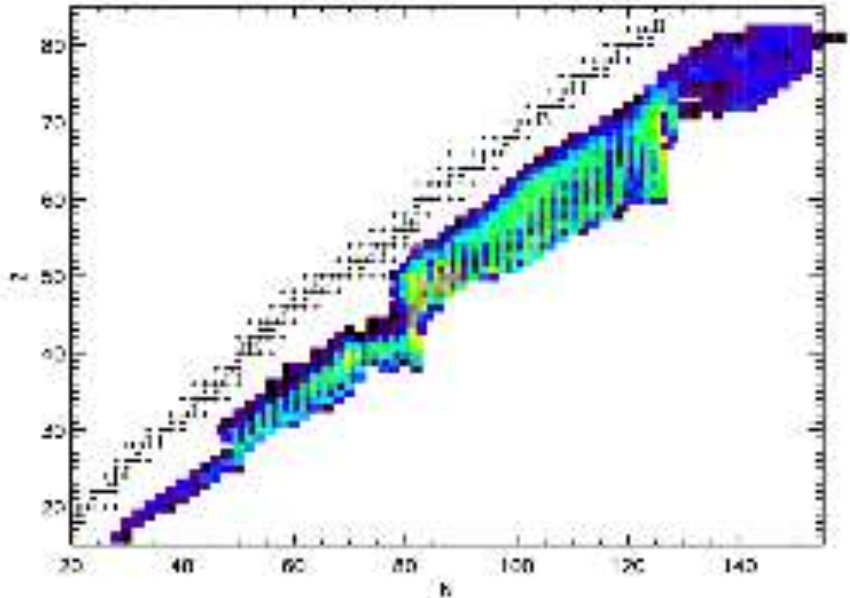

Figure 4: Nuclei produced during the rapid neutron capture process in a model calculation. The nuclei along the reaction sequence are very neutron rich unstable isotopes and most have never been observed in a laboratory. Calculations such as the one shown are now analyzed in detail to identify the critical nuclear physics uncertainties that can then be addressed with experiments at next generation rare isotope beam facilities and new theoretical models based on microscopic theories. (Figure from Rebecca Surman)

The origin of heavy r-process elements remains a challenging problem. The neutrino-driven wind was thought to be the appropriate site, however current simulations show that it is not possible to reach the extreme conditions that the r-process requires. In addition, scatter in observations of Europium abundances at low metallicities indicates that heavy r-process elements cannot be produced in every core-collapse supernova. The origin of these elements must therefore be linked to a rare event. Possibilities include neutron star mergers, jet-like supernova explosions, helium rich layers in stars irradiated by neutrinos from a collapsing core, and accretion disks around neutron stars or black holes or a narrow subset of conventional core-collapse supernovae. The next 10 years are critical to develop better models of these astrophysical environments and (with help of chemical evolution) to constrain the contribution of different r-process sites. This is especially true in the cases of neutron star mergers and collapsars, where the physical fidelity of current models trails that of the iron core and oxygen-neon core collapse models, in part because of the geometric disadvantage of these events being far removed from spherical symmetry (see section 3.8).

**s-process:** Nuclear physics data for n-capture rates (see section 3.4) as well as β-decay rates, both at stellar temperatures are also needed for the important s-process branchings that provide detailed probes of many advanced nuclear production sites in the late stages of stellar evolution. New radioactive beam facilities (see section 3.2), paired with appropriate theory effort need to address this nuclear data need. Predictions of such branchings can be combined with isotopic data from pre-solar grains (see section 3.9.10) to provide powerful validation scenarios.

**p-process:** The p-process is responsible for the origin of 35 neutron-deficient isotopes of elements in the selenium to mercury range. These isotopes are very rare in nature. The favored process is a $\gamma$-induced process that occurs when the outgoing shockwave in a core-collapse supernova passes through oxygen and neon layers of the exploding star. The sudden heating triggers removal of neutrons, protons, and $\alpha$-particles from the heavy nuclei pre-existing in these layers, producing neutron-deficient isotopes. The scenario is unavoidable in a supernovae, but seems produce insufficient nuclei in the $A$=92-98 and $150<A<165$ mass range. While the origin of the latter may be explained through (underestimated) contributions from other processes such as the s-process, neutrino processes, or even the rp-process, the former represents a major challenge for nuclear astrophysics. Either the initial composition of the O-Ne layers is very different from what has been assumed, the $\gamma$-process occurs in a different environment such as thermonuclear supernovae, or an entirely different process such as the $\nu$p-process or the rp-process are responsible for the origin of neutron deficient $A$=92-98 isotopes. Reliable nuclear physics for the p-process and the seed producing s-process are needed to solve this puzzle.

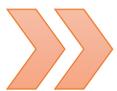

### 2.1.6  Origin of the Elements Strategic Thrust 3: Nucleosynthesis Yield Grids

Over the last decades, many important aspects of stellar evolution and associated nucleosynthesis have been investigated and clarified. A goal for the next decade will be to enhance the fidelity, accuracy and completeness of large, internally-consistent nucleosynthesis yield sets. These yield sets describe the isotopic contribution of a star to the interstellar medium during its evolution and through its final stellar wind or supernova phase as functions of stellar mass, initial stellar composition and other parameters. They are an essential input in models of galactic chemical evolution (see section 2.8) and enable the use of these models to constrain nucleosynthesis sites and galaxy formation and merger processes ("near-field cosmology"). Initial versions of such data sets have become available in recent years. However, completeness in both covered elements and mass- and metallicity-range, as well as the input physics needs to be further improved. These data sets must continually be updated to reflect our current understanding of the events that release these isotopes (see, for example, sections 2.2.8 & 2.3.8 The computational tools for such large-scale yield calculations are available (see section 3.8), and can be used to systematically explore uncertainties in all contributing areas of input physics, including in particular the two most important areas - nuclear physics rates and hydrodynamic mixing processes. This capability will ensure that future progress in both areas can be readily confronted with the increasing body of stellar abundance observations.

The new yield sets need to incorporate progress in the quantitative understanding of stellar mass loss made possible through observations and simulations. However, other basic input physics aspects, such as stellar opacities, need to be updated as well. In the latter case, the way abundance mixtures in stars are approximated is outdated and atomic physics expertise at the national labs could make critical contributions through accurate and flexible microphysics modules that are commensurate with today's computational resources (see section 3.8).

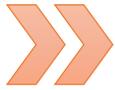

## 2.1.7  Origin of the Elements Strategic Thrust 4: Observations of Element Production Signatures

Observations of ancient, metal-poor stars found in the Milky Way provide the only available diagnostics for studying the nucleosynthesis in the early Galaxy, including the very first stars that lit up the Universe after the Big Bang. Metal-poor stars preserve the chemical fingerprint of the integrated nucleosynthesis events that occurred before these stars formed. Major breakthroughs from such observations have been the discovery of r-process stars that provide information from individual r-process events, the identification of new heavy element producing processes such as the Light Element Primary Process, and the potential discovery of the signatures of the very first supernovae after the Big Bang. However, observational data at the lowest metallicities remain sparse.

Searches for these stars have been ongoing for decades, including as part of large scale surveys that involve millions of stars, such as Hamburg-ESO, SDSS-SEGUE, and LAMOST. Ongoing large scale surveys such as APOGEE, SkyMapper, GALAH, FunnelWeb, Dark Energy Survey, and Gaia will provide abundance and metal content constraints on many more stars (see section 3.9.4 & 3.9.5 It will be important to significantly increase the volume of detailed stellar abundance data at the lowest metallicities through spectroscopic followup studies of promising candidate stars with the world's largest telescopes, such as Magellan, the VLT, Subaru, Gemini, Keck, and HST. A larger sample is critical to discover the full range of nucleosynthesis processes operating in the early Galaxy, and to determine event-to-event variations that provide clues on the nature of the events. These data will also provide large, statistically significant, samples of stellar abundances for the smaller number of key elements that provide information on the compositional evolution of the early Galaxy and are essential for comparison with modern chemical evolution models (see section 2.8)

While the detailed spectroscopic observations of the chemical evolution of the elemental composition of our Galaxy constitute a major accomplishment for the field, ambiguities remain because of the lack of isotopic information. For example, the observed signatures of a new process producing strontium, yttrium, and zirconium can be explained by a novel proton capture process, or a weak r-process – while isotopic composition would be very different, both types of processes can produce similar elemental abundance patterns.

Observations of isotopic abundances are therefore extremely important. In stellar spectra, isotopes can in principle be distinguished through the isotope shift of spectral lines, for example, deuterium abundances have frequently been reported. However, the shifts are very small for heavier elements, and so far only very crude determinations have been made for a few heavy elements such as europium or barium. The observation of characteristic γ-rays from long-lived radioactive nuclides can provide such isotopic abundance information (see section 3.9.7). However, there are only a few isotopes that have half-lives and decay properties suitable for astronomical observation. Furthermore, the field is currently lacking an observatory with enough sensitivity. COMPTEL and INTEGRAL have detected $^{60}$Fe and $^{26}$Al, which were breakthroughs for the field, but both isotopes are too long-lived to be uniquely attributed to specific nucleosynthesis events. COMPTEL made a pioneering detection of $^{44}$Ti decay radiation in one supernova remnant, which has been consolidated by INTEGRAL and led to NuSTAR's unique detailed image of the distribution of $^{44}$Ti dispersed by a supernova (see Fig. 10). This provided extremely valuable constraints on supernova models, and INTEGRAL with followups from NUSTAR are beginning to provide additional $^{44}$Ti data on a wider range of remnants, taking advantage of the particularly low γ-energies emitted by the decay of this isotope.

On the other hand, pre-solar grains can provide precise isotopic information about the environment where they formed (see section 3.9.10). These grains are thought to have formed in extra-solar stellar environments, have travelled through space, and have been incorporated into the pre-solar nebula making it possible to find them today in primitive meteorites. Information is limited to stellar environments that do form such grains, and identifying the origin of the grains is difficult and requires initial knowledge of nucleosynthetic signatures and the chemistry of grain formation. The approach has been very successful in construing s-process nucleosynthesis in AGB stars, where observations support grain formation in stellar winds, and abundance signatures clearly indicate the origin of the grains. Grains from supernovae and novae have also been found, though their identification is less certain.

## 2.1.8 Impact on other areas in nuclear astrophysics

Nucleosynthesis in connection with abundance observations creates a pathway to validate models of stars and stellar explosions. Often nucleosynthesis processes probe conditions deep in a stellar environment where no other probes are available and therefore provide unique astrophysical insights. To use nucleosynthesis as a stellar probe requires a good understanding of the relevant nuclear physics. Examples of future possibilities include CNO neutrinos, which would provide an independent tool to determine the solar metallicity, and the s-process branching point abundances, which can be used to characterize temperature and density conditions, as well as mixing processes, in AGB stars. In connection with chemical evolution, nucleosynthesis offers the opportunity to probe birth and evolution of our Galaxy. With observations of the composition of the universe at very early times, for example, in very old stars, or the interstellar medium of galaxies at high redshift, the Big Bang and the cosmological environment of the first stars can be probed. Nucleosynthesis also affects the formation and evolution of planets. For example, type and properties of planets depend on the chemical composition of the host star and evolve with the chemical evolution of the universe. The incorporation of long lived radioactive nuclei strongly affects evolution of planets, defines the present geodynamic conditions of the Earth, and may make the difference between water and desert worlds.

## 2.2 How do stars work?

### 2.2.1 Introduction for non experts

Stars are the most obvious manifestation of nuclear processes in the universe, deriving their power exclusively from an enormous natural nuclear fusion reactor in their cores. During the course of their evolution they produce the majority of elements found in nature. Despite the first stellar models being created in the 19th century, and the nuclear reaction sequences for hydrogen burning being delineated by the mid 20th century, stars are still poorly understood. This begins with the best studied star, our Sun, where the amount of metals in its core is still an open question. One difficulty in understanding stars are the uncertain rates of the very slow nuclear reactions in their interior. The slowness of the reactions enables stars to exist for extended periods of time, but also makes experimental measurements extraordinarily difficult. Other fundamental, yet poorly understood, aspects of stars are convection, rotation, magnetic fields, and mass loss through winds. It is now becoming clear that these effects cannot be adequately modeled in spherical symmetry (one spatial dimension). As multi-dimensional models are developed, validation of the more complex models using nucleosynthesis and abundance observations becomes critical. This requires a much improved understanding of the underlying nuclear reaction rates, which are determined by the low energy nuclear cross sections and the plasma conditions in the stellar interior.

### 2.2.2 Current open questions

- What are the rates of the capture reactions (e.g. $^3$He$(\alpha,\gamma)^7$Be, $^{12}$C$(\alpha,\gamma)^{16}$O, $^{14}$N$(p,\gamma)^{15}$O, $^{17}$O$(p,\gamma)^{18}$F), $^{25}$Mg$(p,\gamma)^{26}$Al, and $^{26}$Al$(p,\gamma)^{27}$Si)) heavy-ion reactions (e.g., $^{12}$C+$^{12}$C, $^{12}$C+$^{16}$O, $^{16}$O+$^{16}$O), and neutron source and poison reactions (e.g., $^{13}$C$(\alpha,n)^{16}$O, $^{22}$Ne+$\alpha$, $^{17}$O+$\alpha$) in stars, and what are the best means to determine these rates at stellar energies?

- Is there a truly unique nucleosynthetic signature of the first stars, and can we infer the properties of the first stars from low-metallicity stellar abundances?

- How do low metallicity stars, including the first stars, evolve and how do convective processes impact their evolution and nucleosynthesis?

- What is the composition and low energy neutrino flux of the Sun?

- What is the nucleosynthetic production of stars as a function of mass, metallicity, and rate of rotation and how much star-to-star variation is exhibited?

- How do 7-10 solar mass stars evolve and what delineates the boundary between the most massive white dwarf remnants and the lowest mass (and therefore most frequent) supernova explosions?

- How does the binary nature of many stellar systems affect stellar evolution and nucleosynthesis?

- What is the origin of carbon enhanced metal poor stars that do not show any signatures of the s-process?

- How do massive stars lose mass, and how much do they lose?

### 2.2.3 Context

One of the greatest accomplishments of nuclear astrophysics is the precision prediction of the solar neutrino flux owing to detailed laboratory measurements of the relevant nuclear reaction rates. This enabled, in connection with an accurate solar model, the discovery of neutrino oscillations. Beyond this discovery, the quest to improve the precision of the solar neutrino flux measurements and to detect neutrinos from other nuclear processes in the Sun continues. In addition to the improved 3% precision of the $^8$B neutrino flux measurement by SNO, BOREXINO has now measured the $^7$Be neutrino flux with 4.6% precision, and has detected first hints of pep neutrinos. The best 90% confidence upper limit on hep neutrinos from SNO is close to the predicted flux. The accuracy of the corresponding standard solar model flux predictions has steadily improved due to continued progress in measurements of the nuclear reaction rates.

In that context, improved reaction models have been developed which allow a much better interpretation of the reaction mechanisms that contribute to the stellar reaction rate. This significantly reduces the uncertainties in the theoretical extrapolation of the reaction cross sections towards the stellar energy range.

New initiatives have emerged to study the impact of stellar plasma conditions on the reaction rate, an aspect that has been experimentally neglected for decades but will affect in particular fusion rates in high density environments. First experiments at the Nation's high density plasma facilities such as OMEGA and NIF have been successfully performed to directly measure the plasma impact on the fusion rate of light elements.

Over the past decade, several groups have significantly expanded capabilities in stellar evolution and nucleosynthesis modeling. One-dimensional stellar evolution simulations are now routine for all phases of the evolution of stars that are suitable for spherically symmetric simulations. Simulations have enabled many detailed investigations of how nuclear physics uncertainties propagate through multi-physics simulation codes and impact astronomical observables. These investigations have demonstrated the importance of many nuclear reactions, including the $^{14}$N +p, triple-$\alpha$ and $^{12}$C+$\alpha$ reactions, as well as $^{12}$C+$^{12}$C fusion and show in detail the significant, and sometimes dominant, contribution of nuclear physics uncertainties to the overall uncertainty budget. At the same time, progress in accelerators and experimental techniques have enabled many new measurements of these and other reaction rates that have, in some cases, significantly reduced uncertainties. The resulting nucleosynthesis signatures have been successfully compared with the compositions found in certain metal poor stars, providing validation for the stellar models. For example, through this work it was possible to verify that those C-enhanced metal-poor stars which also exhibit enhancements of slow-neutron capture elements were polluted by a former giant star companion that is now a white dwarf.

Considerable effort has been devoted in the past decade towards calculating models for low-metallicity stars, all the way down to zero metal content (i.e. those stars that formed out of pure Big Bang material). As a result, first predictions of the synthesis of elements in the first generations of stars have developed to the point where the confrontation with observed abundances of metal-poor halo or extra-galactic stars starts to provide intriguing constraints on cosmological structure formation models.

The quantitative interpretation of isotopic ratios from measurements of pre-solar grains has also made tremendous progress. This observational access to the nucleosynthesis and mixing processes in the interiors of stars and stellar explosions is complementary to astronomical observations and provides extremely powerful constraints on the fluid mixing and nuclear physics in the interior of stars that directly determine nucleosynthesis. The role of rotation and magnetic fields is now much better understood, although significant uncertainty in this area remains.

Several groups are now investigating convection and convective mixing in stellar interiors through three-dimensional hydrodynamic simulations. This new development has started in earnest only in the past decade, and takes full advantage of the substantially increasing computational resources. The simulation-based findings are converging to the conclusion that the conventional treatment of convection via mixing-length theory (MLT) in stellar evolution models urgently needs to be updated. In particular, the important behavior of fluid motion at convective boundaries is poorly described by MLT. This poses significant limitations to further progress toward predictive simulations of the origin of the elements in stars.

In the coming decade, there is an opportunity for significant advances in stellar models. Multidimensional computational hydrodynamics efforts have now reached a level that they can inform spherically symmetric (1D) stellar models, offering the opportunity to develop hybrid models that use results from 3D models to enhance the fidelity of 1D models and adequately model critical aspects of stars such as mixing, convection, and improve our understanding of the impact of rotation and magnetic field. This is especially important for models of low metallicity stars, which to date have been largely ignored in multi-dimensional studies. At the same time, new experimental capabilities offer the opportunity for significant improvements in the underlying nuclear physics.

As a byproduct of several transient-method based exoplanet search missions a significant amount of data for asteroseismology has been collected over the past decade, notably through the CoRoT and KEPLER missions (see section 3.9.11). This data provides another independent avenue to constrain stellar and nuclear astrophysics simulations. In this way, our understanding of the internal rotation profiles of giant stars and the role of turbulent mixing and diffusion in stars has been significantly improved.

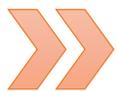

### 2.2.4  Stars Strategic Thrust 1: Constraining the rates of nuclear reactions in stars

The structure and evolution of stars depends very sensitively on the rates of key nuclear reactions. The reaction rates define the timescale for the stellar evolution, dictate the energy production rate, and determine the abundance distribution of seed and fuel for the next burning stage. Recent work has reinforced these well known facts by demonstrating that, in some cases, even changes of reaction rates at the 10% level can dramatically change the location and duration of various convection zones inside stars, affecting the internal structure of the burning zones. Some of the poorly known nuclear reactions that are of particular importance for stellar evolution are capture and fusion reactions (e.g., $^3\text{He}(\alpha,\gamma)^7\text{Be}$, $^{14}\text{N}(p,\gamma)^{15}\text{O}$, $^{17}\text{O}(p,\gamma)^{18}\text{F}$, $^{12}\text{C}(\alpha,\gamma)^{16}\text{O}$) and heavy-ion reactions (e.g., $^{12}\text{C}+^{12}\text{C}$ - see Fig. 5, $^{12}\text{C}+^{16}\text{O}$, $^{16}\text{O}+^{16}\text{O}$) that characterize the subsequent evolutionary stages of massive stars. A number of these reactions involving stable targets (e.g., $^{14}\text{N}(p,\gamma)^{15}\text{O}$, $^{12}\text{C}(\alpha,\gamma)^{16}\text{O}$, $^{13}\text{C}(\alpha,n)^{16}\text{O}$, and $^{22}\text{Ne}+\alpha)^{25}\text{Mg}$, have a tremendous impact on stellar explosions, since these reactions determine the seed abundance distribution for core collapse, thermonuclear runaway, and the p-process. Other reactions, such as the $^{12}\text{C}+^{12}\text{C}$ fusion process, dictate the ignition conditions for thermonuclear explosions in dense environments such as white dwarf core or neutron star crust. Dedicated programs to enable more sensitive measurements using new techniques (see section 3.1) are needed to experimentally determine these and many other stellar reaction rates.

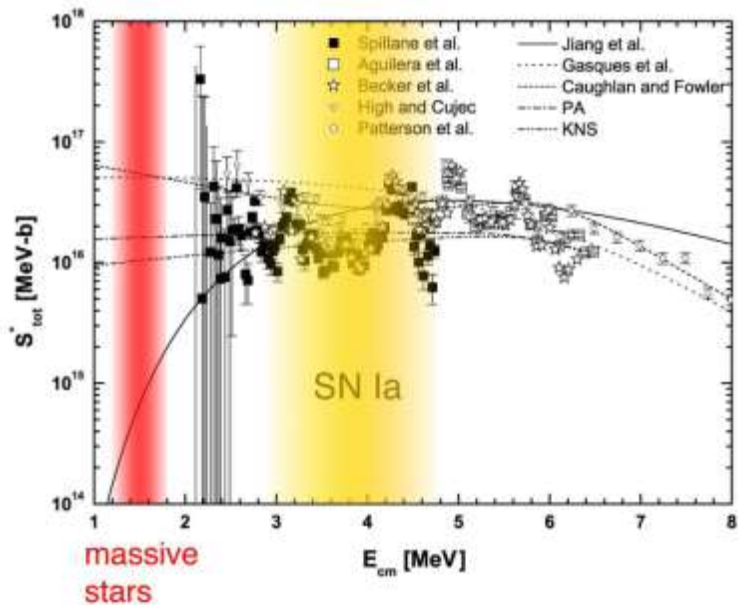

Figure 5: Carbon fusion cross section as a function of energy from various measurements (data points) and theoretical extrapolations to low energy (lines). The energy ranges for peak temperatures in massive stars and type Ia supernovae are colored. For massive stars, the measurement uncertainties and the theoretical uncertainties span several orders of magnitude. This is a key problem in nuclear astrophysics that needs to be addressed in the coming decade. (Figure adapted from F. Strieder et al. Journal of Physics G Nucl. Part. Phys. 35 (2008) 014009)

## 2.2.5 Stars Strategic Thrust 2: Fundamental Advances in Stellar Models

Major progress in modeling convection-induced mixing in all its forms during various stages of stellar evolution can be expected to derive from high-fidelity multi-dimensional simulations that are now becoming possible (see section 3.8). Such simulations are tackling the shell convective zones of C- and O-burning in pre-supernova massive stars and of He-shell flashes in late stages of low- and intermediate mass stars as well as the simmering phase in supernova type Ia explosions. The emerging simulation capabilities provide a huge potential to significantly improve the accuracy of stellar models, and provide an indispensable complement to the upcoming new nuclear physics data for nucleosynthesis away from the valley of stability. Simulations may be performed on new systems, like Blue Waters, on grids with as many as a trillion cells, which has been shown to be more than sufficient to converge key properties of interior convection such as mixing at and across convective boundaries. However, fully demonstrating numerical convergence for all applications, and balancing microphysics detail with the need to manage computational cost in resolved hydrodynamic simulations will remain an issue in this decade. However, with appropriate effort, the basic properties of the stellar interior convection and nuclear burning problem in the late stages of stellar evolution can now be resolved.

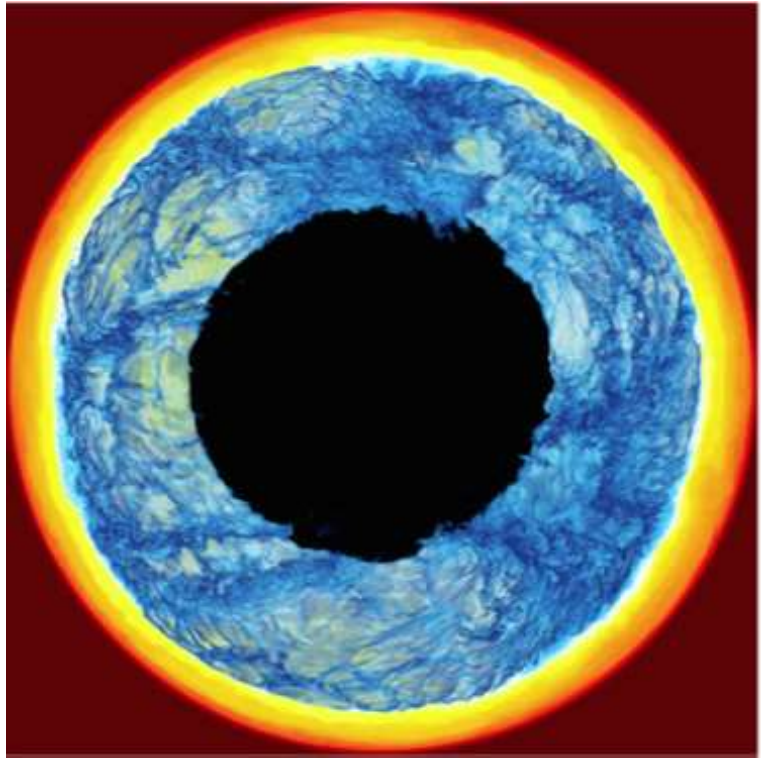

Figure 6: 3D-Hydrodynamics simulation of the inside of a red giant star, where hydrogen rich fluid is ingested via entrainment into a helium rich layer that undergoes an explosion driven by helium fusion reactions (a helium shell flash). The hydrogen reacts with carbon that is present in the helium rich layer, resulting in the production of neutrons that drive novel nucleosynthesis processes. This simulation ran on NSF Blue Waters on 443,232 CPU cores, roughly half the machine, at a sustained rate of 0.42 $Pflop/s$ with 64-bit precision and generating 56.5 TB of data. At this computing rate, it took about 3 min of real time to simulate 1 min of time for the star. Figure from Herwig F, Woodward PR, Lin P-H, et al. (2014) ApJ 792 L3.

Such progress would enable generating more realistic 3D initial models for core-collapse supernova simulations. Another important area for progress would be a thorough investigation of the H-He convective-reactive phases of evolution that are especially prominent in stellar modeling attempts for the lowest metallicity stellar generations (see Fig. 3). In such events, protons and primary $^{12}C$ from a He-burning layer are reacting and releasing energy on the convective time scale ($\sim$ 5 - 60 minutes). Spherically symmetric convection theories (such as the MLT, see above) have been shown to be unreliable under these conditions.

Even in standard convection scenarios, assumptions in the presently-adopted, spherically symmetric, convection theory have to be revisited through simulations of stellar hydrodynamics. The goal is that stellar evolution simulations with such enhanced mixing models will provide more reliable nucleosynthesis predictions for nuclear astrophysics. In the context of mixing processes in stars, rotation also plays an important role. Multidimensional models of fast rotating stars from different groups are needed to investigate this issue.

While the evolution of low- and intermediate mass stars will remain a focus area of stellar evolution, the mass range between 7 and 10 solar masses is most in need of attention at the present time. Some recent investigations have addressed the shell burning properties of super-AGB stars and shown that most uncertainties here result from a lack of understanding of convection and mass loss. The transition from the initial mass range that forms one white dwarfs to the initial masses that provide the progenitors of the lowest-mass supernova remains poorly charted territory. Basic properties of the evolution of convective shells inside electron-degenerate cores, the interaction of turbulence, unusual nuclear reaction chains under extreme conditions and neutrino losses form a delicate balance that allows little room for error. This regime is particularly important because the initial-mass function favors lower-mass supernova progenitors occurring in the greatest numbers. These stars have been suggested to be the nuclear production site for a number of processes, such as the r-process or other high neutron-density processes associated with convective-reactive regimes, which may play a role in some of the poorly understood observational phenomena of the metal-poor universe, such as the C-enhanced metal poor stars with s- and r-process signatures. Light-curves from supernova from this mass range may be constrained by time-domain surveys (see section 3.9.4) and provide pivotal constraints for the underlying physics responsible for the evolutionary fate of stars in this mass range.

Many stars are binaries and many will at some point in their lives interact in one way or another with their companion. These interactions, such as tidal interactions, the common envelope phase, mass transfer and double degenerate mergers, involve hydrodynamic processes far from spherically symmetry, which pose significant challenges to our understanding. Binary evolution may be the most important source of rapidly rotating stars, and despite promising progress in recent years, the quantitative understanding of the detailed evolution, nucleosynthesis and final death of such interacting binaries remains a challenge.

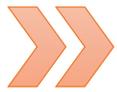

### 2.2.6 Stars Strategic Thrust 3: Nucleosynthesis as validation tool

With major advances in stellar models within reach, and with the expected dramatic increase in model complexity, model validation will take center stage. Direct observations offer one possible pathway. Data from KEPLER, and soon GAIA, are revolutionizing the field (see section 3.9.4). For a very limited set of stars, it is now possible to obtain some constraints on their interior structure with asteroseismology (see section 3.9.11). However, by far the most powerful tool to validate stellar processes are their nucleosynthetic signatures. The available data on the composition of metal poor stars is growing rapidly, providing information on individual nucleosynthesis events and their evolution over time (see sections 3.9.4 and 3.9.5). A prerequisite for this approach are precise nuclear data for the processes involved. Given the increased need for model validation, advances in the understanding of stellar reaction rates are particularly urgent (see section 3.1).

In addition, it will be necessary to establish sound validation procedures for nucleosynthesis predictions so that improvements in modeling accuracy can be measured and assessed. Such a validation framework needs to apply simultaneously the constraints from well-defined and orthogonal observational tests that are directly relevant to the various aspects of input physics. The framework would include multi-wavelength stellar observational data from well-understood astronomical sources, as well as isotopic data from grains. The goal should be to test individual physics models, e.g. of nuclear reaction rate sets or mixing physics simultaneously in different stellar environments with a variety of observational tests to allow breaking the degeneracy that plagues most individual constraints. Without such a systematic validation effort, it is difficult to see how nucleosynthesis predictions based on different uncertain physics assumptions can be evaluated in order to fully support the interpretation of observational data for a more general range of astronomical questions.

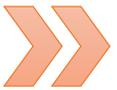

### 2.2.7 Stars Strategic Thrust 4: Solar Neutrinos

Since the detection of the first solar neutrinos in 1965, the field has made formidable progress. With the understanding of neutrino oscillations, and the construction of ever more sensitive detectors the field is now transitioning into a precision era, where solar neutron detection can be used as a tool to precisely understand the interior structure of the sun as well as the physics of neutrinos (see Fig. 7). Recent milestones were the first detection of neutrinos from the rare pep process and a 4.6% precision measurement of the $^7$Be neutrino flux by BOREXINO. SNO has presented its final data on $^8$B neutrinos, which already strongly constrains the solar metallicity and $\tau_{12}$.

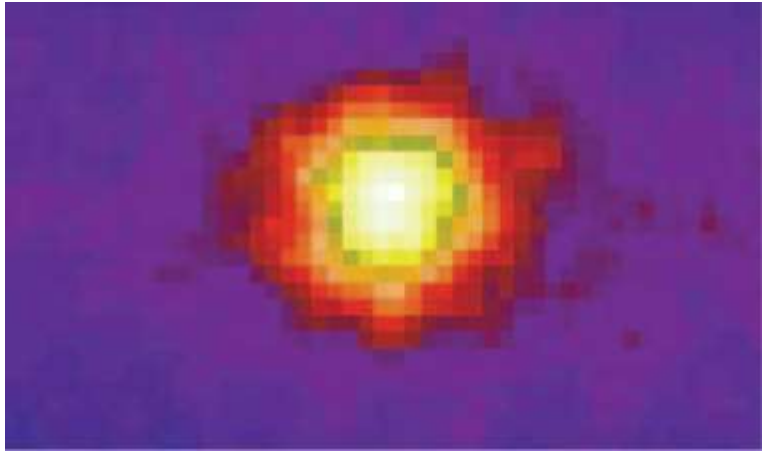

Figure 7: Image of the sun as observed with neutrinos, ushering in the multi-messenger era in astronomy where stellar objects are observed with multiple types of radiation providing complementary pieces of information. Unlike sun light, the observed neutrino radiation does not originate from the surface of the sun but offers a direct view of the nuclear reactions operating in the solar core. Future observations of low energy neutrinos with underground neutrino detectors such as BOREXINO, SNO, or Superkamiokande may reveal the concentration of carbon, oxygen, and nitrogen in the core of the sun, provided the relevant nuclear reactions can be measured with sufficient precision. Image from Superkamiokande Collaboration.

Open questions that remain to be addressed in the future include (1) What is the solar core metallicity? (2) What is the total solar neutrino luminosity (for which, measure of pp and pep neutrinos will be required)? (3) Are there non-standard neutrino oscillations? (4) What is the Sun's CNO neutrino flux? (5) Can we see a day/night effect due to interactions with matter?

Major new opportunities for the next decade emerge from advanced neutrino detection experiments (see section 3.9.13). Superkamiokande and Borexino continue to operate. Plans at Borexino include improvements in purity to reduce backgrounds and enable the first detection of neutrinos from the CNO cycle. SNO+ is nearing completion and promises detection of pep and CNO neutrinos. The detection of CNO neutrinos, together with further constraints from precision measurements of the $^8$B neutrino flux, offers the most promising pathway to determine the metal content of the sun, which is still under debate. The LENS and CLEAN detector projects, which are currently in the exploratory stage with small scale prototypes, have the potential to measure the pp solar neutrino flux and therefore the total solar neutrino luminosity to better than a few percent.

To exploit these opportunities, the critical task for nuclear astrophysics is to improve the precision of the underlying nuclear reaction rates that connect neutrino observations with solar and neutrino physics (see section 3.1). Important reactions include the $^3$He+$^4$He reaction, and the radiative capture reactions in the CNO cycle that affect the abundance of the neutrino emitters $^{13}$N, $^{15}$O, and $^{17}$F. In addition to providing better experimental data, a consistent theoretical analysis of various data sets is crucial to better understand the strength of the various reaction components and mechanisms that affect the low energy extrapolation of the laboratory reaction cross sections.

## 2.2.8  Impact on other areas in nuclear astrophysics

Understanding the lives of stars is an essential component of understanding the origin of the elements. Beyond their direct contributions of nucleosynthesis products to the interstellar medium, stars are also the progenitors of supernovae, and supernova models are sensitive to the structure of the progenitor star resulting from its evolution prior to the explosion. This applies to core collapse supernovae and to thermonuclear supernovae, the latter being sensitive to the structure of the exploding white dwarf, which depends on the nuclear processes that occurred in the early evolutionary stages of the white dwarf's progenitor star. Understanding all types of supernovae therefore requires reliable progenitors with better nuclear physics and astrophysics. The Sun plays a special role as a neutrino laboratory, providing insights into particle physics that also impact our understanding of the role of neutrinos in nucleosynthesis, supernovae, and neutron stars.

## 2.3  How do Core-Collapse Supernovae and Long Gamma Ray Bursts Explode?

### 2.3.1  Introduction for non experts

Massive stars develop in their centers a core of nuclear ashes. This core eventually collapses under its own weight. What happens then is subject of intense research efforts. Observations indicate that in many, but not all, cases a supernova explosion is triggered. Just how a core collapse initiates an explosion remains unknown, and so are the outcomes of the core collapse event for various types of stars. Possibilities include the formation of black holes or neutron stars, bright or faint supernovae, or long gamma ray bursts. The understanding of core collapse supernovae is of central importance in nuclear astrophysics, as they are the main sites responsible for the origin of the elements, and as their energy output is the primary driver of the galactic mixing processes that are fundamental for chemical evolution.

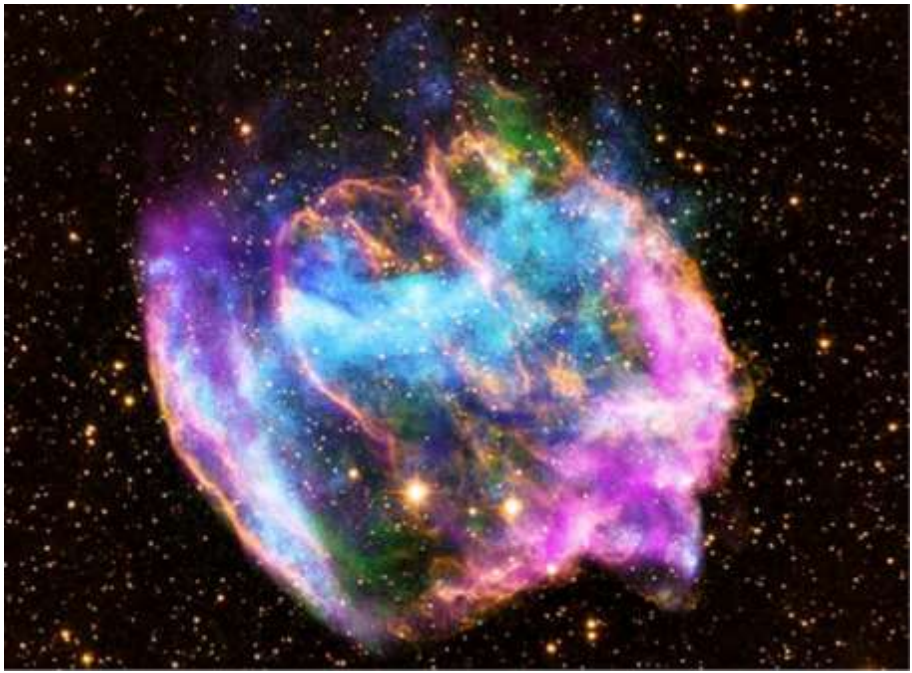

Figure 8: Supernova remnant W49B. One of the major goals of the field is to understand the mechanisms by which stars explode as supernovae. Chandra X-ray data reveal the distribution of elements in the ejecta (iron in blue, silicon in green) that can be compared to infrared (yellow) and radio data (pink). Comparison with multi-dimensional supernova models will enable the use nucleosynthesis as a diagnostics of what happens inside a supernova. (X-ray: NASA/CXC/MIT/L.Lopez et al; Infrared: Palomar; Radio: NSF/NRAO/VLA)

### 2.3.2  Current open questions

- What is the core collapse supernova (CC SN) mechanism?

- Which stars become failed, subluminous, normal, or hyper CC SNe, and which ones result in long Gamma Ray Bursts (GRB)?

- What is the neutrino and gravitational wave signal from CC SNe?

- Are CC SNe the sites of the r- or LEPP nucleosynthesis processes?

### 2.3.3  Context

There has been much recent progress in the modeling of massive star collapse. The community agrees that spherically-symmetric (1D) models of core-collapse supernovae (CCSNe) do not lead to explosions regardless of their level of sophistication. The challenge is to find a mechanism that is able to transfer about a percent of the enormous energy released in the collapse to the outer layers of the infalling matter, with the remainder of the energy being emitted as neutrinos. Various groups agree that multi-dimensional effects, in particular convection and the standing accretion shock instability (SASI), are crucial for driving an explosion. Two-dimensional (2D; axisymmetric) simulations with spectral (i.e., energy-dependent) neutrino transport are now available and have demonstrated that the explosion mechanism based on neutrino heating can work for 2D CCSNe, if all relevant multi-physics components are included, in particular, Boltzmann neutrino transport, general relativity, and a detailed treatment of electron capture and a neutrino interactions with coupling of energy bins. However, resulting explosion energies are generally lower than observed.

Progress has also been made in identifying key physics ingredients that affect supernova explosion models: instabilities, neutrino-matter interactions, neutrino oscillations and transport, general relativity effects, progenitor, nuclear equation of state, and magnetic fields. The goal of the next decade is to improve the understanding of this input physics, and to find ways to incorporate the critical aspects into the most sophisticated supernova models.

Astronomical observations of core-collapse supernovae are rich in variety, and extend over many messengers, from radio through infrared/dust, optical, X-ray, gamma-ray line, and cosmic-ray and even neutrino observations (see Fig. 8). Furthermore, observational constraints on core-collapse properties derive from global/cosmic supernova rate or star formation rates, or from compositional studies in galaxy and star clusters.

The connection of long GRBs and extreme CCSNe is now well established observationally, but how and under which conditions a GRB central engine forms in a dying massive star is uncertain. Modeling such extreme events and understanding their nucleosynthetic consequences is tremendously difficult. It will require bringing together CCSN simulation techniques with the methods of numerical relativity to address the general-relativistic dynamics associated with black hole formation, rapid rotation, and ultra-strong magnetic fields important in GRB central engines.

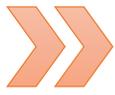

### 2.3.4 CCSNe Strategic Thrust 1: Towards adequate 3D Models

Computational advances are expected to make true 3D simulations of core collapse supernovae possible in the next decade. Studies indicate that this will be an important if not decisive step towards identifying the supernova explosion mechanism. Computational advances will also allow modelers to implement the full underlying nuclear physics that significantly affects supernova model characteristics and observables (see Fig. 9).

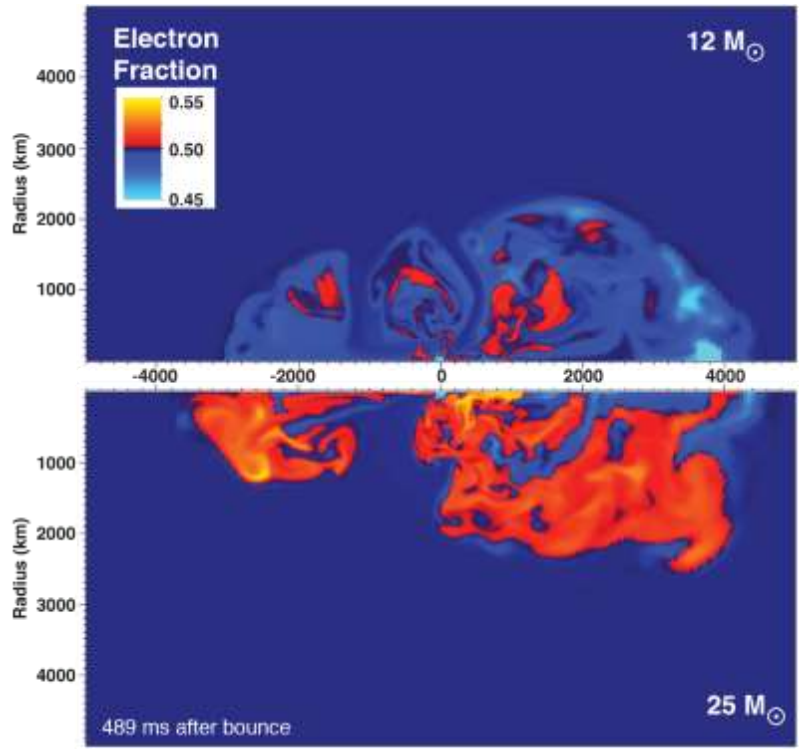

Figure 9: Neutron richness (Electron Fraction $Y_e$) of the material near the center of a core collapse supernovae explosion for the explosion of a 12 (upper panel) and 25 (lower panel) solar mass star. Neutron richness is a critical parameter that determines the elements that are synthesized. The distribution is clearly not spherically symmetric and indicates that multi-dimensional models are needed to reliably determine the production of elements in supernovae. (Figure from Raph Hix)

First 3D hydrodynamical simulations of core-collapse supernovae with a simple neutrino transport are becoming available. One of the big challenges of next decade is to improve the microphysics in such simulations which will provide new insights about supernova explosions. Taking detailed first-principles neutrino radiation-hydrodynamics simulations to three dimensions (3D) is a major challenge, but will be necessary to robustly model the turbulence behind the shock and fully assess the roles of convection, Standing Accretion Shock Instability (SASI), magnetic fields, and rotation. While first 3D simulations with gray (i.e. energy-averaged) transport have been carried out, spectral transport simulations will be crucial to ascertain the explosion mechanism, its multi-messenger signatures, and nucleosynthetic impact. Developing, running, and validating these simulations will require a broader workforce with interdisciplinary training, access to the necessary computational resources (a single spectral-transport simulation requires ~200 million CPU hours), and code comparisons between groups (see section 3.8).

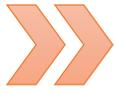

### 2.3.5 CCSNe Strategic Thrust 2: Improved Nuclear Physics

Neutrino interactions, electron capture reactions, high temperature partition functions of nuclei and the nuclear equation of state play an important role in CC SNe. CC SNe are primarily neutrino explosions (99% of the energy is emitted as a neutrino burst). Understanding the properties and interactions of neutrinos with nuclei, electrons, or other neutrinos, is essential for interpreting the observable neutrino signal from CC SNe. Neutrino interactions also play a central role in transferring energy to the ejecta and in modifying the ejecta composition, especially the neutron richness, enabling (or prohibiting) particular nucleosynthesis processes such as the νp-process or the r-process (see section 2.1). Fundamental neutrino properties such as mixing parameters, mass hierarchy, and the existence or non existence of additional sterile species, have shown to influence supernova explosions and nucleosynthesis. Of particular importance is the flavor evolution of the neutrino flux throughout the supernova, and during travel to terrestrial neutrino detectors, as neutrino energy transport, deposition, and detection is flavor dependent. Studies show that despite the small neutrino mass differences collective flavor transformation can take place deep in the supernova.

The nuclear equation of state directly determines the dynamics of the core bounce and the resulting outward bound shock wave that is thought to be the initial step of the explosion mechanism. It also affects neutrino interactions within the proto-neutron star, and therefore influences neutrino energetics, which is a supernova observable and a key ingredient in the explosion mechanism and nucleosynthesis (see sections 2.6, 3.2, and 3.6).

Electron capture reactions play an important role during the initial collapse phase, where electrons provide the pressure support that controls the core collapse (see Fig. 2). Improved electron capture rates and realistic estimates of remaining nuclear uncertainties are now available for the important stable nuclei up to the iron/nickel region. This has been achieved by a coordinated effort of systematic shell model calculations and targeted experiments that measure cross sections for charge exchange. Total absorption spectroscopy experiments can be used to constrain the electron-capture rates far from stability, not accessible to charge exchange reactions. The challenge for the next decade is to extend this improvement to heavier nuclei and nuclei far from stability, which are also present in significant quantities in collapsing supernova cores (see section 3.2).

In the next decade many groups will begin implementing true higher-order multi-dimensional neutrino transport schemes. These higher order schemes will be able to use more a more detailed description of neutrino interactions and their coupling to the nuclear equation of state. Advances in computing power will allow to more accurately follow compositional changes and to implement more detailed weak interaction processes with various nuclei. There is therefore a growing need for better models of neutrino interactions, weak interaction processes with nuclei, and the nuclear equation of state.

Improved nuclear physics is also needed in the stellar models used to calculate the structure of the pre-supernova progenitor stars (see section 2.2).

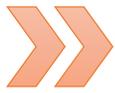

### 2.3.6 CCSNe Strategic Thrust 3: More realistic progenitor models

A major deficiency of current CCSN models is the reliance on spherically-symmetric models of the presupernova star as initial condition for collapse simulations. Progress in the 3D modeling of convective burning suggests that presupernova stellar structure may be significantly different from what current purely 1D models predict. In addition, the rotational structure of the pre-supernova star, its magnetic fields, and its mass, determined by mass loss processes during its previous evolution, turn out to be crucial parameters determining the outcome of the supernova explosion. Robust 3D CCSN models will thus crucially depend on advances in multi-D stellar evolution (see section 2.2).

Connected to this is the challenge to predict the ultimate outcome of stellar collapse (and its various observational signatures, including nucleosynthetic yields) as a function of zero-age-main-sequence mass, metallicity, and rotation. This will require the generation of an extensive grid of stellar pre-supernova models as a function of various initial parameters, and taking into account astrophysical and nuclear uncertainties.

An often neglected aspect of this problem is the fact that most stars are part of a stellar binary system. The evolution of such stellar binary systems, such as common envelope phases where one star is located inside the other star, and various mass transfer episodes must therefore be taken into account when determining the pre-supernova properties of a star.

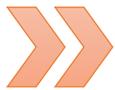

### 2.3.7 CCSNe Strategic Thrust 4: Multi-messenger observations

Multi-messenger studies imply observing supernovae not only with photons at different wave lengths but also with gravitational waves and neutrinos. The combination of these observations and their associated communities is key to make advance in understanding these complicated and fascinating environments.

The advent of wide-field optical surveys, with spectroscopic follow-up, has greatly increased the observational catalog of supernovae (thermonuclear SNe and CCSNe), revealing peculiar events and rare features. This includes timely observations of shock breakout, which provides powerful constraints on the progenitor star's structure. Comparisons with archival data from Hubble and other sources has revealed several progenitors in their presupernova state, allowing correlations between CCSN observables and progenitor features, thereby furnishing extremely valuable constraints on stellar evolution and CCSN modeling. Optical spectropolarimetric observations of supernovae and X-ray compositional maps of supernova remnants have revealed wide-spread, but moderately strong, asymmetry in the ejecta, providing insight into the morphology of the central engine. Studies of the afterglows of $\gamma$-ray bursts have confirmed the association of long duration bursts with CCSNe while establishing that short duration bursts do not share this association. Our ability to simulate the photon radiative transfer that produces the visual display of all of these events has also improved, with 3D time-dependent light curve and spectrum calculations now possible, including the far-from-equilibrium conditions that occur during shock breakout.

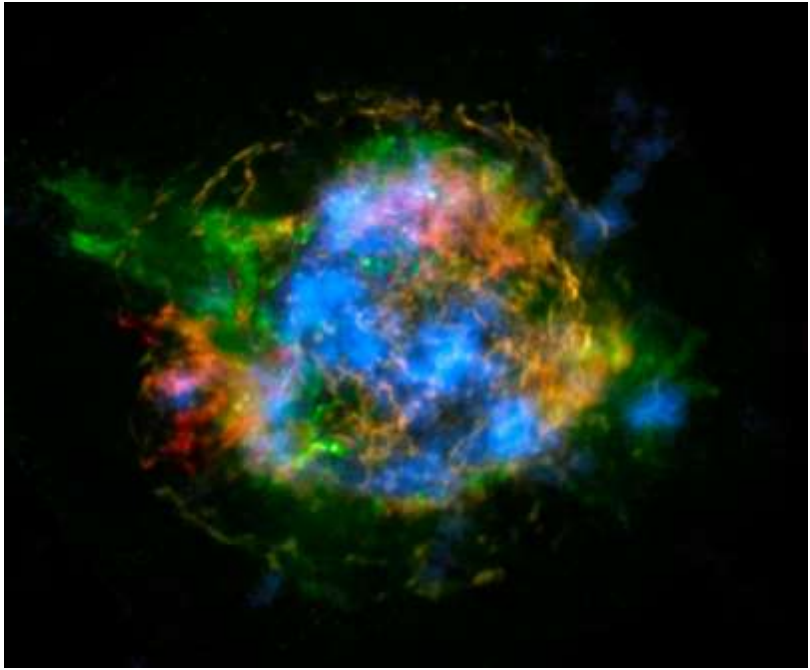

Figure 10: $^{44}$Ti in the remnant of the supernova Casseopaia A that exploded about 300 years ago. New high energy X-ray data from NUSTAR image for the first time the distribution of the radioactive isotope $^{44}$Ti produced in the explosion (blue). The distribution can be compared with stable elements such as silicon and magnesium (green) and iron (red) that had been previously observed. The amount and distribution of $^{44}$Ti can be used to probe the innermost boundary between material falling back and being ejected, once nuclear physics uncertainties are addressed. (Image Credit: NASA/JPL-Caltech/CXC/SAO)

Despite this myriad of new observations that provides crucial constraints of CCSNe, observations of explosive astrophysical events have been limited in terms of sky coverage and cadence. Improving our supernova census and determining the a more complete morphology of all types of supernova events will require high cadence surveys with near full sky coverage and through follow-up across all wavelengths (see section 3.9). Rapid follow-up at X-ray and UV wavelengths are particularly important to maximize the information we can glean from shock breakouts and γ-ray bursts (in particular in combination with gravitational-wave observations). Expanding the catalog of presupernova progenitor observations will require highly-resolved, deep multi-band imaging of a multitude of galaxies. To bridge the gap between these new observations and modeling of the central engines and nucleosynthesis, further improvements in light curve and spectral modeling are needed.

γ-ray observations allow one to determine the yields of radioactive isotopes produced in supernovae offering unique opportunities to probe the deepest layers of nucleosynthesis. Model predictions of $^{44}$Ti yields do vary significantly. INTEGRAL and NUSTAR have been able to provide $^{44}$Ti yields and spatial distributions for a number of historic supernova remnants in the Galaxy, including 1987A (see Fig. 10). There is also a ≈50% chance to directly observe a core collapse supernova during their expected instrument lifetime. INTEGRAL will continue to observe other radioactive isotopes expected to be produced in supernovae, which cannot be detected directly with NUSTAR owing to the higher γ-ray energies. NUSTAR may provide indirect, model dependent constraints on these isotopes by detecting their contributions to the low energy γ-ray flux via down scattering processes.

Much progress has been made in predicting the neutrino signal seen by neutrino detectors and the gravitational-wave signal seen by Advanced LIGO and its partner observatories from the next galactic CCSN. Understanding these signals is crucial for observationally probing the dynamics and thermodynamics in the core, but most predictions still come from 1D (for neutrinos) and 2D (for neutrinos and gravitational-waves) simulations. Detection of a Galactic supernova with Advanced LIGO would provide unique information about asymmetries, rotation, and convective flows in the supernova engine.

Observing supernova neutrinos is currently restricted to events in our Galaxy or nearby satellites such as the Small or Large Magellanic Clouds. A number of detectors are currently operating and would provide detailed information on the neutrino signal from a Galactic supernovae. The most sensitive detector is Super Kamiokande, with KamLand, Borexino, LVD, ICEcube, and MiniBoone also being expected to detect of the order of 100 neutrino events. Planned larger detectors such as the LBNE water Cherenkov detector or Gadzooks would provide much improved sensitivity to Galactic supernova neutrinos. With such detectors it might also be possible to detect a few events from a supernova in Andromeda, slightly increasing the chance of detecting a supernova during the detector lifetime.

In addition, neutrino detectors are sensitive to the cumulative relic neutrino background from all supernovae having occurred in our Galaxy to date. Detecting this relic neutrino flux would provide information on the Galactic supernova rate and the average neutrino fluxes and energies of core collapse supernovae. So far only upper limits have been detected, and a next generation of neutrino detectors will be needed to provide useful constraints, or a detection.

### 2.3.8 Impact on other areas in nuclear astrophysics

Advances in supernova modeling directly impact our understanding of the origin of the elements. The r-process has been shown to be sensitive to "fallback", where material might be almost ejected only to fall back onto the proto-neutron star, where it can become re-energized to finally be fully ejected. Clearly the nucleosynthesis conditions are very different for such a scenario, compared to a simple straight ejection in a one dimensional model. Similarly, multi-D supernova models indicate that the electron fraction, a key parameter determining nucleosynthesis outcomes, evolves highly anisotropically in the deepest layer of the supernova. Clearly 1-D models are not adequate to predict the nucleosynthesis of the deepest ejected layers where the electron fraction plays a critical role. Multi-D models and observations also indicate that supernova ejecta are asymmetric and anisotropic, including the formation of clumps and filaments. Existing and future observations of element distributions in supernova ejecta can be a powerful tool to validate supernova models, but require advanced supernova models in 3D with reliable microphysics, in particular the nuclear physics that determines the nucleosynthesis outcomes.

The study of how supernovae and massive stars altogether return ejecta with new nuclei into next-generation stars is key to chemical evolution models. This can also be observationally studied through $\gamma$-ray line observations from long-lived radioactive isotopes such as $^{26}$Al and $^{60}$Fe. Examples are the discovery of superbubbles around massive-star groups in our Galaxy identified with INTEGRAL from $^{26}$Al gamma-rays, and the discovery of $^{60}$Fe in terrestrial ocean crust and lunar material that is attributed to the debris of a supernova explosion near the Sun about 2-3 million years ago.

Realistic supernova models are also needed to predict the properties of the remnants, in particular mass and velocity distributions of neutron stars.

## 2.4 Compact Object Binary Mergers and Short GRBs

### 2.4.1 Introduction for non experts

Compact stellar objects such as white dwarfs, neutron stars, and black holes are remnants of stars that have run out of nuclear fuel. As many stars exist in pairs, so called stellar binary systems, pairs of compact objects that orbit each other are also common. Gravitational wave emission and other mechanisms lead to a reduction of the distance over time, eventually ending in the collision and merging of the two compact objects. Predictions are uncertain but indicate that such mergers should occur at least a few times to a hundred times per Million years in our Galaxy alone, making it a very common phenomenon in the universe. Of particular interest for nuclear astrophysics are the mergers that involve white dwarfs or neutron stars. White dwarf mergers are associated observationally with thermonuclear supernovae (and are in more detail discussed in section 2.5), while neutron star mergers are associated with the short gamma ray bursts. During a neutron star merger event neutron rich neutron star material is ejected or blown off by neutrinos, providing a possible site for the r-process (see section 2.1). The decay heat produced by the large amount of r-process radioactivity may explain the so called 'kilo novae', afterglows that are sometimes observed following a short gamma-ray burst. Finally, mergers that involve neutron stars are prime candidates for emitting detectable levels of gravitational waves, and the observations of such waves during the merging process with future gravitational wave detectors may provide insights into the structure of neutron stars.

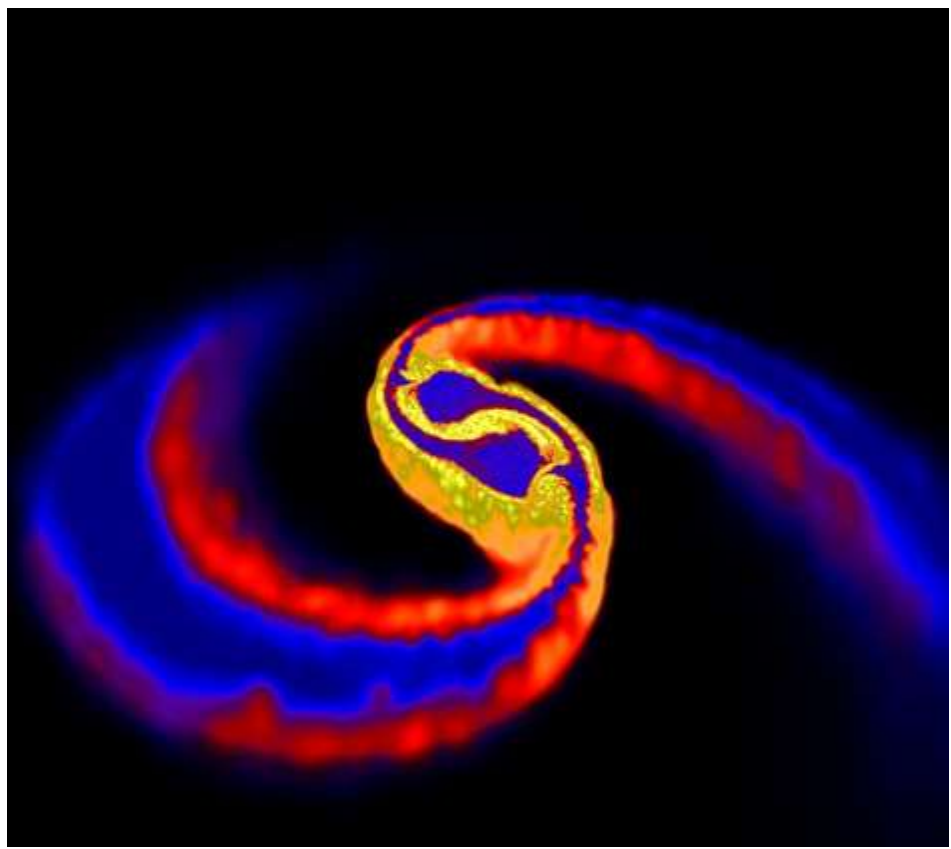

Figure 11: Multi-dimensional simulation of the merging of two colliding neutron stars into a black hole. Some material is ejected and provides a possible site of the r-process. (Daniel Price (U/Exeter) and Stephan Rosswog (Int. U/Bremen))

### 2.4.2  Current open questions

- What are the rates of white dwarf mergers, and how do those relate to thermonuclear supernovae?

- What are the rates of neutron star neutron star and neutron star - black hole mergers and how has the merger rate changed over the history of the Galaxy?

- What are the gravitational wave signals from merging neutron stars, and what would they tell us, if detected, about the properties of neutron stars?

- What is the nucleosynthesis output of neutron stars - neutron star and neutron star - black hole mergers and how is it affected by neutrino physics?

- How long does the hyper massive neutron star formed by the merger of two neutron stars survive and does its neutrino-driven wind delay or prevent a GRB due to baryon loading?

- How massive is the accretion disk that is formed when the hyper massive neutron star collapses and how massive is it in BHNS systems?

- Can, and if so, how does the postmerger state evolve to a short GRB?

- What is the photon signature of the early merger afterglow?

### 2.4.3 Context

The past decade has seen a breakthrough in numerical relativity, enabling for the first time long-term simulations of merging binaries of black holes and neutron stars in full general relativity with microscopic equations of state and neutrino emission (see Fig. 11). These events are thought to be the progenitors of the frequently observed short gamma ray bursts, and are also a candidate site for the rapid neutron capture process (r-process) (see section 2.1). r-process nucleosynthesis in neutron star mergers is now modeled by multiple groups using trajectories of material ejected in multi-D merger models. The robustness of the resulting r-process has renewed interest into neutron star mergers as sites of the r-process.

The connection between gamma ray bursts and heavy element synthesis has recently been strengthened through the observations of optical and near infra red (near IR) afterglows following a short gamma ray burst, so called kilo novas. The properties of the kilo nova event agree well with predictions based on the decay of ejected long lived r-process nuclei. In addition the first generation of laser interferometer gravitational-wave (GW) observatories reached design sensitivity. While no detection was made, interesting upper limits were obtained. This has triggered progress in theory linking neutron star properties such as the neutron star radius and the nuclear equation of state to the expected gravitational wave signals.

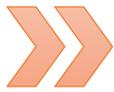

### 2.4.4 Compact Mergers Strategic Thrust 1: Advanced Models

Merger simulations using Newtonian gravity or various approximations to general relativity already include realistic nuclear EOS, but use treatments of neutrino interactions and transport too crude for reliable predictions of nucleosynthetic yields. More importantly, they lead to unreliable predictions of the survival time of the HMNS and ejecta masses. Full numerical relativity simulations, on the other hand, capture the general-relativistic aspects of the problem, but generally employ polytropic NS models, which are not useful to study the postmerger evolution. Since Advanced LIGO will be taking data within a few years, relativistic merger simulations must urgently be improved to theoretically underpin Advanced LIGO observations and the follow-up observations in the electromagnetic spectrum. Collaborations between numerical relativists and CCSN modelers will be necessary to incorporate nuclear EOS, spectral and angle-dependent neutrino transport, and neutrino and nuclear interactions into relativistic merger models. This effort will therefore directly benefit from advances in our understanding of the nuclear equation of state through experiment (see section 2.6) as well as X-ray and radio observations of neutron stars (see section 3.9). The structure of the neutron star crust may also play an important role (see section 2.6). Methods need to be developed to accurately follow the long-term evolution of the merger remnant and robustly predict its nucleosynthetic yields and determine the viability of mergers as short GRB central engines and r-process sites. To predict the lightcurve and spectrum of the early merger afterglow in photons, non-LTE radiative transfer simulations will be needed. These, in turn, will require reliable photon opacities for the range of possible exotic and neutron rich nucleosynthesis products of the merger. Resulting advanced merger models need to be extended to cover the critical phases of nucleosynthesis and must then be coupled to nuclear reaction networks to predict the composition of the freshly synthesized nuclei that are ejected and contribute to the chemical evolution of the Galaxy.

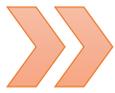

## 2.4.5   Compact Mergers Strategic Thrust 2: Multi-messenger observations

Advanced LIGO is expected to turn on in 2015 (see section 3.9.14). When it reaches its design sensitivity within a few years, it will have horizons of $\sim 200 Mpc$ and $\sim 600 Mpc$ for neutron-star – neutron-star (NSNS) and neutron-star – black-hole (NSBH) binaries, respectively (see section 3.9.14). Depending on population synthesis models, these ranges correspond to conservative estimates of *tens of observations per year*. Observations at low signal-to-noise ratio will yield constraints on the total system mass and possibly the mass ratio of the components. Rarer closer events will yield detailed information on the individual masses of NSNS and NSBH binaries and the spin of the BH in BHNS binaries. In the last hundreds of orbits before merger, the tidal interaction (which depends on the nuclear equation of state [EOS]) of the components has an influence on the GW signal. In BHNS systems, the NS may be disrupted and the GW frequency at which this occurs can be connected to NS structure. In NSNS systems, a hypermassive NS (HMNS) is formed whose long-term survival and the GW signal emitted in the postmerger phase depend on the system mass, on the nuclear EOS, and neutrino cooling.

More observations of optical/near IR afterglows of short gamma ray bursts, so called kilo novas, with current and future wide-field observatories such as Pan-STARRS, DECam, Subaru, or LSST will also be important (see section 3.9). These observations may directly constrain the amount of radioactive r-process nuclei ejected in a neutron star merger event.

### 2.4.6   Impact on other areas in nuclear astrophysics

Neutron star-neutron star, or neutron star-black hole mergers may play a central role in the origin of the heavy elements as sites for the r-process. A better understanding of the merger process and the associated neutrino physics is therefore important. In addition, neutron star mergers offer an opportunity to probe neutron star structure and the nuclear equation of state through future gravitational wave observations.

## 2.5 Explosions of White Dwarfs

### 2.5.1 Introduction for non experts

White dwarf stars are the cores of nuclear ash left over at the end of the lives of low mass stars. If they are formed in a stellar binary system, mass transfer can induce a very wide range of nuclear phenomena, ranging from explosions of accreted fuel on the surface, giving rise to novae, to the thermonuclear explosion of the entire white dwarf, giving rise to a type Ia supernova. Many other phenomena, some possibly still undiscovered, some currently classified as anomalous or sub luminous type Ia supernovae, may occur. The most common explosions - novae and regular type Ia supernovae are observed frequently, but what exactly determines their evolutionary pathways and which pathway is the one leading to a type Ia supernova remains unclear. Novae and type Ia supernovae are important contributors to the elements in the cosmos - SN Ia produce many elements around iron, and novae may be responsible for the origin of a number of rarer isotopes such as $^{15}$N, $^{17}$O as well as radioactive $^{26}$Al. A successful model for type Ia supernovae is also important to address cosmological questions about the expansion of the universe, as calibrated SN Ia are the prime distance indicators for cosmological distances. Models are needed to reliably estimate systematic errors that might arise, for example, from a dependence of the explosion mechanism on age or environment.

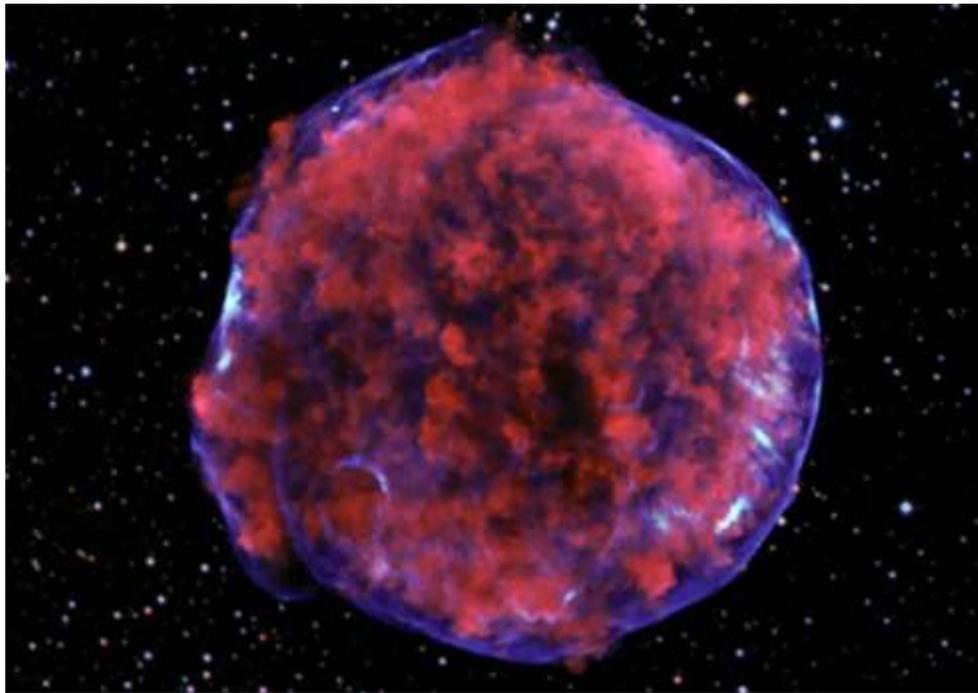

Figure 12: The Tycho supernova remnant thought to be the result of a type Ia explosion of a white dwarf that was observed in 1572. Type Ia supernovae play a critical role in cosmology and are responsible for the origin of significant amounts of iron group elements in the universe. The understanding of ignition and nucleosynthesis is still plagued by nuclear physics uncertainties. (Image credit: X-ray: NASA/CXC/Rutgers/K.Eriksen et al.; Optical: DSS)

### 2.5.2 Open questions

- What are the most common origins of type Ia supernovae? What determines whether an accreting white dwarf becomes a type Ia supernovae, and/or a nova system?

- What is the exact composition of white dwarfs material and how does it effect the nuclear reaction pattern in classical novae?

- What is the reason for the diversity of type Ia supernovae? How common are unusual events?

- How does the nature of type Ia supernovae depend on the environment, and how can type Ia supernovae be used as probes of galactic environments at very high redshifts?

- How robust is the calibration of type Ia supernovae as standard candles?

- What are best potential observational indicators of type Ia environment or explosion mechanism?

- What are suitably accurate computational methods for modeling combustion in full-star explosion simulations, and how can their uncertainty be quantified?

- Does the Urca electron capture / positron decay process play an important role in single degenerate ignition of type Ia supernovae?

- Why do classical Novae eject more mass than expected? Is there a problem with the expectations, the measurements, or both?

- How is interior white dwarf material mixed into a classical nova explosion? And what is its role as an accelerant and in nucleosynthesis of distinctive nuclides?

- What is the contribution of classical novae to the origin of stable and radioactive nuclei in the Galaxy?

- Are there pre-solar grains from Novae?

- What other explosive or non-explosive thermonuclear phenomena occur on white dwarfs, how do they relate to the variety of observed types of explosions, and can we predict so far unobserved phenomena and their observational signatures?

### 2.5.3 Context

Over the past decade numerous supernova surveys have amassed a huge observational data set on type Ia supernovae (see Fig. 12), that spans from nearby explosions to the most distant objects at redshifts around 1.9. An important conclusion is the large diversity of type Ia supernovae with large variations in maximum brightness and numerous peculiar subclasses. There are also hints of a correlation of the observed properties with the characteristics of the respective host galaxies.

However, these observations have not allowed conclusive identification of the correct progenitor model and very few observations have identified pre-explosion objects. While based on nucleosynthesis arguments it is robustly believed that type Ia supernovae originate from the incineration of a white dwarf star, it is unclear how the thermonuclear explosion of the white dwarf is triggered. All proposed scenarios have severe shortcomings. Among the most studied scenarios is the so called "single degenerate" model, where a single white dwarf accretes matter from a companion star until gravitational collapse sets in, which in turn triggers the thermonuclear explosion. In the "double degenerate" approach that has gained more interest recently, two white dwarfs merge or collide (possibly explaining the fact that in some cases the observationally inferred mass exceeds the maximum mass of a single white dwarf, and explaining the general absence of companion star relics). Various variations involving different details of ignition and explosion within these types of progenitors have also been explored.

One challenge of the single degenerate models is that under most circumstances matter accreted onto a white dwarf results in the thermonuclear explosion of a surface layer after 10s to 100,000s of years of accretion, giving rise to a classical nova explosion. These explosions are thought to eject more material than has been accreted, hence leading to a decrease of white dwarf mass over time. Classical novae are critical testing grounds for stellar physics and are regularly observed at all wavelengths both within our galaxy and in nearby galaxies. This has allowed measurements of approximate ejected masses as well as abundances of elements in the ejecta. Recent radio observations show multiple mass loss episodes in novae, perhaps indicating multiple outbursts.

One dimensional nova models (see Fig. 13) have been very successful in explaining the typically observed composition of the ejecta that comprise of a mixture of mixed in white dwarf material and the ashes of thermonuclear burning. However, the inputs necessary (in particular, enrichment of the H envelope) do not match observations. There is also a significant discrepancy between inferred and predicted ejecta masses. Multi-dimensional models have focused on the role of convection in producing the necessary enrichment, and have shown some success. Nova models predict nova ejecta to be enhanced in some otherwise rare isotopes such as $^{15}$N, $^{17}$O, and radioactive $^{26}$Al. Novae may therefore be at least in part responsible for the origin of these isotopes in nature. Enhancements of these isotopes are found in a number of pre-solar grains extracted from meteorites, which may have been formed in nova ejecta. If confirmed, such nova grains could be analyzed in terms of other isotopic characteristics that may help to constrain nova models.

While most studies of supernovae and novae attempt to explain observed events, there has been some work on predicting rare and novel event types based on the observed population of binary systems. An example are so called .Ia ("point one a") events where a thick helium layer explodes on the surface of an accreting white dwarf without triggering the ignition of the white dwarf star itself. Another example are rare classical nova events, where relatively cold white dwarfs are expected to accumulate more massive hydrogen and helium rich layers that lead to novae that may produce heavier elements through a more extended rapid proton capture process.

There has been tremendous progress in the understanding of the underlying nuclear physics of explosions related to accreting white dwarfs. This is critical for constraining and validating various theoretical possibilities, for predicting compositional and nucleosynthetic signatures, and for exploring environmental effects. For type Ia supernovae carbon and oxygen induced fusion reactions are critical and have been shown to affect ignition conditions, the strength of the detonation, and the ejected composition. Experiments have shown that the uncertainty on fusion reactions is much larger than previously thought. In addition to the well known effect of sub-barrier fusion enhancement, there is now evidence from fusion reactions with heavier nuclei for significant sub-barrier fusion suppression. It is not clear to which extent such a suppression is present, for example in the $^{12}C + ^{12}C$ fusion, but together with the possibility of rate enhancement through unknown resonance, the possibility of this effect dramatically increases the rate uncertainty. Weak interaction rates, such as electron captures, also play an important role in type Ia supernovae. Experiments with charge exchange reactions in the last decade have provided enough experimental data in weak interaction strength in nuclei to guide the development of effective interactions in shell model calculations, and to characterize quantitatively the remaining uncertainties near the valley of stability. In the example of core collapse supernovae, however, additional measurements of (d,2He) charge exchange on neutron rich rare isotopes are required to constrain the relevant weak reaction rate.

Progress in understanding the underlying nuclear physics has been even more impressive in the case of Nova explosions. These conditions represent a fortuitous case in that temperatures are high enough for reaction cross sections to be sufficiently large to be measured directly. On the other hand temperatures are still low enough so that the reaction sequence only includes stable nuclei, or unstable nuclei located close to stability where reasonably intense radioactive beams have been developed in many cases. As a consequence, Classical Novae are the only astrophysical explosions where most of the nuclear reaction rates are known experimentally. A few exceptions remain, where beam development has been particularly difficult.

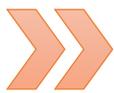

### 2.5.4  WD Explosions Strategic Thrust 1: Advancing the models

**Type Ia Supernovae:** Chandrasekhar mass single degenerate models for type Ia supernovae are relatively mature. Prediction of nucleosynthetic yields of sufficient quality for detailed observational comparison to sequences of spectra still requires further refinement of computational techniques for both detonation and deflagration modes of combustion (see section 3.8). Gross yields of, for example, $^{56}Ni$ are modestly accurate (~10%), but this may not be sufficient to distinguish between alternative models. Large parameter studies with 3D simulations are possible, which allows observational questions relating to the population of supernovae to addressed, but this must be pursued in parallel with improvements in computational techniques. Performing radiative transfer calculations is only possible for a few groups currently, and suffers from computational challenges to run effectively on modern massively-parallel supercomputers.

Models of double degenerate systems are also fairly mature though they received less attention in the early 2000's than Chandrasekhar mass models. Some areas of necessary refinement are similar, such as accurate computation of the dynamics and products of detonation, while others are different, such as the treatment of the mass transfer process, which utilizes a particle rather than a grid method for computing hydrodynamics. The division between violent mergers, which may lead to explosions, and non-violent ones that likely lead to collapse to a neutron star requires further investigation as computational methods tailored for this problem continue to be refined (see section 3.8).

Relevant to all channels, theoretical work on binary stellar populations, and specifically short-period binaries containing white dwarfs, is important for understanding the frequency at which various possible scenarios might be realized. Producing the right rate of events is typically a challenge for candidate scenarios. Computing these rates for comparison to observed rates depends critically on the understanding of the short period white dwarf binary population.

A particularly important challenge is the incorporation of the latest nuclear physics advances into supernova models. This does not just require updating of nuclear data inputs, but for example in the case of weak interactions extensive nuclear model calculations that use experimental constraints to determine improved rate sets with uncertainties that can be used in supernova models (see section 3.6). Extensive post processing calculations are needed to take advantage of progress in nuclear physics to obtain nucleosynthetic signatures, that can be used to validate (or falsify) models. Of special importance is the incorporation of nuclear uncertainties that now become available for various reaction rates (see section 3.10).

A well understood physical model of type Ia supernova would be a major breakthrough and would open up possibilities to explore potential systematic errors in standard candle calibration, to understand the contribution of SNIa to the chemical enrichment of galaxies, and to exploit environmental dependencies to constrain the properties of the oldest, most distant galaxies such as their star formation properties.

**Classical Novae:** Multi-dimensional models of Novae have so far not been possible, though aspects such as the onset of the explosion have been modeled in 3D. This is a severe limitation as many of the current open questions likely require the reliable modeling of multi-D effects (see section 3.8). One critical open question is the mixing of white dwarf material into the explosion. Not only does the mixing process affect ignition conditions, and nature of the explosion and nucleosynthesis, but it is also critical in determining whether continued nova explosions lead to growth or erosion of the underlying white dwarf. One challenge here is the low speed of convection compared to the sound speed, which necessitates more sophisticated and complex numerical hydrodynamics methods. Refined multidimensional models may be key to understanding the dynamics of the mixing, but algorithms can have trouble right at the H envelope boundary since numerical mixing may artificially enhance the burning. Algorithmic sensitivity studies are needed to understand what is real and what is unphysical. Other important questions that require multi-D modeling include the role of the companion in the ejection process, the impact of the explosion on the accretion process, or the impact of magnetic fields.

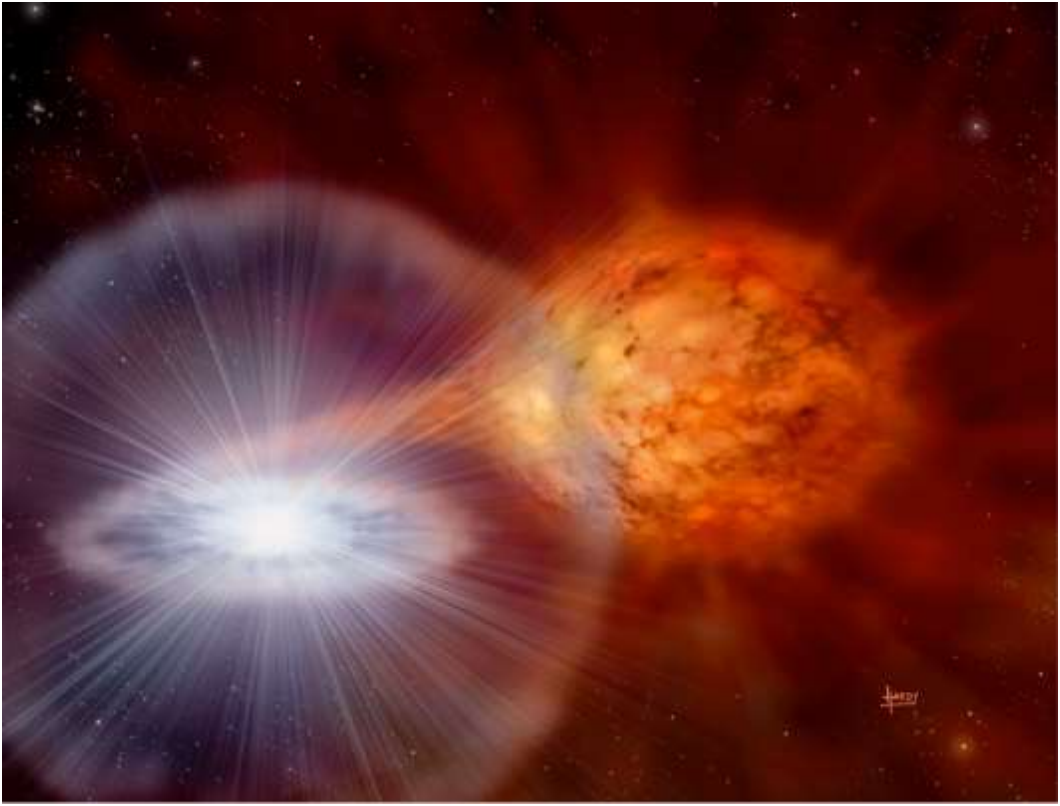

Figure 13: Artists view of a white dwarf accreting matter from a companion star in a close stellar binary system. The accreted matter accumulates on the white dwarf and every few 100,000 years gives rise to a thermonuclear explosion that is observed as a Nova. The hottest novae can produce elements up to Ca or even higher, and may also produce significant amounts of longer lived radioactive isotopes, but predictions are uncertain due to unknown nuclear reaction rates. Credit: David A. Hardy http://www.astroart.org & Science and Technology Facilities Council.

With the complexity of models the need for validation will also increase. The produced abundance distribution provides an important pathway for such validation. Reliable nuclear physics will be essential to take advantage of this opportunity. In addition it will be important to constrain the basic model parameters such as accretion rate and white dwarf mass independently through observational means.

**Other types of explosions:** In general it will be important to expand the parameter space considered in supernova and nova modeling to capture the full variety of observed phenomena and to be able to predict unknown types of transients. In particular for novae it has been shown that low accretion rates and low core temperatures can lead to much more powerful explosions with drastically changed nucleosynthesis. Another explosion regime are thick helium layers that lead to Ia supernovae. So far, self consistent sensitivity studies to explore the critical nuclear physics, that take into account mixing and the impact of reaction rate changes on energy generation through the production of radioactive isotopes have not been carried out - neither for standard classical nova models, nor for the range of rarer phenomena. Such sensitivity studies will be critical to guide future efforts in nuclear physics. This is especially important as most of the nuclear physics can probably be determined experimentally as long as it is clear what is needed. Similar efforts will be needed for type Ia supernova models.

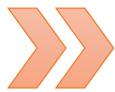

### 2.5.5 WD Explosions Strategic Thrust 2: Multi-wavelength observations

The biggest observational need related to type Ia supernovae is to provide constraints on the progenitor. This is difficult due to the rarity of these events. New surveys such as the Palomar Transient Factory (PTF) find large numbers of transients, including more events close to Earth that provide richer observables. Rapid followup observations at all wavelengths are key (see section 3.9). For example, radio observations with EVLA only two days after PTF found a type Ia supernova have led to significant constraints on total ejected mass and therefore on the possible nature of progenitors. New telescopes will also open windows in the IR. This is interesting because SN Ia are more similar in the IR and distinct lines can be seen there. More observations of polarization can help distinguish between models (for example, mergers may be expected to have high polarization). It is also possible that isotopic information can be obtained from these observations.

For novae, more detailed observations of the composition of ejecta and radio constraints on ejected masses will be needed to validate models. A severe limitation in this context is the lack of UV spectroscopic capabilities (see section 3.9). Of critical importance are also observational constraints on system parameters such as accretion rate, nature of the companion star, and mass of the white dwarf. Such observational constraints would greatly tighten nova model validation through abundance observations, enabling modelers to take full advantage of the advances in our understanding of the nuclear physics of novae.

The observation of nuclear $\gamma$-rays is of particular importance for both, type Ia supernovae and novae (see section 3.9). $\gamma$-rays provide information on the produced abundance of a specific isotope, and are largely free of radiation transport issues that plague the traditional elemental abundance observations. The time evolution of $\gamma$-radiation can also provide a handle on distinguishing various progenitor scenarios. INTEGRAL's recent detection of $\gamma$-rays from the $^{56}$Ni decay chain in SN2014J demonstrates the potential of direct $\gamma$-ray observations at MeV energies. NUSTAR is expected to detect one to two thermonuclear supernovae per year, albeit only in lower energy $\gamma$ radiation that does not allow for the direct isotope specific tracing of radioactive ejecta other than $^{44}$Ti. Combining all SNIa messengers for such nearby events, we may be able to learn as much about thermonuclear supernovae as we learned about core-collapse supernovae from nearby SNI987A. A future $\gamma$-ray mission with enhanced sensitivity in the few MeV energy range compared to existing instruments would provide powerful diagnostics for type Ia supernovae and novae. In novae, detection of decay $\gamma$-rays from $^{22}$Na and, if the instrument can be pointed early, $^{18}$F may also be possible.

New time domain surveys (like PTF and soon LSST) will find a large number of transients, possibly in novel recurrence time regimes (see section 3.9). Questions that will arise will include: Are these new objects or extreme examples of our existing models? Detailed observations on these new transients will be needed. This will have tremendous impact on nuclear astrophysics, with new types of objects likely requiring new nuclear physics investigations.

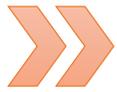

## 2.5.6 WD Explosions Strategic Thrust 3: Pinning down the nuclear physics

Novae and type Ia supernovae are thermonuclear events and nuclear physics naturally plays a critical role for ignition, energy generation, and nucleosynthesis. An important problem are the huge uncertainties associated with carbon and carbon-oxygen fusion reactions. These are among the most uncertain thermonuclear rates in astrophysics limiting models of the ignition process of type Ia supernovae. The carbon and oxygen fusion reaction rates also affect the strength of the detonation in the most common deflagration - detonation models impacting directly the composition of the supernova ashes. Multiple strategies are needed to address this problem (see section 3.1). While measurements need to push to lower energies to reach the relevant astrophysical energy range, a better understanding of fusion reactions in general is also needed. This requires work in reaction theory (see section 3.6), which then has to be validated with cross section measurements for a range of different types of reactions. Improved rates for charged-particle rates on $^{48}$Ca are also needed-this can be a possible discriminator of progenitor models.

While much progress has been made over the past decade in measuring weak interaction strengths mainly through the use of charge exchange reactions on stable nuclei, the steps of converting this experimental data into the products that can be used by modelers is incomplete. Systematic studies of weak interaction strength on unstable nuclei, in conjunction with shell model calculations, will become possible at facilities that offer fast radioactive beams, such as FRIB in the US, FAIR in Germany, or RIBF in Japan (see section 3.2).

Temperatures in novae are much more moderate than in SN Ia. In fact, in terms of the feasibility of experimentally determining reaction rates on unstable nuclei in stellar explosions, novae provide a somewhat ideal case. Temperatures are high enough to make cross sections accessible with moderate beam intensities (unlike reaction rates in stars) while temperatures are still sufficiently low to confine the reaction sequence to lighter nuclei up to $Z\sim20$ close to stability. As a consequence, many of the important rates in Novae are already experimentally determined. There are a few remaining reactions, that remain unmeasured and strongly limit predictions of radionuclide production or interpretation of abundance signatures to validate models. Key reactions include $^{18}$F(p,$\gamma$)$^{19}$Ne, $^{25}$Al(p,$\gamma$)$^{26}$Si, and $^{30}$P(p,$\gamma$)$^{31}$S. Measurements with low energy beams at the Facility for Rare Isotope Beams should be able to determine the majority of the nuclear reaction rates needed to understand nova nucleosynthesis (see section 3.2).

Much work has been done on identifying the critical nuclear physics in standard SNIa and nova models. Such sensitivity studies provide critical guidance for the improvement of the underlying nuclear physics and also provide estimates for the nuclear uncertainty component of observables. In the future it will be important to carry out such sensitivity studies along with changes and improvements of astrophysical models. For novae it will be important to carry out sensitivity studies that take into account feedback from reaction rate changes on the hydrodynamical evolution and nuclear uncertainties (see section 3.1).

In addition, sensitivity studies are needed for the rarer events, such as Ia supernovae, other sub luminous SNIa, as well as for rarer classes of classical novae, such as the recently predicted explosions on white dwarfs with particularly low temperatures and accretion rates. For the latter systems it has been shown that breakout from the CNO cycles and a much stronger rapid proton capture process into the iron region and beyond ensues. In general, reaction sequences can vary widely depending on the system parameters, and for each case reliable nuclear physics is needed to predict observational signatures.

### 2.5.7  Impact on other areas in nuclear astrophysics

Type Ia supernovae are important nucleosynthesis sites - they are responsible for the origin of a significant fraction of iron and near iron elements, and may also be a site of the p-process. Novae may contribute to the origin of $^{17}$O, $^{15}$N, and $^{26}$Al. In addition, type Ia supernovae play a primary role in cosmology as standard candles on very large cosmic distances and are one of the pillars of the modern paradigm of an accelerating, dark energy driven universe. Progenitors and properties of type Ia supernovae need to be understood to gauge for example systematic errors in distance measurements through environment dependencies of the luminosity calibration.

## 2.6  Neutron Stars

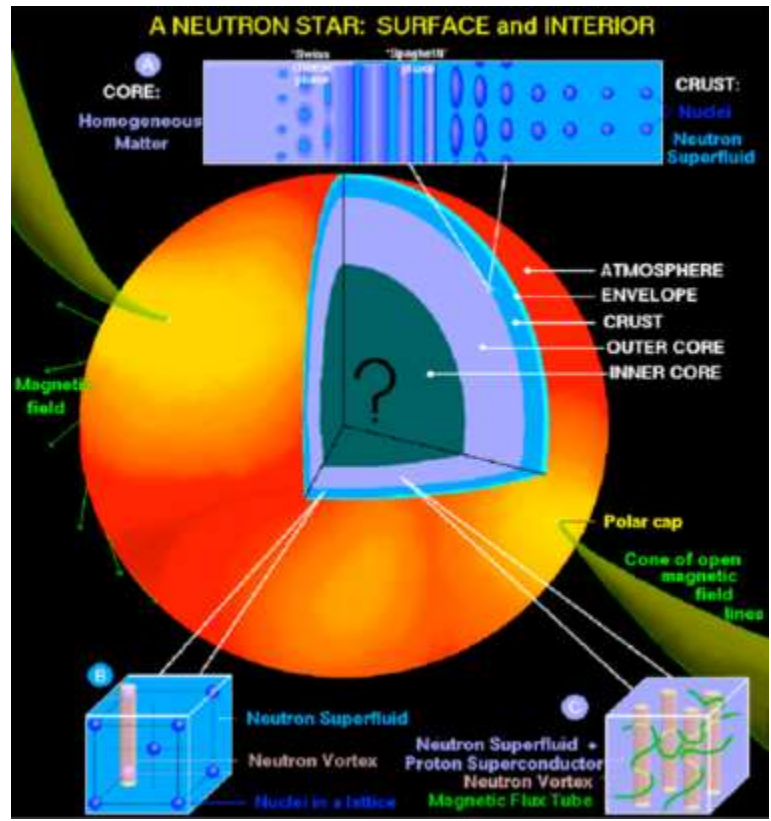

Figure 14: Schematic of the possible interior structure of a neutron star. The crust is composed of neutron rich rare isotopes, mixed with superfluid neutrons at depths beyond neutron drip. The crust core interface may be characterized by exotic so called pasta phases of nuclear matter. While the outer core is thought to consist mainly of superfluid neutrons, superconducting protons, electrons, and muons, the inner core may consist of more exotic forms of matter such as hyperons, meson condensates, or a color glass condensate. Unraveling this interior structure by comparing observations with nuclear physics based models is a major goal of the field. (Image from Dany Page)

### 2.6.1  Introduction for non experts

Neutron stars are formed in supernova explosions as the remnants of the collapsed cores of massive stars. They are only about 20 kilometers across but with a mass similar to the mass of the sun and their interiors contain more extreme forms of dense matter than that found inside terrestrial nuclei (see Fig. 14). Neutron stars offer the unique opportunity to study matter at extreme density ($\varrho 10^{10}$-$10^{15}$ g/cm$^3$), which is hot by terrestrial standards ($T \simeq 10^6$-$10^9$ K) but still cold enough for quantum properties to manifest on macroscopic scales. Understanding the properties of matter under these extreme conditions through astronomical observations, advances in theory and simulations and terrestrial experiments continues to be a grand challenge for nuclear physics and nuclear astrophysics. The goals are to interpret and motivate multi-wavelength and multi-messenger observations of neutron stars through advanced modeling and improved theoretical description of the properties of dense matter. Experimental nuclear physics continues to be a key element as it provides critical data needed to benchmark theories of dense matter. Some of the key physical ingredients required to model and interpret neutron star observations are: the equation of state of nuclear matter (the relationship between energy density and pressure), the nature of neutron superfluidity and proton superconductivity, the possibility of exotic degrees of freedom in dense nuclear matter (hyperons, pions, kaons, or deconfined quarks), and the interactions of neutrinos with nuclear matter. Neutron stars exhibit a wide range of phenomena producing unique photon, neutrino and gravitational wave signals. These include common phenomena such as pulsations in radio and X-ray, and X-ray bursts observed daily, less frequent phenomena such as giant flares in magnetars, superbursts and thermal relaxation in accreting neutron stars that recur on the timescale of few years, and rare phenomena such as supernova and neutron stars mergers. Although the basic paradigm exists for each of these phenomena, a quantitative understanding is only now beginning to emerge due to recent advances in nuclear and neutrino physics, computational astrophysics, and nuclear experiments. The prospect of connecting neutron star observations directly to open questions in nuclear, neutrino, condensed matter, and plasma physics at extreme conditions is now very real.

## 2.6.2 Current Open Questions

- What is the neutron star mass-radius relation and the maximum neutron star mass? How do they change with magnetic field strength, temperature, and rotation?

- How do neutron stars cool and how can we use observations to determine the composition of the core and to identify specific microscopic processes? Is there evidence for cooling in neutron stars beyond the so called minimal cooling model?

- What are the constraints on the equation of state of nuclear matter provided by astronomical determinations of the mass-radius relation and cooling rates of neutron stars? How do such astronomical constraints compare to those extracted from laboratory measurements?

- What is the origin of the intense surface magnetic fields as large as $10^{15}$ Gauss found in magnetars, and the origin of observed flares and their quasi-periodic oscillations?

- What does the phase diagram of dense matter at low temperatures look like? How can we combine neutron star observations, laboratory measurements, and theoretical developments to learn about those phases? What are the best effective field theories which describe cold and dense matter?

- What is the nature of absorption features detected from isolated and accreting neutron stars?

- Is there a limit to the spin frequency of milli-second pulsars? If so, why?

- What precisely controls the durations, shapes, and frequency of X-ray bursts and why is the transition to stable burning at a much lower accretion rate than expected? To what extent can better laboratory data on nuclear masses and reaction rates reconcile differences between theoretical calculations and astronomical observations?

- What is the origin of burst oscillations, and what do they tell us about the underlying neutron star?

- Is unstable burning of carbon the cause of all superbursts? What is the inferred unknown shallow heat source in the neutron star ocean that seems to be required to ignite superbursts and to explain the early light curve of cooling transients?

- What are glitches and why do they occur? What is the trigger that couples the superfluid to the crust over a timescale of less than one minute? What are the relevant dissipative processes?

- How does one link the microphysics of transport, heat flow, superfluidity, viscosity, vortices/flux tubes to neutron star phenomenology? How can we probe these quantities in the laboratory? Can we confirm the presence of "nuclear pasta" in neutron stars?

### 2.6.3 Context

**Mass and radius measurements:**

Neutron stars are observed in a variety of different environments at a wide range of wavelengths. Significant progress has been made in quality and quantity of observational data, in developing the theoretical tools to interpret them, and in obtaining experimental data on the underlying nuclear physics. Rapidly spinning neutron stars with magnetic fields are observed as radio pulsars, and precision observations of the pulse trains arriving on earth allow for rather accurate neutron star mass measurements. A major breakthrough was the recent discovery of two 2 solar mass neutron stars, PSR J1614−2230 and PSR J0348+0432, using NRAO's Green Bank Telescope (GBT). The existence of these objects provides a lower limit for the maximum mass of a neutron star providing strong, model-independent lower limits on maximum pressure of the nuclear matter equation of state and thus on the behavior of QCD at high densities and low-temperatures.

Several observations have been made which provide information on neutron star radii. There are several X-ray observations of thermal emission from the neutron star surface, in particular in globular clusters where distances are known. X-ray bursts that exhibit photospheric radius expansion and the analysis of pulsations in neutron stars which pulsate in the X-ray band have also led to radius constraints. These radius measurements are currently constrained by systematic uncertainties and are not yet model-independent, but suggest that all neutron star radii are between 9 and 14 km. Recent advances in theoretical work on describing neutron matter have converged to suggest a very similar range for neutron star radii, indicating some concordance between observations and theory.

**Cooling neutron stars:**

Cooling of neutron stars can be observed with X-ray telescopes and puts strong constraints on the neutron star interior structure as many neutrino emission processes are sensitive to the existence of superfluidity, exotic phases and the isospin dependence of nucleon effective masses. Typically neutron stars are relatively old and cool slowly, so X-ray observations of their thermal emission together with an age determination provide a single data point on a particular neutron star cooling curve. However, it has now become possible to observe an actual change in temperature in a few systems. The most spectacular case is the recent observation of rapid cooling of the 330 year old neutron star in Cassiopeia A from $2.12 \times 10^6$ K to $2.04 \times 10^6$ K over a period of 10 years, which was interpreted as confirmation of the occurrence of neutron superfluidity and proton superconductivity in the dense interiors of neutron stars.

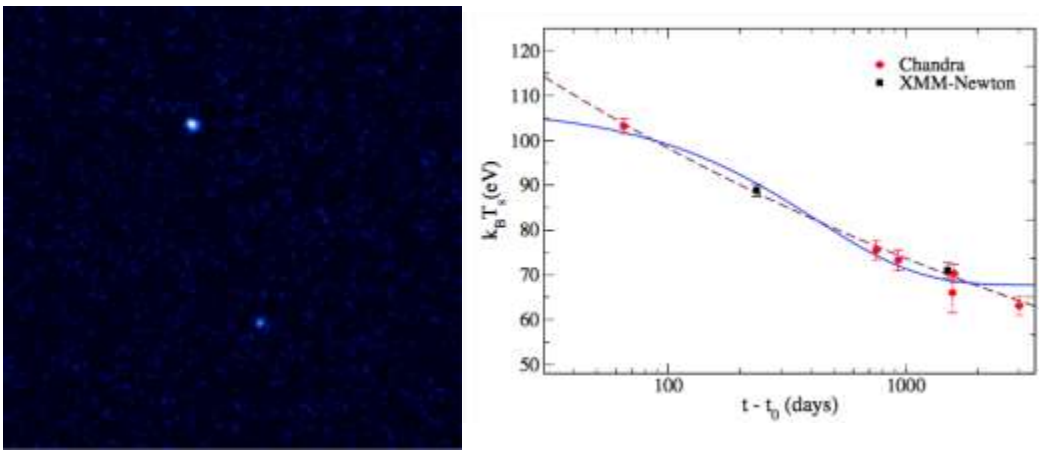

Figure 15: Transient KS1731-260, an accreting neutron star for which accretion turned off in early 2001. The faint X-ray afterglow (left) is believed to be from the heated neutron star crust, and its decrease in temperature over many months has been observed (see graph on the right). Such observations can reveal the inner structure of the neutron star crust, once the nuclear processes that heat and cool the crust are understood. This is a major goal of the field. (Image credit left: NASA/Chandra/Wijnands et al., Figure right from A. Turlione et al. arXiv:1309.3909 )

In accreting neutron stars the crust is heated separately from the core and can cool on timescales of years once accretion shuts off. This has now been observed in a number of X-ray transients using long term observations with Chandra and XMM (see Fig.15). The results confirmed the theoretical prediction that neutron stars have crusts, and are also beginning to shed light on the elastic and transport properties of crystalline structures in neutron star crusts. Significant progress has been made in identifying the nuclear processes responsible for crustal heating during accretion, but accurate predictions require improved knowledge about weak interaction strengths, masses, the location of the neutron drip line, and fusion cross sections of extremely neutron rich-nuclei.

**X-ray bursts:**

Accreting neutron stars exhibit a range of thermonuclear bursting behavior - ranging from regular relatively short X-ray bursts lasting 10-100 s with recurrence times of hours to days, to intermediate long bursts, to the rare, but 3 orders of magnitude more energetic superbursts, which can last hours to days. The frequent regular X-ray bursts have been found to exhibit a variety of phenomena, including variations in burst duration, multi-peaked light curves, and unexpected changes in recurrence time. Mapping out this phenomenology with X-ray observatories such as XMM, Chandra, RXTE, Swift, and INTEGRAL is one of the major accomplishments of the last decade. The MINBAR burst archive now contains thousands of bursts setting the stage for future efforts to understand the observations theoretically.

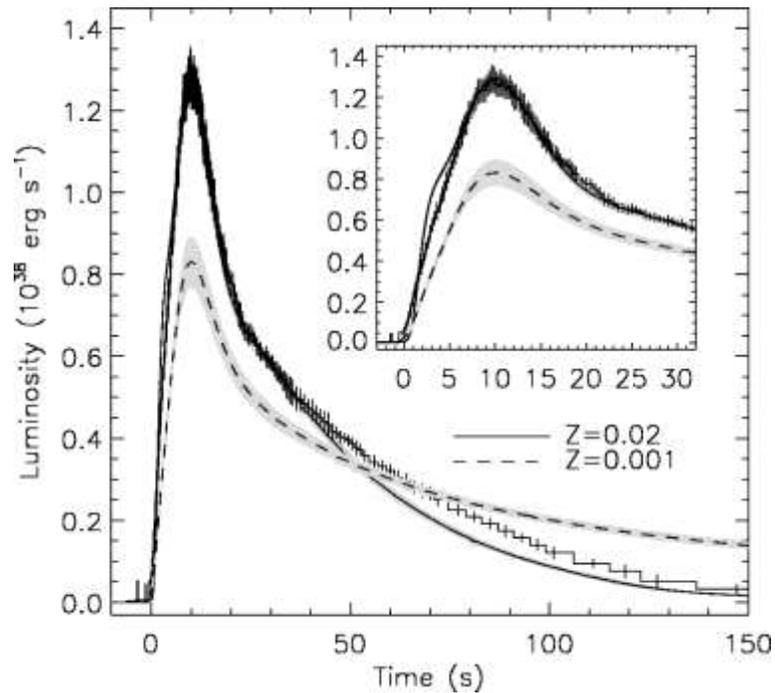

Figure 16: Observed X-ray bursts from the surface of an accreting neutron stars, compared with predictions of explosion models where various parameters such as the metallicity $Z$ of the accreted material have been adjusted. However, the model predictions suffer from nuclear physics uncertainties that will be addressed by FRIB in the coming decade. (Figure from A. Heger et al. 2007 ApJ 671 L141)

Detailed and systematic comparisons between model bursts and observations (see Fig. 16) require improved understanding of the nuclear physics underlying the $\alpha$p- and rp-processes that power X-ray bursts, and determine light curve shapes, recurrence time behavior, and the peak luminosity of sub-Eddington bursts. While huge progress has been made in measuring $\beta$-decay half-lives and nuclear masses, in particular through high precision mass measurements with Penning traps and storage rings, experimental information on important proton and helium induced reaction rates is still very limited.

This experimental information is within reach of future experimental programs at existing and planned rare isotope facilities such as FRIB (see section 3.2).

Progress has also been made in further developing X-ray burst models. 1D hydrodynamic models of regular bursts have now been extended to consistently simulate the deeper occurring super bursts shedding light on the interplay between the two burst modes. Ultimately 3D models may be necessary, in particular, to describe anisotropies in burning that are often observed as quasi periodic oscillations during bursts. The necessary codes are under development and indicate the importance of rotation and magnetic confinement in localizing the burning.

These advances have inspired a number of groups to revisit the question whether burst observations can be used to constrain the mass and radius of the underlying neutron star. Radius expansion bursts are affected by surface gravity and thus offer together with spectral information the possibility to extract mass and radius constraints. X-ray bursts can also result in hot spots that create oscillations in the burst flux due to the rotation of the neutron star. Modeling the profile of oscillations embedded in both the early rise and the tails of X-ray bursts can lead to accurate mass and radius constraints. New, higher quality observational data are critical for these efforts and are a major motivation for the proposed LOFT mission (see section 3.9.6).

**The nuclear matter equation of state:**

Matter within a neutron star is supported against gravitational collapse by the pressure provided by the nuclear matter equation of state (EOS). An important component of this pressure comes from the symmetry energy. The symmetry energy is the energy difference between isospin asymmetric matter and matter composed of equal numbers of neutrons and protons. The density dependence of the nuclear symmetry energy determines the neutron star pressure and, consequently, its structure, radius, and observed thermal evolution through various neutrino emitting processes that relentlessly cool the star.

Many recent laboratory experiments have provided initial constraints on the density dependence of the nuclear symmetry energy and on the EOS of neutron-rich matter at near-saturation and sub-saturation densities. These constraints have been extracted from measurements of nuclear structure properties, such as nuclear masses; excitation energies of isobaric analog resonances; energies and strength functions of giant monopole and dipole resonances; electric dipole strength functions and electric dipole polarizability sum rules; energies and strengths of pigmy electric dipole resonances; and measurements of neutron skin thicknesses. Experiments have also probed the differential flows of neutrons and protons during nuclear collisions to constrain the density and momentum dependence of the symmetry energy at sub-saturation densities. Relevant observables from these collisions are the diffusion of isospin between projectile and target nuclei in binary collisions, and comparisons of the spectra and flows of mirror nuclei, such as neutrons vs. protons, or tritons vs. helions ($^3$He). Combining both laboratory and astrophysical data should lead to the tightest possible constraints on the EOS.

Impressive accomplishments have been made on the theoretical front both in first principle calculations and in applications to neutron star phenomenology (see section 3.6.1). Areas of significant progress include ab-initio calculations of neutron matter highlighting the role of three-nucleon forces on the density dependence of the nuclear symmetry energy; the application of effective field theory methods to low density superfluid and solid matter with implications on transport (particularly thermal conductivity) and elastic (shear and bulk moduli) properties of the neutron star crust; and molecular dynamical simulations of neutron star crusts and their transport properties.

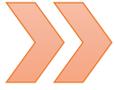

### 2.6.4 Neutron Star Strategic Thrust 1: Observations

The maximum neutron star mass has profound implications for the minimum black hole mass, the mass of the progenitor star that gave birth to it, and the equation of state of hadronic matter. The minimum neutron star mass raises questions about stellar evolution and how low mass neutron stars can be formed within the current paradigm of core collapse supernovae. A concerted effort in astronomy to increase the sample of neutron stars with measured masses will provide very important information about the range of possible neutron star masses and their formation mechanisms. Radio observations of pulsars require sensitive radio telescopes such as the Green Bank Telescope (GBT) and Arecibo; future facilities such as the Square Kilometer Array (SKA), MeerKAT, ASKAP, and ATA are expected to provide many more mass measurements (see section 3.9). Observations with these advanced radio telescopes will also give a more detailed picture of pulsar glitches, including time resolved data on the glitch, to the point where models can be tested. Continued measurements of the double radio pulsar system (PSR J0737–3039) over the next 5-10 years offer the prospect of a direct measurement of the moment of inertia of pulsar A in this system.

In order to delineate the interior composition of neutron stars and the associated neutrino emission processes, we need a future generation of time and energy sensitive X-ray observatories (see section 3.9.6). In particular, precise measurements of the masses and radii of several individual stars would revolutionize this field. Obtaining such measurements from X-ray pulsar observations will be significantly facilitated by larger X-ray telescopes with excellent timing resolution such as NASA's Neutron Star Interior Composition ExploreR (NICER) on the International Space Station (ISS) and the proposed ESA mission LOFT. The GAIA mission will greatly improve the accuracy of such studies by providing vastly improved distance measurements, significantly reducing one of the main uncertainties in determining neutron star radii from quiescent neutron stars, and increasing the number of systems with reasonably known distances that can be used for such measurements.

Open questions concerning neutron stars in X-ray binaries and related phenomena such as X-ray bursts, superbursts, and quasi-periodic oscillations can be addressed by extending X-ray observations to increase statistics and time domain coverage, especially on burst oscillations and their time-resolved spectral evolution, on accretion flows onto milli-second X-ray pulsars, and on phenomena related to super bursts. In addition, such measurements can address exciting new predictions that result from a better understanding of the nuclear physics and astrophysics of bursts, such as predicted oscillations triggered by the $^{15}O(\alpha\gamma)$ reaction in the tails of super bursts, or absorption edges in X-ray spectra due to ejected material. Such observations can be continued in the near future with existing instruments such as XMM/Chandra, Swift, INTEGRAL, while burst catalogues can be concurrently expanded and more fully analyzed. In the future LOFT (ESA), NICER (NASA), ASTROSAT (India), ASTRO-H (Japan), and eROSITA (Germany/Russia) (see section 3.9.6) will provide important new insights. A particularly important feature of LOFT is the timing resolution that will enable the study of various oscillatory modes on the neutron star surface that are directly related to composition and therefore to nuclear processes, as well as other time dependent effects like the spreading of the nuclear burning front during the early stages of an X-ray burst. The Athena mission will provide high sensitivity and high resolution X-ray spectroscopy of accreting neutron stars. This will allow one to identify the composition of material burned by X-ray bursts or superbursts, and will provide very stringent constraints on the surface redshift and therefore on neutron star compactness. Because many systems show rapidly rotating hot spots, time resolved spectroscopy has a particularly strong discovery potential.

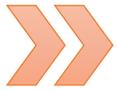

### 2.6.5 Neutron Star Strategic Thrust 2: Physics of Bursts and Crusts

Nuclear processes on accreting neutron stars involve a broad range of nuclei ranging from extremely neutron deficient isotopes during the rp-process in X-ray bursts to crustal electron capture and fusion reactions of extremely neutron rich nuclei at and beyond the neutron drip line.

**X-ray bursts:**

In X-ray bursts, nuclear properties affect the shape of the burst light curve, the composition of small amounts of material that might be ejected, and the composition of the burst ashes which sets the stage for re-ignition in superbursts and the composition of the neutron star crust. The critical nuclear physics governing the rp-process in X-ray bursts are $\beta$-decay half-lives, nuclear masses, and the rates of proton and helium induced reactions along the proton drip line up to $A \approx 108$ (see Fig. 17). All the $\beta$-decay half-lives and many of the relevant masses are now measured. Precision mass measurements with Penning traps or storage rings are needed for a number of remaining nuclei in the path of the rp-process, many of which can be reached with some improvements at current accelerator facilities (see section 3.2).

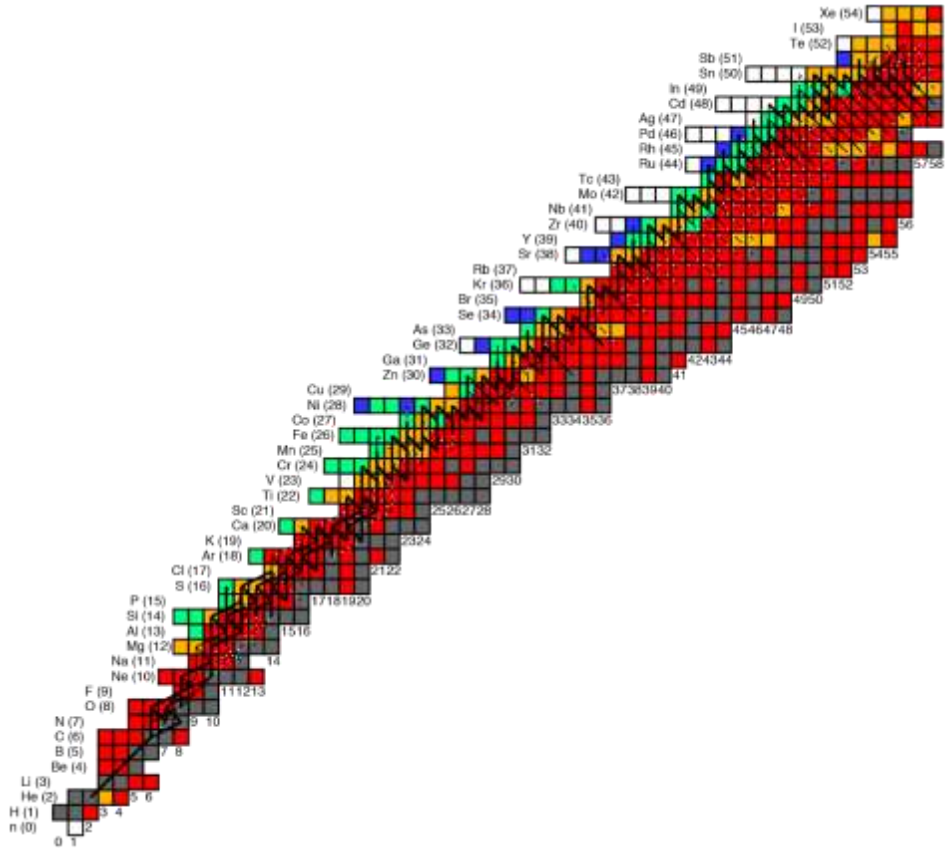

Figure 17: Predicted rp-process reaction sequence in a X-ray burst. The colors indicate nuclei with masses known to better than 10 keV (red), masses known (orange), half-lives known (green) and observed (blue). Mass measurements for nuclei along the rp-process path are needed along with radiative capture reaction cross section measurements to determine the reaction rates.

The biggest challenge for the future are the measurements of the reaction rates. Here a multi-accelerator approach is essential. Of highest importance are direct measurements of critical (p,γ) and (α,p) reactions with intense, low energy radioactive beams on windowless hydrogen and helium gas targets using recoil separators and detector arrays. A major limitation is the difficulty of realizing such beams in the laboratory and as a consequence only a handful of reactions have been measured so far, mostly at ISOL facilities. This problem is being addressed in the US through the development of a new, complementary, radioactive beam production technique, the reacceleration of radioactive nuclei that have been produced by fragmentation at intermediate energies, and have then been stopped in a gas cell system. This technique will be available at the new ReA3 facility at NSCL, and later at FRIB, and will enable in concert with the SECAR recoil separator and the JENSA gas jet target the direct measurement of many critical rp-process reaction rates (see section 3.2).

Indirect studies of reaction rates on unstable nuclei, such as measurements of direct capture Asymptotic Normalization Coefficients (ANCs) and the properties of resonances, are complementary to direct measurements. They are needed as preparation to the more difficult direct measurements, and they provide information on reaction rate components that are too weak to be measured directly. In some cases, transfer reactions with stable beams can reach the unstable nuclei in the rp-process, especially below calcium where the rp-process proceeds closer to stability. Stable beam facilities with high resolution spectrometers such as TU-Munich (Germany), RCNP (Japan), and i-Themba (South Africa), or high resolution gamma arrays such as GAMMASPHERE at ATLAS are necessary for such experiments (see section 3.1). These types of measurements need to be exploited to the fullest extent possible, so the more difficult and more resource intensive radioactive beam experiments can be optimized.

For most rp-process reactions, transfer reactions with unstable beams are needed to reach the relevant nuclei. At the next generation radioactive beam facilities, and especially at FRIB, it will be possible to reach all rp-process nuclei to study excitation energies and other properties of resonances. Proton transfer reactions with radioactive beams such as (d,n) or ($^3$He,d) (or neutron transfer reactions on mirror nuclei) as well as $\alpha$-transfer reactions such as ($^6$Li,d) have the advantage to populate preferentially the critical states and offer the additional opportunity to directly constrain proton and $\alpha$ spectroscopic factors, ANCs and widths (see section 3.2).

While theoretical predictions of rp-process reaction rates are not sufficiently reliable for X-ray burst models, the combination of theoretical information with indirect experimental data on excitation energies and other level properties can result in reaction rate uncertainties that are acceptable for some of the less critical reactions, and that provide much more reliable data until further experiments become feasible. Expanding shell model spaces and developing new effective interactions so that calculations in the region of the deformed nuclei $^{64}$Ge - $^{74}$Sr become possible would be important.

**Crust processes:**

The nuclear properties of the neutron star crust determine its mechanical, electrical, and thermal structure. This is especially true for accreting neutron stars, where the crust is replaced by X-ray burst ashes, greatly expanding the range of possible non equilibrium compositions, and where ongoing accretion continuously drives electron capture reactions, $\beta$-decays, neutron captures and releases, and fusion reactions. These reactions determine crustal heating and cooling during accretion and are therefore directly related to observations of cooling transients. Electron capture induced density discontinuities have also been predicted to give rise to gravitational wave emission provided the neutron star spins rapidly and exhibits temperature anisotropies.

The relevant nuclei are neutron rich and range from near stable isotopes to rare isotopes beyond the neutron drip line (see Fig. 2). Most of these nuclei have never been observed in a laboratory. The location of the neutron drip line itself is of prime interest as it defines the depth in the crust where neutrons appear. Currently the neutron drip line is only known experimentally up to oxygen. Here FRIB with its unique reach to produce the most neutron rich isotopes will have a significant impact (see section 3.2). It is possible that FRIB experiments will delineate the neutron drip line up to around $A \approx 100$ (depending on where the drip line turns out to be located) covering essentially the range needed for modeling of accreted crusts. Mass measurements of neutron rich nuclei with $A < 100$ at next generation radioactive beam facilities are needed to define the location of electron capture transitions, and charge exchange reactions and β-decay studies can constrain the weak interaction strengths, in particular the excitation energies of the lowest lying transitions that can directly affect heat deposition and cooling. Experiments with low energy beams of very neutron rich nuclei can be used to explore the dependence of fusion reaction cross sections on neutron richness.

For the foreseeable future, neutron star crust models will have to rely on a combination of experimental and theoretical data (see section 3.6). Especially modifications to masses and effects such as superfluidity, pasta phases, and neutrino emissivity will have to be calculated using nuclear theory. An important development are mass and drip line predictions by modern density functional theory, which can provide estimates for theoretical uncertainties that can be taken into account in astrophysical models. Shell model calculations can provide relatively reliable electron capture and β strength, but the effective interactions need to be tested with data on neutron rich nuclei, especially in the electron capture direction. A major challenge are predictions of weak interaction strength in neutron rich nuclei that are heavier than current shell model spaces. QRPA approaches are frequently employed in astrophysical model calculations, but have been shown to be unreliable in predicting the strength in a few isolated low-lying states needed for neutron star crust models.

**Magnetar flares:**

The properties of the neutron star crust may also be directly observable through the flares of high-energy X-rays and gamma rays which are emitted from magnetars. These flares are thought to be the result of seismic activity in the crust; the relationship between the energy of the flares and the time between flares follows a log-normal distribution similar to that of earthquakes. In a few of the most energetic flare events, quasi-periodic oscillations have been detected. While the precise mechanism connecting seismic events in the crust with observed oscillations is unclear, many recent modeling efforts suggest that the oscillation frequencies are potentially connected to the nuclear properties of the crust. This is an exciting connection between nuclear theory and observations of oscillations.

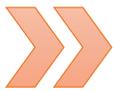

## 2.6.6   Neutron Star Strategic Thrust 3: The nuclear matter equation of state

The nuclear matter equation of state is a fundamental aspect of matter, yet, it is not well known. Neutron star properties depend sensitively on the equation of state of cold nuclear matter in a density range of 0.1 to 10 times the nuclear saturation density ($2.7 \times 10^{14}$ g/cm$^3$). Particularly uncertain is the density dependence of the symmetry energy - the energy difference between nuclear matter with protons and neutrons, and pure neutron matter. The symmetry energy determines a range of neutron star properties such as cooling rates, the thickness of the crust, the mass-radius relationship, and the moment of inertia. The better we can constrain the symmetry energy in laboratory measurements and using theoretical approaches, the more we can learn from neutron star observations, and the more meaningful will the constraints be that neutron star observations impose on the remaining aspects of the nuclear matter equation of state that are not accessible in laboratories.

Laboratory measurements that can constrain compressibility and symmetry energy are studies of masses, giant resonances, dipole polarizabilities, and neutron skin thicknesses of neutron rich nuclei. Extending such measurements to more neutron rich nuclei, and increasing the precision, especially of neutron skin thickness measurements, at existing and next generation radioactive beam facilities will be important (see section 3.2). Experimental constraints on the symmetry energy at supra-saturation densities can only be achieved in laboratory-controlled experiments by colliding and compressing nuclei in central collisions. Calculations predict that comparisons of the emission and flows of different members of isospin multiplets such as $(K^0, K^-)$, $(\pi^+, \pi^-)$, $(p,n)$ and $(^3He,t)$ in collisions between neutron-rich nuclei can allow such constraints to be extrapolated to neutron rich matter in astrophysical environments such as neutron stars and core collapse supernovae. Constraints on the isospin splitting between the neutron and proton effective masses, which is key to understanding the thermal properties of dense neutron rich matter, will be obtained at FRIB.

In general experiments do not provide direct data on the symmetry energy or the equation of state. An increased and comprehensive theory effort therefore has to accompany experimental programs to extract the relevant quantities with greater precision and to understand systematic uncertainties (see section 3.6). There are significant challenges for nuclear theory. Nuclear interactions and many-body approximations remain uncertain. Advances in the phenomenological description of nuclear reactions can significantly reduce the model dependence of experimental constraints on the density dependence of the symmetry energy. Advances that combine computational methods, Density Functional Theory and Effective Field Theory ideas can prove useful if experiments can be used to validate them.

We do have available a comprehensive strong-force theory for cold matter: QCD. For example, at supra-nuclear density hyperon-nucleon and hyperon-hyperon interactions can potentially be constrained by lattice QCD. Further developments for the application of QCD to cold matter are beginning to emerge. For these developments to be "useful" will require advances both in ideas (the sign problem) and computational efforts. In the mean time, plausible guesses for the phases of dense matter and calculations of their properties with improved techniques are continuing.

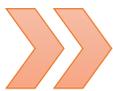

### 2.6.7  Neutron Star Strategic Thrust 4: Comprehensive models of accreting neutron stars

The surface and crust of accreting neutron stars exhibit a wide range of observable phenomena of current interest. Timescales are relatively short on astronomical scales, ranging from ms oscillations over burst phenomena to crust cooling timescales of years, making these systems unique laboratories to study astrophysical explosions and the properties of dense matter. To address the open questions, progress in astronomy and nuclear physics has to go hand in hand with progress in theoretical modeling. Current models are not adequate. The puzzling transition from unstable to stable burning at relatively low accretion rates, the appearance of burst oscillations, very short burst intervals indicating incomplete burning of fuel in X-ray bursts indicate that multi-dimensional models that can follow the accretion flow, hot spots, and the spreading of the burning front during X-ray bursts are needed to understand observations even on a qualitative level. This requires low Mach number modeling codes that are currently under development (see section 3.8 and 3.7).

The problems with explaining superbursts – neither can the correct amount of carbon fuel be produced, nor can ignition depth be understood - strongly indicate that effects such as rotation, gravitational settling of heavy elements, diffusion and phase separation at the liquid to solid transition at the bottom of the surface ocean have to be taken into account. While progress has been made in identifying the nuclear reaction sequences in the crust of accreting neutron stars, the interplay between the nuclear physics and magnetic field, lattice, plasma, and superfluidity has not been included systematically.

An additional complication is the interconnectedness of the different processes that require treatment with vastly different physics. Superbursts in the ocean depend sensitively on the composition created by the shallower X-ray bursts, and the ashes of the superbursts defines the possible nuclear reactions that can occur in the crust. Similarly, heat generated in the crust affects superburst ignition, and superbursts strongly affect the behavior of normal bursts triggering normal burst precursors, and quenching normal bursts for some time after a superburst. It is therefore important to treat all surface and crust processes in accreting neutron stars self consistently in a comprehensive way.

## 2.6.8 Impact on core-collapse supernovae, neutron star mergers, and the r-process

Neutron stars play a central role in nuclear astrophysics as the engines of core collapse supernovae and neutron star mergers and as the likely providers of neutron rich matter for the r-process. In core-collapse supernovae, strong-interaction physics at the neutrinosphere in the neutron star dictates the neutrino spectra and influences the r-process abundances. In neutron star mergers, the r-process material originates in the neutron star crust, and the composition of the crust and how it responds to stress caused by the merger dictates the amount of r-process material which is ejected. The nuclear equation of state and neutrino interactions in supernovae affect the dynamics of both collapse and mergers, but their full role remains to be systematically explored. The lower limit of the neutron star mass distribution constrains fall back, which is an important aspect of the explosion mechanism and supernova nucleosynthesis. In summary, the goal will be to arrive at a consistent picture of neutron stars and the nuclear physics that governs them, informed by X-ray observations, gravitational wave observations, and laboratory experiments.

## 2.7 Big Bang Nucleosynthesis

### 2.7.1 Introduction for non experts

Standard Big Bang nucleosynthesis (SBBN) is one of the most successful theories in nuclear astrophysics. As the relevant nuclear reactions are relatively well known, the amount of hydrogen, helium, and lithium produced about 3 minutes into the Big Bang can be predicted with just one free parameter, the cosmic baryonic matter density. With this parameter now fixed from observations of directional variations in the cosmic microwave background, the electromagnetic echo of the Big Bang, SBBN becomes essentially a parameter free theory. At the same time, the amount of hydrogen, helium, and lithium produced in the Big Bang can be inferred from observations, for example from observations of relatively unmodified gas clouds at large distances, or from observations of the composition of very old stars. Comparing the observationally inferred composition with the parameter free SBBN predictions provides stringent constraints on any Beyond the Standard Big Bang scenarios, including non standard Big Bang models, or Beyond the Standard Model particle physics such as dark matter decays and/or annihilations, non standard neutrino physics, the existence of new particles, and supersymmetry. SBBN light-element constraints on Supersymmetry are powerful and are complementary to accelerator probes, extending to parameter regimes inaccessible to the LHC. Overall the agreement between SBBN and the observed Big Bang element composition is good within observational uncertainties, with the exception of a puzzling discrepancy for lithium, the so called lithium problem.

### 2.7.2 Current open questions

- Does the discrepancy between predicted Big Bang lithium production and the primordial lithium abundance inferred from observations point to new physics?

- Will the combination Big Bang nucleosynthesis studies with new precision measurements of anisotropies in the cosmic microwave background and improved nuclear reaction rates reveal new particle physics?

### 2.7.3 Context

Since the 1999 White Paper, cosmology and particle physics have seen major progress and revealed profound surprises. Big-bang nucleosynthesis (BBN) has been cast in a new light in the era of precision cosmology: BBN has become a much sharper probe of physics beyond the Standard Model of particle physics and of cosmology. For example, measurements of the cosmic microwave background (CMB) will soon open the way to powerful and clean tests of neutrino physics using BBN. Moreover, the primordial 'lithium problem' increasingly seems to point to new physics at play in the early universe.

The simplest, 'standard' version of BBN (Standard BBN or SBBN) has only one free parameter: the cosmic baryon density $\Omega_b h^2$, or equivalently the baryon-to-photon ratio $\eta = n_b / n_\gamma$. SBBN produces only the lightest nuclides, and consequently requires a far smaller and thus simpler nuclear reaction network than typically found in stellar nucleosynthesis calculations; indeed, the light-element abundances are sensitive to only 11 reactions as well as the neutron lifetime. Moreover, the cross sections for these reactions are all measured in the laboratory at the relevant energies. Consequently, SBBN makes tight predictions for light element abundances, and it is feasible to do rigorous error analysis of the predictions and to express the results in terms of likelihood distribution functions.

BBN has a strong interplay with the precision determination of cosmological parameters via measurement of CMB anisotropies. Since the first WMAP data release in 2003, the CMB has provided the best cosmic "baryometer," independently of BBN. The SBBN and CMB determinations of the cosmic baryon density are in broad agreement; this rough concordance represents a great success of the basic hot big-bang model and a triumph for nuclear astrophysics. Furthermore, the CMB holds the promise (through measures of anisotropies at high multipoles) of also determining the cosmic helium abundance and the number of relativistic degrees of freedom (e.g., neutrinos). Indeed, the South Pole Telescope has presented determinations of these parameters, but with large uncertainties; with the *Planck* data release, these measurements are becoming competitive.

Using the precise CMB-determined cosmic baryon density, SBBN becomes a zero-parameter theory, and makes tight predictions for the light-element abundances. How do these compare with observations? High-redshift deuterium observations agree quite well, and low-redshift helium observations are also consistent. But the predicted abundance of $^7$Li is *higher* than that observed in Galactic halo stars by a factor 3–4, or 4–5$\sigma$. This discrepancy marks the primordial 'lithium problem.' Yet this 'problem' is in fact a success story for nuclear astrophysics: we have only become aware of this discrepancy due to the close interplay among theory, observation, and experiment.

Taken at face value, the lithium problem suggests new physics at play in the early universe, making a large perturbation to the lithium abundance but not to the abundances of deuterium or helium. Such perturbations can arise due to dark matter decays or annihilations; minimal Supersymmetry models can provide such perturbations as well. Changes in the fundamental constants can also solve the lithium problem.

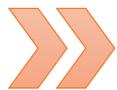

### 2.7.4 Big Bang Strategic Thrust 1: Improving BBN models

The precision of BBN predictions–both within and beyond the standard picture–directly depends on the nuclear inputs. While these are quite well-studied, a few important reactions remain outstanding and merit experimental study to further increase the reliability and precision of predictions to make BBN a more powerful probe for physics beyond the Standard Model. (1) The neutron lifetime remains an important component of the error budget for all light elements. (2) Some otherwise subdominant reactions could become important if their rates were enhanced due to resonances: $^7\text{Be} + d \xrightarrow{9} \text{B}^*$, $^7\text{Be} + t \xrightarrow{10} \text{B}^*$, and $^7\text{Be} + {}^3\text{He} \xrightarrow{10} \text{C}^*$. For these light species, direct calculation of nuclear structure is now possible using quantum monte carlo techniques, which can give important guidance in the reaction rates and possible resonant behaviors (see section 3.6). (3) Calculations for nonstandard BBN require accurate determinations of spallation and photodisintegration reaction cross sections and their uncertainties, spanning energies up to 1 GeV (see section 3.1). (4) For some key reactions such as d(pγ)$^3$He improved cross section measurements would further tighten SBBN particle physics constraints (see section 3.1).

As extragalactic and CMB determinations of the cosmic helium abundance improve, there is a need for improved precise calculations of the primordial $^4$He abundance and its uncertainties; this involves numerous subtle effects that are challenging to compute accurately and completely. The effects of neutrino oscillations and their interplay with effects of nonstandard neutrino properties (e.g., *CP* violation, nonzero chemical potential) must similarly be characterized accurately and completely.

A systematic comparison of BBN codes and their uncertainties would be a great service to the field.

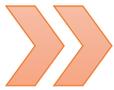

### 2.7.5 Big Bang Strategic Thrust 2: Astronomical observations.

The constraints that Big Bang nucleosynthesis provides are only as good as the primordial element abundances inferred from astronomical observations.

The possibility remains that the lithium problem could reflect systematic errors in the inference of the primordial lithium abundances from observations of Milky Way metal-poor stars. Indeed, the "Spite plateau" in lithium abundances recently has been shown to "melt down" at very low metallicity ([Fe/H]<-3), indicating that destruction of lithium occurs in at least some of these stars. On the other hand, no lithium abundances are found above the plateau. An understanding of these trends is critical. Fortunately, an independent method has very recently been found to testing lithium observational systematics. Lithium has been detected in the interstellar medium of the Small Magellanic Cloud. The SMC measurement marks the first extragalactic lithium measurement and gives an abundance comparable to that of Milky Way stars with similar metallicity, suggesting that large depletion and/or systematic errors are not plaguing the stellar observations. Future observations of interstellar lithium in low-metallicity galaxies hold great promise of independently determining the primordial $^7$Li abundance, and possibly separately determining the lithium isotopes and thus providing unique and robust new information on the $^6$Li abundances.

Deuterium observations in high-redshift quasar absorption line system provide a strong probe of the cosmic baryon density and test of BBN. Unfortunately, suitable systems are rare: after nearly two decades of effort, ~10 systems give solid D/H abundances. Recent D/H measurements have improved the D/H observational precision to better than 2now H measurements are more accurate than the theory predictions, pointing the need for improved cross section measurements. At this level of accuracy the D/H now becomes an important probe of N$_{eff}$ and thus new physics.

$^4$He observations in extragalactic HII regions are also dominated by systematic uncertainty. Fortunately, *Planck* CMB measurements have given a competitive new measurement of primordial helium. In addition *Planck* provided new data on the cosmic baryon density, and the number of relativistic degrees of freedom (expressed, e.g., as the effective number N$_{ν,eff}$ of light neutrino species), based on the clean determination of

CMB anisotropies. BBN predicts light-element abundances as a function of the baryon density and–going beyond the standard model–of $N_{\nu,\text{eff}}$. Thus the *CMB data alone* will provide a zero-parameter test of the consistency of the BBN prediction, and will thus probe new neutrino physics and/or new physics of any other "dark radiation."

## 2.7.6 Impact on other areas in nuclear astrophysics

The lithium problem currently hinges on the evolution of lithium in halo stars, and its spectral signature. Better theoretical and observational signatures of lithium depletion and diffusion, and therefore improved understanding of the processes in stellar interiors are critical to determine a stellar modeling contribution to the lithium problem.

## 2.8  Galactic Chemical Evolution

### 2.8.1  Introduction for non experts

One of the goals of nuclear astrophysics is to explain the relative abundances of elements and isotopes in the cosmos. Observations of the composition of the solar system or other stars provide a snapshot of the composition of the interstellar medium at the time and location the respective stellar system formed. In general, observations don't provide insights on individual nucleosynthesis events, but instead inform us about the cumulative effects of many, in the case of the solar system perhaps hundreds, of nucleosynthesis events that occurred in our Galaxy at earlier times (an exception are the compositions of rare, extremely old stars that formed out of the debris of very few nucleosynthesis events). In order to confront nucleosynthesis models with observations it is therefore of critical importance to model the gradual enrichment of the galaxies through various nucleosynthesis events over time. The results do not only depend on the nuclear physics and astrophysics of the individual events that affect their contribution to the galactic inventory of nuclei, but also on how our Galaxy formed, how gas and dust are transported, and how stars form. While this greatly complicates the modeling of Galactic Chemical Evolution, it also offers a new opportunity to investigate fundamental questions on star and galaxy formation through nucleosynthesis. With an explosion of observational data on the evolution of the chemical composition of the Galaxy, and with a new generation of chemical evolution model frameworks that rely on high performance computing, the field is now at the verge of a major advance.

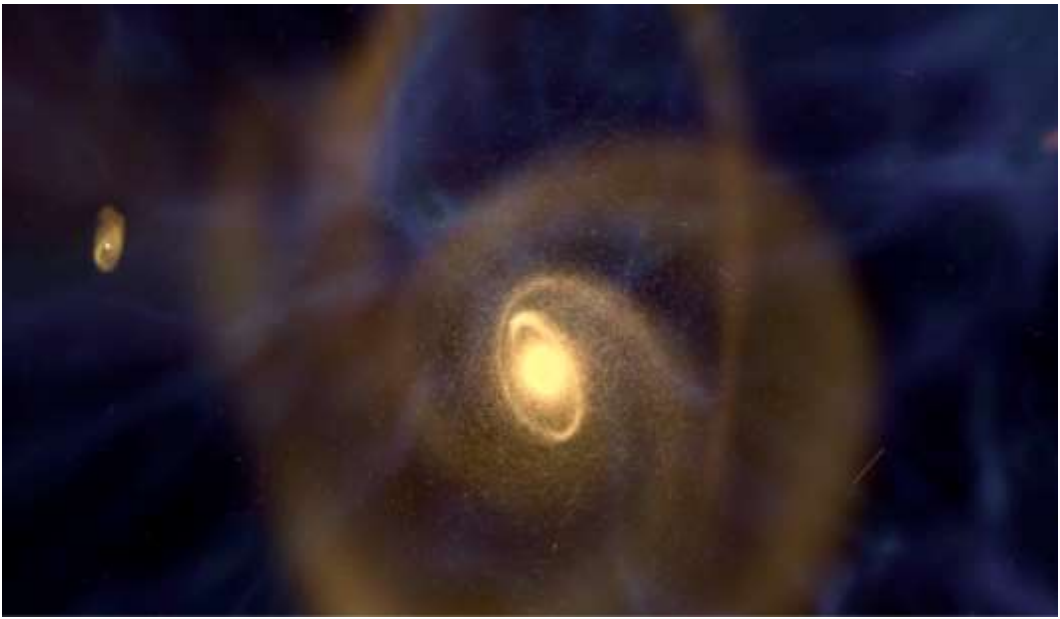

Figure 18: N-body simulation of the formation of a Milky Way like galaxy by collisions and merging of numerous smaller galaxies. Such calculations can be coupled with chemical evolution models that trace the creation and destruction of elements by nuclear processes in stars and stellar explosions contained in each yellow pixel, to determine the resulting

distribution of the elements. By comparing with observations of the chemical composition of the stars's in the Milky Way, one can then learn about the sequence of events that led to the formation of our Galaxy after the Big Bang, provided the element formation processes are well understood. (Image Credit: Brian O'Shea (MSU) and Donna Cox (NCSA Advanced Visualization Laboratory)

## 2.8.2 Current open Questions

- **IMF:** Is the stellar initial mass function invariant over time, with metallicity, and galaxy type?

- **ISM mixing:** How effective is the mixing of the interstellar medium? In other words, how much of the observed variation in nucleosynthesis at low metallicity is due to distinct progenitor populations vs. inhomogeneous mixing? Similarly, at later times, what is the role of radial mixing in the Milky Way disk?

- **Pop III stars:** What is the initial mass function of Population III stars, and does it evolve over time? What is the typical multiplicity and distribution of rotation rates for primordial stars? Is there a truly unique nucleosynthetic signature of Population III stars, and can we infer Pop III stellar properties from low-metallicity stellar abundances?

- **Galaxy formation:** To what extent can the Milky Way be regarded as a template for galaxies of its type? How did the components of the Milky Way stellar halo form? What are the signatures of galaxies merging into the Milky Way, and what do they tell us? Is there a fundamental explanation for the differences in enrichment of the Milky Way vs., e.g., the Magellanic clouds? Can hydrodynamic GCE models reproduce the mean trends and the range of variability seen in the solar neighborhood and in star clusters?

- **Stellar populations:** What are the progenitors of the carbon-enhanced metal poor (CEMP) stars, particularly those at the lowest metallicities, and why are there more CEMP stars as one goes to lower [Fe/H]? What are Type Ia supernovae, and how do their populations evolve? How varied can supernova explosions be in their nucleosynthetic outputs? What is the site or sites of the r-process? What are the limits of chemical variability for stars hosting exoplanets?

- **Nucleosynthesis calculations:** What are the key ingredients for modeling the evolution of Population III stars, and for the most massive stars? What is the contribution of intermediate-mass (1-8 Msun) stars to the galactic/extragalactic neutron-capture elements? What are the effects of binary stars and cosmic rays on nucleosynthesis? What is responsible for the behavior of neutron-capture elements at low metallicities?

- **Cosmic Rays:** What is the impact of spallation and cosmic ray propagation on the composition of the interstellar medium.

### 2.8.3 Context

There have been several important observational advances. Perhaps most profoundly, the Sloan Digital Sky Survey, its follow-on programs, and high-resolution spectroscopic surveys have enabled incredible observational advances with regards to Galactic stellar populations. SDSS has enabled the discovery of two kinematically and chemically distinct components of the Milky Way stellar halo, a population of ultra-faint dwarf galaxies that have extremely low-metallicity stellar populations whose formation was truncated very early in the age of the Universe, and clear signatures of galaxies merging into the Milky Way (including stellar streams and "ringing" of the Galactic disk). Furthermore, these surveys have produced tremendous insights with regards to the early evolution of the Milky Way's progenitors: several interesting low-metallicity populations have been discovered, including carbon-enhanced metal poor stars, and other chemically peculiar stellar populations that have distinct neutron-capture signatures. Nebular spectroscopy has enabled the study of the nucleosynthesis of neutron-capture elements in both galactic and extragalactic planetary nebulae and HII regions. And, finally, low-metallicity, high-redshift Damped Lyman-Alpha systems that are metal-poor, but contain both enhanced carbon and r-process elements have been discovered. All of these observations provide crucial clues to galactic chemical evolution.

Theoretically, ever-growing computational capabilities have resulted in tremendous insights. This includes high-resolution simulations of Population III star formation demonstrating that Population III stars can form in binary or higher-multiple systems. A critical step forward has been made by using semi-analytical models that couple N-body simulations stellar evolution models (see Fig. 18), as well as full chemodynamic simulations that include both the hydrodynamical and stellar history of a galaxy and its progenitors. These models incorporate chemical evolution models with detailed yields, and enable the comparison of both kinematic and chemical behavior of stellar populations.

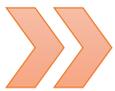

### 2.8.4 Chemical Evolution Strategic Thrust 1: Observations

Photometric and medium-resolution spectroscopic surveys of large numbers of stars are well underway with the prospect of dramatically increasing the number of old stars with known composition into the millions (see section 3.9.4). The most crucial need is for high resolution spectroscopic follow-ups of metal-poor field stars, dwarf galaxies (particularly ultra-faint dwarf galaxies), open clusters, and globular clusters. This will help to establish in detail the star formation history of these objects, as well as the frequencies of various nucleosynthetic phenomena (e.g., carbon-enhanced metal poor stars, r-process enhanced stars, etc.), and their orbital and binary properties. A critical (and related) theoretical advance that is required is for more careful modeling of stellar model atmospheres, particularly in 3D, to study the effect that this has on hard-to-measure elemental and isotopic abundances.

Observations of isotopic abundances in interstellar gas are key to determine the mixing scales in space and time, from nucleosynthetic production of new metals towards their ingestion into next-generation stars. Here, advances in the detection of diffuse $\gamma$-ray line emission from long-lived radio-isotopes ($^{26}$Al, $^{60}$Fe) address the My time scale, while the observation of sub-mm lines addresses the time scales preceding the next star formation events. Such observations are complemented by studies of the composition of primitive meteorites that determine the abundances of specific radioactive isotopes in the early solar system.

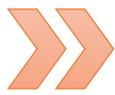

### 2.8.5  Chemical Evolution Strategic Thrust 2: Theoretical Advances

The most critical theoretical advance needed is for libraries of stellar models for GCE evolution using consistent (and possibly open-source) modeling tools and nuclear reactions, densely covering mass and metallicity space. In particular, the field needs improved nucleosynthetic yields for metal-poor and metal-free stars and for intermediate-mass (i.e., asymptotic giant branch) stars. Related to this, there is a critical need for a better understanding of mixing physics in stars and of stellar mass loss, and for an r-process nucleosynthetic models that can explain both the actinide boost phenomenon and the r-process observations in metal-poor stars. On a larger scale, there is a clear need for detailed models of the formation and star formation history of the fundamental building blocks of the Milky Way, including dwarf-like galaxies, and particularly the ultra-faint dwarf population. A related need is for the development of chemodynamical predictions for large spiral galaxies such as the Milky Way and M31, and for techniques to make detailed comparisons between such models and upcoming large astronomical surveys such as LSST, GAIA, HERMES, etc.

Modern chemical evolution studies require aggregating, analyzing, and curating massive amounts of data from observations, experiment, and simulations. The planned introduction of 100Gb internet nationwide will be critical to collaboration and to making progress in the era of Big Data (see section 3.8).

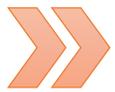

### 2.8.6  Chemical Evolution Strategic Thrust 3: Nuclear physics advances

At the heart of chemical evolution are the nuclear processes that create new elements. The advances in nuclear physics (and other areas such as supernova models or the sites of s-, r-, and p-processes) described in section 2.1 are therefore essential for progress in understanding chemical evolution. Of particular importance are key nuclear reactions, where relatively small changes in the rates affect nucleosynthesis broadly, such as the triple alpha and $^{12}C(\alpha,\gamma)^{16}O$ rates.

### 2.8.7  Impact on other areas in nuclear astrophysics

The strong dependence of chemical evolution on galaxy formation and star formation history opens the door to probing these processes once advances in modeling, nuclear physics and observations have been achieved.

# 3 EXPERIMENTAL, OBSERVATIONAL, AND THEORETICAL TOOLS FOR NUCLEAR ASTROPHYSICS

## 3.1 Stable and γ-beam facilities

Stable beam experiments have been a major contributor to our understanding of nuclear astrophysics since the field began in the 1930s. Photon beams are a relative newcomer, but also promise significant impact on our understanding of stellar evolution and nucleosynthesis. Specifically we ask: where are the elements produced that are the building blocks of life? Where are 50% of the elements beyond iron produced (see section 2.1)?, How do stars evolve (see section 2.2)? What was the early history of the solar system? Are there nuclear signatures to probe the stellar interior, e.g., CNO neutrinos for determining the metallicity of the sun (see section 2.2.7)? Beyond this, we recognize that almost all questions in astrophysics ultimately require a detailed understanding of stars and stellar properties.

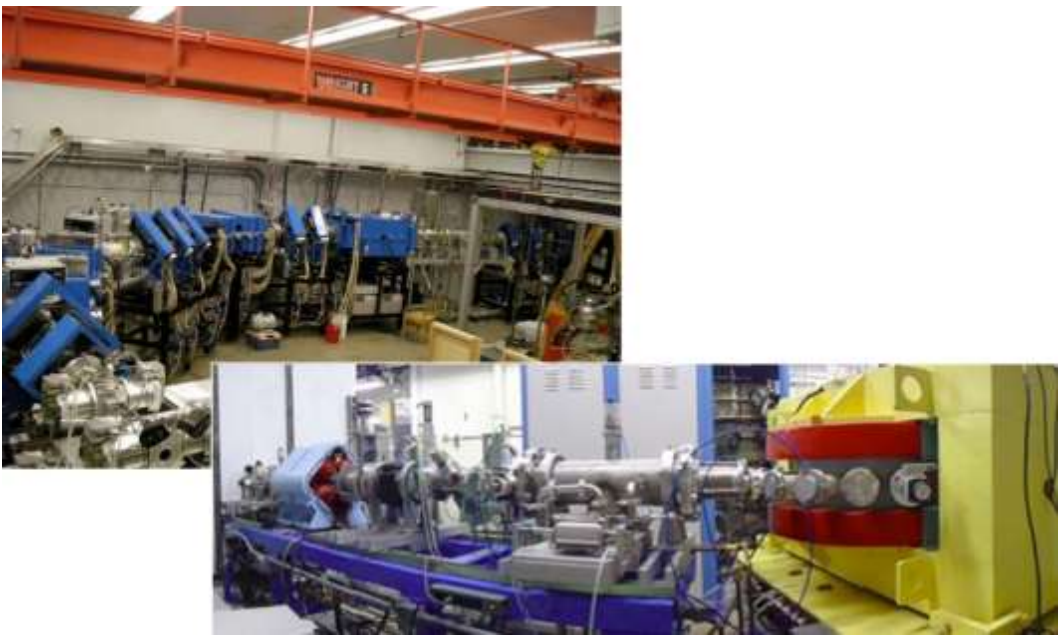

Figure 19: University accelerator facilities, organized in the ARUNA network, provide stable, neutron, and gamma beams that are critical for measurements of astrophysical reactions that involve stable nuclei and govern the evolution of stars. (Upper left: University of Notre Dame Nuclear Science Laboratory, Lower right: LENA accelerator at TUNL, Credit: Christian Iliadis)

Some of the poorly known nuclear reactions impacting these issues are capture and (p,α) reactions, for example $^3$He(α, γ)$^7$Be, $^{14}$N(p,γ)$^{15}$O, $^{17}$O(p,γ)$^{18}$F, $^{17}$O(p,α)$^{14}$N, $^{12}$C(α,γ)$^{16}$O, the NeNa and MgAl cycles, etc.; heavy-ion reactions such as $^{12}$C+$^{12}$C, $^{12}$C+$^{16}$O, $^{16}$O+$^{16}$O; neutron sources and neutron poisons that include $^{13}$C(α,n)$^{16}$O, $^{17}$O(α,n)/(α,γ), and $^{22}$Ne(α,n)/(α,γ). Note that this is not meant to be an exhaustive list, but merely indicates that a significant amount of work remains to be done (see also sections 2.1.4 and 2.2.4). These and other reactions involving stable targets (e.g. $^{14}$N(p,γ)$^{15}$O, $^{12}$C(α,γ)$^{16}$O, and $^{22}$Ne+α) also have a tremendous impact on stellar explosions, since they determine the seed abundance distribution for shock-induced nucleosynthesis in core collapse supernovae, such as the p-process, and for the nucleosynthesis patterns in thermonuclear runaways, such as the αp- and rp-processes (see section 2.3.6). Progress on these fundamental questions in science will require investment in advancing the state of the art in experimental sensitivity.

### 3.1.1  Experimental Methods and Techniques

Direct measurements of reactions at the low energies of interest require high beam currents, efficient and low-background detection methods, and long running times. Backgrounds can be reduced by utilizing pulsed beams, ultra-pure targets, passive and active shielding, coincident detection techniques, and/or by performing the measurements deep underground. The existing Laboratory for Experimental Nuclear Astrophysics (LENA) at the Triangle Universities Nuclear Laboratory (TUNL) is presently performing such measurements with normal kinematics and is completing an upgrade that will increase the beam intensity into the tens of mA range. An underground accelerator, CASPAR, is presently being developed by a collaboration involving the University of Notre Dame, South Dakota School of Mines and Technology, and Colorado School of Mines. This facility located at the Sanford Underground Research Facility (SURF) is designed to study the s-process neutron source reactions, which are cases where the background reduction resulting from being underground is particularly compelling. Another approach to direct measurements is inverse kinematics, where a heavy ion beam bombards a gas target of hydrogen or helium and the heavy reaction products are detected in a recoil separator with high efficiency. The recently-completed heavy-ion accelerator and recoil separator (St. Ana and St. George) at the Notre Dame Nuclear Science Laboratory (NSL) is an example of this approach.

Indirect methods are necessary in situations where direct measurements are not possible. Transfer reactions are particularly useful for the determination of excitation energies, spins, parities and partial width information for astrophysically-important states. These approaches provide crucial complementary information for estimating reaction rates, and are also helpful for guiding difficult direct measurements with stable and radioactive beams. Specific implementations, such as the Asymptotic Normalization Coefficient or the Trojan Horse methods, use nuclear-reaction models to link spectroscopic information to the stellar reaction. Such experiments are performed, for example, at the Cyclotron Institute at Texas A&M University (TAMU). Indirect approaches often require some reaction theory modeling, and hence need careful validation in order to obtain defensible reaction rate uncertainties. The best resolution for detecting charged-particle reaction products is provided by magnetic spectrometers, a capability which is presently limited in North America to the MDM spectrometer at TAMU. However, the situation will improve in the near future: the John D. Fox Accelerator Laboratory at Florida State University (FSU) is planning to install the Enge split-pole spectrometer that was previously utilized at Yale University and TUNL is upgrading their Enge split-pole spectrometer. These developments together with the planned upgrade of the focal plane detector at TAMU will provide a robust set of tools for high resolution charged-particle spectroscopy. For the case of neutrons in the final state, high-resolution neutron spectroscopy is carried out using pulsed beams and long flight paths at Ohio University's Edwards Accelerator Laboratory.

Apart from "direct" and "indirect" measurements using hadron beams, photon beams have become increasingly important for nuclear astrophysics. The world's premier laboratory in terms of photon beam intensity and resolution is the High-Intensity $\gamma$-ray Source (HI$\gamma$S) at TUNL. Measurements can be performed by directing a quasi-monoenergetic $\gamma$-ray beam on a suitable sample. The cross section for a reaction of interest can be obtained by measuring the reverse reaction and by applying the reciprocity theorem. A particularly important example is the possibility of measuring the $^{16}O(\gamma,\alpha)^{12}C$ reaction in order to estimate the $^{12}C(\alpha\gamma)^{16}O$ rate, which is crucial for the evolution of stars, at low energies. This would require the development of new techniques, which may include a further increase in photon beam intensity. Similar measurements are also in progress, using bremsstrahlung photons produced at the Jefferson Laboratory. Near threshold levels that are important as sub-threshold state or resonant contributions to the reaction cross section can also be studied via nuclear resonance fluorescence, i.e., $(\gamma, \gamma')$, in order to measure precise compound level energies and quantum numbers.

### 3.1.2  Opportunities and Experimental Needs

The past decade featured the construction and commissioning of dedicated stable- and photon-beam facilities for nuclear astrophysics in the U.S (Fig.19). All of these facilities are at university-operated accelerator laboratories: TUNL (high-intensity proton beams, mono-energetic photon beam), Ohio University (neutron time-of-flight), Texas A&M University (indirect transfer studies), and University of Notre Dame (normal kinematics and inverse kinematics with a recoil separator). The unique and diverse training of graduate students at university facilities cannot be overemphasized: by the time of graduation, students working in nuclear astrophysics have become experts in radiation detectors, high-voltage and vacuum systems, electronics, computer programming, and the physics associated with their projects. Consequently, they are highly attractive for academic, federal and corporate employers in an increasingly competitive job market.

Certainly, some important stable-beam and $\gamma$-ray-induced reactions are already within reach and will be measured at these facilities over the next decade. However, to take full advantage of the capabilities of these laboratories for advancing the field, future equipment upgrades are essential. These needs include: (a) development of higher ion beam intensities and implementation of pulsed low-energy beams; (b) next-generation $\gamma$-ray detector arrays with an increased detection efficiency; (c) construction of ultra-pure low-background neutron detectors, based on the extensive expertise of the neutrino and dark matter community; (d) increase of the photon-beam intensity available at HIγS.

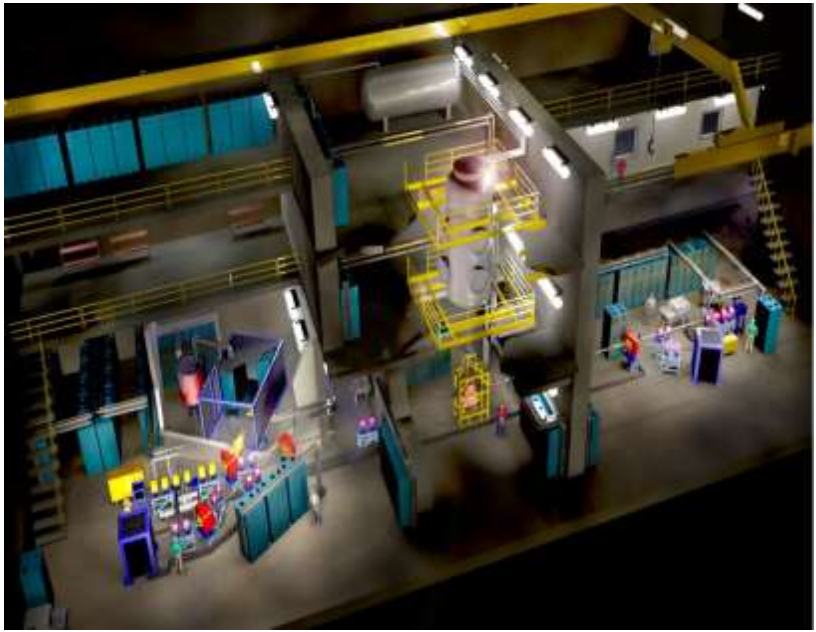

Figure 20: A low energy underground accelerator for intense stable beams is a high priority for nuclear astrophysics. The cosmic ray background shielding provided at an underground location enables measurements of very slow astrophysical reaction rates at, or very close to astrophysical energies. (DIANA concept study).

We expect that these experiments will approach the astrophysically-important energy region and, thereby, significantly improve knowledge of crucial thermonuclear reaction rates. However, stable-beam measurements at stellar energies ultimately require a dedicated underground laboratory, where the cosmic-ray muon background is reduced by orders of magnitude. The only underground accelerator facility at the moment is the Laboratory for Underground Nuclear Astrophysics (LUNA), operated for the past 20 years by a European collaboration in the Laboratori Nazionali del Gran Sasso, Italy. LUNA has limitations in terms of beam intensity, ion beam type, and detection versatility for measuring the key reactions discussed above and so in response to the urgent scientific need for a next-generation underground accelerator facility in the U.S., a pilot project, the Compact Accelerator System for Performing Astrophysical Research (CASPAR) is presently under construction at the SURF. CASPAR will initially focus on the measurement of stellar neutron sources using intense helium beams that are not available at LUNA. However, it is also limited in beam current and detection systems. Thus, we propose the development of a high-intensity accelerator laboratory located deep underground to ensure a broad range study and analysis of critical reactions in stellar helium, carbon and oxygen burning, that govern stellar evolution and provide the seeds for all subsequent nucleosynthesis processes (Fig. 20). Such a facility would be based on the existing DIANA concept.

In summary, the highest long term priority for this subfield is the construction of the high current underground facility based on the existing DIANA concept. In the near term, the priority is to maintain operation and invest in upgrades of the existing facilities and instruments dedicated for nuclear astrophysics research with stable beams.

## 3.2  Radioactive beam facilities

Radioactive isotopes play a critical role in the reaction sequences that create heavy elements in nature (see section 2.1.4 and Fig. 2), that govern supernova explosions and their synthesis of observable $\gamma$-ray emitters (see section 2.1.7 and 2.3.7) that power novae (see section 2.5.6), that power X-ray bursts (see section 2.6.5), and that drive nuclear transformations in neutron star crusts (see section 2.6.5). Laboratory studies of the structure and reactions of neutron rich nuclei are also important as probes of the equation of state of neutron rich nuclear matter that defines the properties of neutron stars (see section 2.6.6). Radioactive beam facilities are therefore critical for addressing many of the open questions in nuclear astrophysics.

### 3.2.1  Existing radioactive beam facilities in North America

The highest nuclear physics priority of the nuclear astrophysics community is the expeditious completion of the planned FRIB facility which is scheduled to be operational by the end of the decade (Fig. 22). Until then it is crucial to pursue an active nuclear astrophysics program at existing radioactive beam facilities which will allow us to advance the science, to continue to make discoveries, to develop and test new equipment and to train the young researchers that will do the experiments at this next generation radioactive beam facility. All existing RIB facilities in North America are presently involved in a vibrant experimental program and at the same time undergo important upgrades that will increase the available beam intensities or give access to new nuclei that are critical to nuclear astrophysics (Fig. 21). These facilities are briefly summarized below.

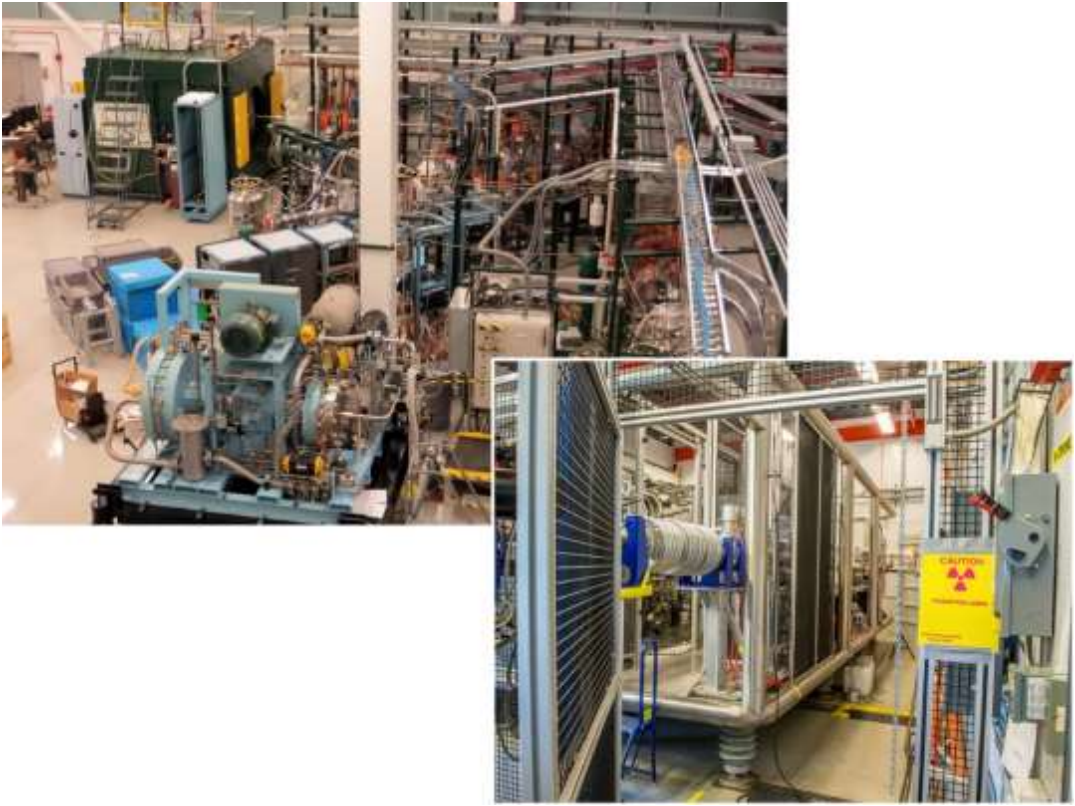

Figure 21: Upgrades at existing radioactive beam user facilities such as ReA3 at Michigan State University's NSCL (upper left, credit: Thomas Baumann) or CARIBU at Argonne National Laboratory (lower right) offer significant new opportunities to determine critical properties of unstable nuclei that govern the creation of elements in stellar explosions.

**ANL:** At Argonne National Laboratory, the CARIBU project has recently been completed providing access to new nuclei on the n-rich side of the mass valley. First experiments with a 400 mCi Cf source have started measuring masses with the Canadian Penning Trap. Transfer reaction studies using the HELIOS spectrometer and measurements of $\beta$-delayed n-branching ratios will start soon. The In-Flight radioactive beam program is planning to install a dedicated separator which together with liquid production targets will increase the beam intensities of experiments with radioactive isotopes by factors of $10^2$-$10^3$. With this upgrade $(\alpha,p)$ and $(^3He,d)$ reactions (surrogates for $(p,\gamma)$) which are critical for novae and X-ray bursts will be measured.

**FSU:** At Florida State University, the RESOLUT facility has started a new program with radioactive beams produced with the In-Flight technique. With the planned installation of additional resonators from KSU the range of radioactive beams will be extended towards heavier nuclei. An active target detector system (ANASEN) has been developed and plans for a neutron detector (ResoNeut) are being pursued. This will allow for $(d,n)$ measurements as surrogate reactions of important $(p,\gamma)$ processes.

**University of Notre Dame:** The TWINSOL facility is one of the earliest designs to produce and separate light radioactive ions by the In-Flight technique using two superconducting solenoids. TWINSOL has been used for the study of critical reactions for understanding Big Bang nucleosynthesis and is presently focused on reactions associated with the hot CNO cycles. Besides studies in nuclear astrophysics TWINSOL has been instrumental in studying low energy fusion reactions with proton and neutron skin and halo nuclei to study the impact on low energy cross section behavior. The facility focuses on light radioactive beams because of the energy limitations of the FN tandem driver machine. Presently TWINSOL is being modified to operate also as a helical spectrometer for transfer studies using the first solenoid as collecting element and the second as helical separator. Further, TWINSOL is being coupled with unique detectors such as VANDLE and the prototype AT-TPC for additional capabilities.

**NSCL:** At the National Superconducting Cyclotron Facility, nuclear astrophysics is presently being pursued with fast, reaccelerated and stopped beams. The experiments include mass measurements, direct reaction and structure studies, weak interaction studies as well as half-live and decay experiments with astrophysically-important isotopes. A first campaign of the Gamma Ray Energy Tracking Array (GRETINA) was recently performed at the S800 spectrograph, yielding several results with high-impact on nuclear astrophysics. Also of great importance to nuclear astrophysics are radioactive beams from the ReA3 project which, together with a dedicated recoil separator (SECAR) and a gas target (JENSA), will allow us to measure the resonance parameters of some of the critical $(p,\gamma)$ reactions that play a role in explosive nucleosynthesis. Initially, the lower energy limit of beams available at ReA3 will be about 300 keV/u. It is foreseen that lower beam energies will be made accessible at a later stage, for example by placing the target at a high-voltage platform. The experience obtained in these experiments will be particularly important for similar studies with this device at the future FRIB facility, as discussed below. Continued improvements to the experimental techniques used with stopped beams, for example to increase the sensitivity and reduce the systematic uncertainties for mass measurements in ion traps, also have a strong impact on nuclear astrophysics.

**TAMU:** At Texas A&M University, the radioactive beam program, which in the past has used the In-Flight technique at the Mars recoil separator, is presently going through an upgrade. Light stable beams from the K150 cyclotron will produce secondary particles which are stopped in a gas stopper and then transported to the K500 superconducting cyclotron for acceleration. Planned experiments will investigate ground state properties of nuclei, carry out decay spectroscopy of astrophysically-important isotopes, and measure nuclear reactions.

**TRIUMF:** The Isotope Separator and Accelerator (ISAC) facility, based on the 500 MeV 50kW beam from the TRIUMF cyclotron, is currently the only operational ISOL facility in North America and the highest power ISOL facility worldwide. The low-energy beams from ISAC-I (20 keV to 1.7 AMeV), are used to measure relevant ground state properties, e.g. masses, half-lives, beta delayed neutron emission, as well as direct reaction rates for astrophysically-important reactions using the DRAGON and TUDA facilities. Furthermore, the superconducting ISAC-II linear accelerator provides Coulomb barrier energy beams for indirect reaction rate studies. With the addition of proton induced fission of actinide targets ISAC has started to explore opportunities for nuclear astrophysics involving neutron-rich nuclei. The Advanced Rare Isotope Laboratory (ARIEL), currently under construction, will provide additional opportunities for nuclear astrophysics. The 50 MeV, 10 mA electron linear accelerator (e-linac) will significantly extend the reach towards neutron-rich nuclei using photo-fission of actinides. Through the addition of the e-linac and a further proton beam line from the cyclotron ARIEL will provide full multi-user capabilities with the delivery of three parallel rare isotope beams. This will enable the development of the most challenging rare isotope beams for direct reaction rate measurements as well as the extended beam times for the measurement of these very low rates at astrophysical energies.

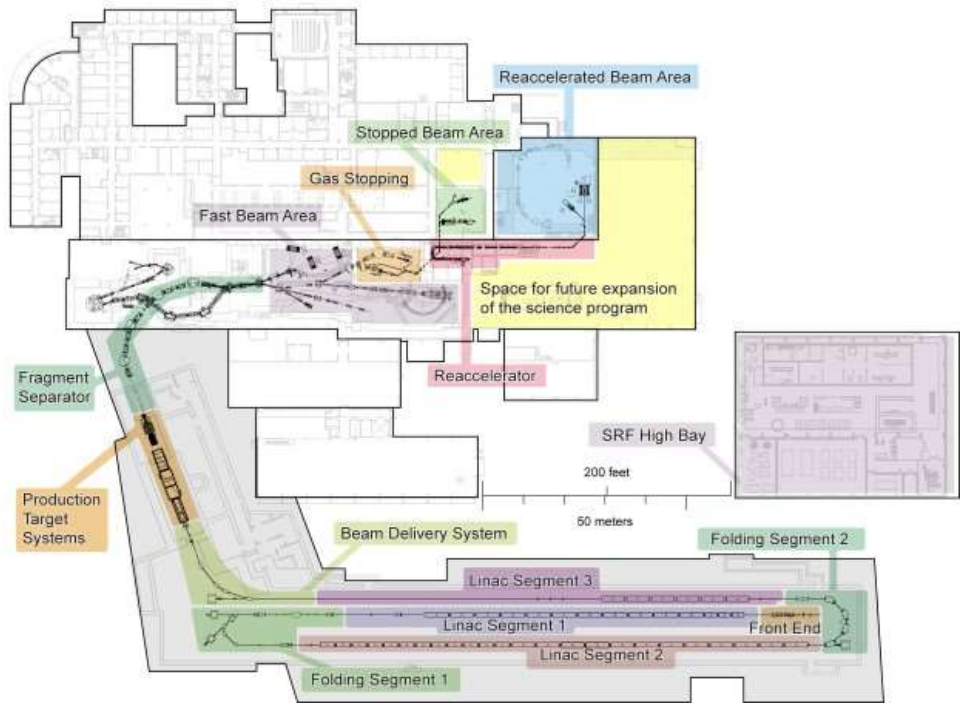

Figure 22: The highest nuclear physics priority of the nuclear astrophysics community is the expeditious completion of the Facility for Rare Isotope Beams FRIB. FRIB offers game changing opportunities to produce fast, stopped, and reaccelerated beams of unstable nuclei that govern the synthesis of elements in novae, supernovae, x-ray bursts, neutron stars, and neutron star mergers.

## 3.2.2  Facility for Rare Isotope Beams (FRIB)

FRIB will produce a wide variety of short-lived isotopes with sufficient intensity to unlock the fundamental nuclear physics for understanding stellar explosions and the origin of the elements. At FRIB, the number of isotope species that can be produced is roughly double of what is known at present and approximately 80% of the isotopes that are are estimated to exist. These species include a very large fraction of those needed to accurately model the r-process path in the very neutron-rich regions of the table of isotopes, and all of those needed to extract detailed information on the rp-process and p-process paths in the very proton-rich regions (see figure 2). It will be possible to demarcate the neutron drip line up to sufficiently heavy masses to model the crusts of accreting neutron stars. And it will be possible to measure the masses and determine the decay and reaction rates of most of the isotopes that play significant roles in a wide variety of astrophysical phenomena. Given the unparalleled discovery potential of FRIB, it is no surprise that expeditious construction of FRIB is the highest nuclear physics priority for the nuclear astrophysics community.

At FRIB, unstable isotopes of interest for astrophysics will be available in the form of fast, stopped and reaccelerated beams. In addition, longer-lived unstable isotopes can be harvested and used in offline experiments, either at FRIB, or at other facilities. The experimental tools to most efficiently and accurately determine the nuclear physics properties of the unstable nuclei of interest for astrophysics have been developed, or will have been developed when FRIB comes online, at a wide variety of institutions by a wide range of collaborations, as briefly summarized in the section on existing facilities.

The availability of reaccelerated beams of unstable isotopes at astrophysical energies is of critical importance for the study of the rp-process in X-ray bursts (see section 2.6.5), novae (see section 2.5.4) and supernovae (see section 2.3.5), as well as explosive Silicon burning in supernovae (see section 2.3.5) . Direct measurement of (p,γ) and (α,γ) reaction rates will be possible using the SECAR recoil separator (see Fig. 23), in combination with the JENSA gas-jet target. The construction of these devices is of acute importance for advancing the goals of the nuclear astrophysics community.

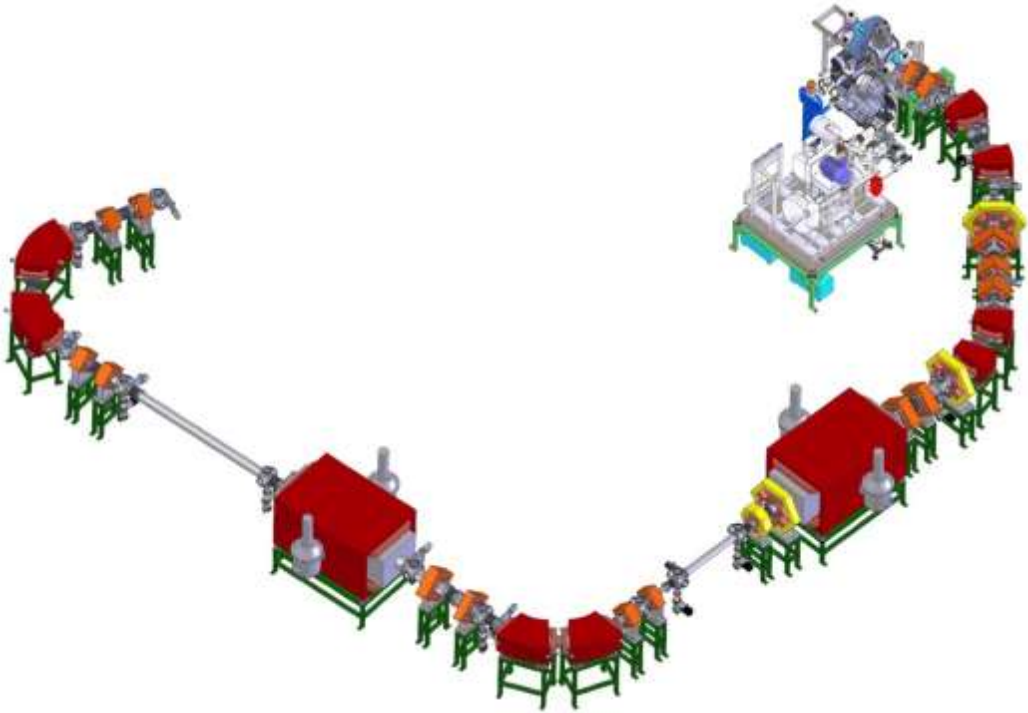

Figure 23: The proposed Separator for Capture Reactions (SECAR) will enable the direct measurement of astrophysical reaction rates with unstable nuclei at FRIB.

Active-target detectors, such as the AT-TPC and ANASEN have been developed and are available for measurements of astrophysical reaction rates with charged-particles in the exit channels, such as ($\alpha$,$p$) reactions. The ability of these detectors to measure energies and angles with good resolution and high luminosity makes them well suited for use with reaccelerated radioactive beams. A HELIOS-type spectrometer will also play an important role in the study of transfer and other direct reaction studies. And a variety of $\gamma$-ray spectrometers such as GRETINA (and the future full 4$\pi$ array GRETA), SeGA, HAGRID, CAESAR, SuN, and Apollo (which has been coupled to HELIOS) will be used for studying reactions involving $\gamma$-ray emission.

Reaccelerated beams of slightly higher energy (up to 15 MeV/u) also play an important role for extracting nuclear physics information for astrophysical purposes, mainly by using transfer reactions, e.g. (d, n) and (d, p), that serve as surrogates for proton and neutron transfer studies. Information about charged-particle direct-capture reactions can be obtained by measuring asymptotic normalization coefficients with unstable beams in inverse kinematics. A recoil separator, such as ISLA, for analyzing the forward-going recoils (at ~6-15 MeV/u) is required, in combination with charged-particle, and neutron detectors such as VANDLE and LENDA.

Longer lived harvested nuclei could be used for the determination of (n,$\gamma$) rates on unstable nuclei and provide data for branching points in the s-process (see section 2.1.4) by creating samples that can be irradiated at neutron beam facilities. Plans for the implementation of isotope harvesting are being developed, both for primary user experiments as well as for secondary harvesting from the primary beam dump, which enables true multi-user capability at FRIB. The availability of such long-lived isotopes for secondary experiments is critical - in neutron capture, for instance, existing neutron facilities have reached a point where measurements on many s-process branch points are possible, but samples are not available. The accurate determination of masses of neutron-rich isotopes is of critical importance for understanding the r-process (see section 2.1.4). High-precision mass measurements can be performed at the LEBIT facility using stopped beams. Further developments, such as the single-ion Penning trap, will be of tremendous value to further push the limit out to which the masses of neutron-rich nuclei can be measured.

The highest yields of the most exotic unstable isotopes at FRIB will be achieved for fast beams. Experiments utilizing fast beams will, therefore, be critical in providing nuclear structure and decay information for isotopes furthest from the valley of stability. Decay spectroscopy ($\beta$,$\gamma$,n and p) of implanted fast beams will be critical for measuring half-lives and decay properties of astrophysically-important nuclei. Ongoing improvements to the efficiency and accuracy of the detectors (such as the Beta Counting System, 3HeN, NERO for neutrons, and SuN, MTAS and a community-supported Clover Array for $\gamma$-rays emitted by stopped isotopes) used to characterize the various decay products will be important to optimize decay experiments performed at FRIB. In-flight decay spectroscopy using knock-out, Coulomb excitation, pickup and transfer reactions provides high-precision information on the structure of unstable isotopes. This information is required to improve theoretical models used in astrophysical calculations. More central heavy-ion collision studies provide information on the equation of state of nuclear matter, which is important for understanding dense environments such as neutron stars. Charge-exchange reactions with fast beams provide the only way to effectively constrain theoretical models used for estimating weak interaction rates, which are critical for understanding core-collapse and thermonuclear supernovae. Time-of-flight mass measurements with fast beams will be used for mass regions where the life-time is too short for high-precision trap measurements. The measurement of fission excitation functions of importance for astrophysics can be performed with both fast and reaccelerated beams.

A wide variety of detection systems for particle and gamma detection for experiments with fast-beams are, or will become available for early experiments at FRIB. At present, a large fraction of such experiments are performed with the S800 spectrometer at NSCL. However, in order to reach and perform experiments with nuclei furthest from stability, many of which play important roles in astrophysical phenomena, a spectrometer with a higher bending capability (up to 8 Tm, compared to the existing 4 Tm of the S800) is required. Such a High Rigidity Spectrometer will be very important to reach the long-term objectives of the nuclear astrophysics community.

## 3.3 High density plasma facilities

### 3.3.1 Existing high density plasma facilities in North America

A critical aspect of reaction rate calculations is the impact of the plasma environment in stars. Effects like electron screening affect in particular not only the decay and production rates through electron capture processes but also charge particle cross sections, which can be significantly enhanced by the reduction in Coulomb repulsion. These effects have so far only been computed using the Debye-Hückel theory, an approach that has not been tested experimentally for stellar plasma conditions. High-energy-density facilities provide unique opportunities to access dense, hot plasmas, and environments with extraordinarily high neutron fluxes. The short-lived high density and temperature plasma conditions resemble conditions in the interior of stars. This opens up the opportunity to study these effects for the first time and test the theoretical predictions that affect all reactions associated with processes in high density conditions from stars to thermonuclear explosions.

At OMEGA and NIF, inertial confinement fusion experiments have been used to observe charged particle reactions produced in plasmas at temperatures ranging between 2-20 keV and densities from 10 mg/cc to 100 g/cc.

**OMEGA:** Considerable efforts have resulted in the successful measurements of a number of light ion, deuteron and tritium induced fusion reactions at low temperatures collecting the reaction products by an external magnetic separator device. The results can be directly compared to accelerator based data to determine the low temperature rates in order to study the impact of plasma screening. This program is scheduled to be expanded to $^3$He induced processes that are critical for the solar pp-chain reactions.

**NIF:** Besides efforts to utilize the strong neutron flux for nuclear astrophysics related studies, a major goal is also the study of light isotope fusion reactions. Because of the higher temperature and density conditions achievable (compared to OMEGA) it is planned to study screening processes through selected high cross section reactions such as $^{10}$B(p,$\alpha$)$^7$Be through the collection and measurement of the characteristic $^7$Be activity.

In addition to these inertial-confinement fusion (ICF) facilities, the US nuclear astrophysics community has already taken advantage of short-pulse laser facilities such as the Texas Petawatt Laser to study select stable isotope reactions critical for stellar burning that are strongly affected by electron screening in the stellar plasma. The development of high-intensity laser systems with the chirped-pulse amplification (CPA) technique has opened up new possibilities for astrophysical research on nuclear interactions in laser produced-plasmas. Existing facilities in the US and future high intensity facilities such as the Extreme Light Infrastructure (ELI) in Europe can be employed for these studies that are complementary to what can be studied at traditional accelerator beam facilities.

## 3.4 Neutron Beam Facilities

In nature, the production of the elements beyond iron proceeds primarily through neutron-capture reactions, either in explosive environments (via an r process) or in stars (via s-process nucleosynthesis) (see section 2.1.4). Further, neutron-induced reactions, including (n,γ) and (n,α), are critical for modeling p-process environments as well as for providing the nuclear physics underpinnings for γ-ray astronomy and solar system formation. While measurements are still needed in select cases on heavy, stable isotopes, the most critical present needs are measurements on abundant, light isotopes, even-even isotopes in the mass $60 < A < 90$ region, and on heavy, unstable isotopes where often no measurements exist. First, abundant light isotopes may act as neutron poisons in the stellar environments where s-process nucleosynthesis takes place. Second, recent studies have shown that historic measurements of the cross sections of isotopes in the Fe-Ni-Zn regime suffer from systematic uncertainties that were not characterized adequately; these isotopes serve as bottlenecks that affect the full mass $60 < A < 90$ region of the weak s process. Finally, unstable isotopes on the s-process path serve as potential branching points in the s-process reaction sequence and define the characteristic isotopic abundance features in meteoritic inclusions. Unstable isotopes in the s-process that have long half-lives can also be observed directly with γ-ray observatories. In most cases no experimental data for neutron capture cross sections on these unstable isotopes have been obtained.

Because of the lack of a Coulomb barrier for these reactions, the energies of interest are significantly lower than for competing charged particle reactions—in fact, the energies reflect directly the thermal particle distributions. This provides additional challenges for theoretical efforts to calculate neutron-induced reaction rates. While the reactions sample regions of relatively high excitation, the details of the nuclear structure at these energies strongly impact the cross sections.

As a result, the measurements of cross sections for neutron induced reactions have focused primarily on direct reaction measurements with neutron beams. The last 15 years have seen significant changes in the experimental facilities for these measurements. The two major facilities, Forschungszentrum Karlsruhe (FZK) and the Oak Ridge Electron Linear Accelerator (ORELA), have both closed. Taken together, these two facilities represented roughly 80% of the world activity for neutron capture measurements for nuclear astrophysics. While the closing of these facilities is an obvious loss, the measurement techniques they developed and pioneered have been incorporated in the new facilities which have begun operation in the last 15 years. In addition, new techniques have been developed to use indirect methods that can provide the nuclear physics inputs for theory-based calculations of neutron induced reaction rates. In that respect the photon strength function has emerged as a particularly influential input. Techniques for measurements to constrain these inputs require a broad range of facilities, including neutron facilities, but also stable, gamma, and radioactive beam facilities.

### 3.4.1 Present Neutron Facilities

There are a number of neutron beam facilities of importance for nuclear astrophysics currently operating, where the US nuclear astrophysics community is presently involved:

**n TOF:** The neutron time-of-flight facility at CERN (Switzerland) couples both C6D6 and BaF2 detector arrays at the end of a ∼200 m flightpath to a neutron source driven by 20 GeV proton-induced spallation on a Pb target. The combination of a long flight-path with an intense production mechanism provides high neutron energy resolution, intense peak neutron flux, and moderate average neutron flux. A second experimental area at ∼20 m has recently been commissioned, with flux gains of almost a factor of 30.

**LANSCE:** The Los Alamos Neutron Science Center at Los Alamos (USA) is a facility with multiple neutron flightpaths driven by 800 MeV proton-induced spallation on tungsten. The primary capability for nuclear astrophysics comes from the Detector for Neutron Capture Experiments (DANCE), an 160 element, $BaF_2$ calorimeter located 20 m from the spallation target. DANCE was designed specifically with the goal of performing neutron capture cross section measurements on short-lived (>100 d) isotopes. The combination of short flight-path with high-intensity neutron source offers a high neutron flux with moderate neutron energy resolution. In addition to DANCE, LANSCE offers several other neutron flightpaths for the measurement of (n, n'), (n, p), and (n, α) reactions.

**FRANZ:** Frankfurter Neutronenquelle am Sternâ€"Gerlachâ€"Zentrum at the University of Frankfurt is under construction and first neutron beams are expected by the end of 2015. It will focus on the production of intense neutron beams specifically for nuclear astrophysics with keV time-of-flight beams of $>10^7$ $n/cm^2/s$ and activation beams of $10^{12}$ n/s. Detection systems will include a 4π $BaF_2$ array.

**GELINA** The Geel Electron Linear Accelerator in Belgium offers multiple flightpaths ranging from 10-400 m from the spallation target. Neutrons are produced via photo-neutron production from Bremsstrahlung from the 150 MeV electron beam. Detection capabilities include C6D6 detectors, ionization detectors for neutron-induced charged particle reactions, and $^6$Li-glass detectors for total cross section measurements. GELINA offers modest flux, but very high energy resolution, which is needed for total cross-section measurements.

**JPARC** The Japan Proton Accelerator Research Complex at Ibaraki, Japan includes the capability for neutron-induced reaction measurements with an HPGe array and NaI(Tl) spectrometer at the Accurate Neutron-Nucleus Reaction Measurement Instrument (ANNRI). An intense flux of time-of-flight neutrons is available from the subthermal up to the keV energy regime.

**SARAF:** The Soreq Applied Research Accelerator Facility (Israel) couples a high current proton accelerator to a liquid lithium target to provide high intensity Maxwellian-averaged neutron fluxes for activation measurements.

**NIF**: The National Ignition Facility (USA) is a plasma facility for light isotope fusion (d+d, d+t, t+t) reactions that produce as by-product intense nano-second bursts of neutrons in a 1-20 keV plasma environment. While the neutron spectrum is not matched to a stellar energy distribution, the combination of extreme peak flux with a plasma environment offers a new method of measuring nuclear reactions. The environment is challenging and presents a new set of systematic limitations. Development is underway to determine how to best exploit this new resource for nuclear astrophysics.

Taken together, the future of neutron facilities is quite bright. As an additional benefit to the nuclear astrophysics community, the operation and construction of many of these facilities has been funded by agencies that are not traditional funding sources for US nuclear astrophysics.

## 3.5  Neutrino Facilities

Direct experiments with neutrinos and nuclear astrophysics have been closely intertwined from their beginnings when nuclear physics underground experiments, detailed stellar modeling, and laboratory measurements of the fusion reactions in the sun revealed the solar neutrino problem, and, ultimately, the existence of neutrino oscillations. Today, the detection of astrophysical neutrinos is an important part of the multi-messenger observational approach to astrophysical phenomena that will play a critical role in the coming decade in addressing a number of open questions in nuclear astrophysics. Future measurements of solar neutrinos across the full energy spectrum open the opportunity to probe the thermal stability of the sun and determine the composition and structure of its core (see section 2.2.7). The observation of neutrinos from a supernova exploding in our Galaxy would offer a unique probe of the explosion mechanism and the conditions for the synthesis of new elements in the innermost regions of the explosion (see section 2.3.7). In addition, measurements of supernova neutrinos offer unique opportunities to probe important aspects of neutrino physics, including neutrino oscillations and mass hierarchy.

A broad range of neutrino detectors is required to make the necessary measurements. For solar neutrinos, low thresholds that enable future measurements of low energy neutrinos are of particular importance, while for supernova neutrino detection, continuous operation and a large detection volume to achieve good statistics are critical. For all measurements, low backgrounds and therefore well shielded underground locations are essential. The following list summarizes the main detection systems that are either available, being developed, or planned, and that are of particular importance for nuclear astrophysics. More information can be found in the white paper of the neutrino and fundamental symmetries Town Meeting.

### 3.5.1  Large water Cherenkov detectors

Large-scale water Cherenkov detectors such as Super Kamiokande in Japan's Mozumi mine or new planned megaton range detectors such as Hyper Kamiokande will play a critical role for the detection of supernova neutrino signals. The large detection volume will provide much higher statistics than other types of neutrino detectors, up to $10^5$ events for a supernova at 10 kPc distance. This will enable detailed mapping of the neutrino light curve, which will provide unique physics insights. While water Cherenkov detectors have played a key role in past solar neutrino observations, their high energy threshold makes them sensitive to only a small fraction of the solar neutrino flux. IceCube uses natural Antarctic ice instead of liquid water. Even though IceCube is designed to detect high energy neutrinos in excess of 10-100 GeV, it will be sensitive to low energy neutrinos from a Galactic supernova, and is predicted to detect of the order of $10^5$ events from an explosion at 10 kPc distance.

### 3.5.2 Argon TPC detectors

Time projection chambers filled with liquid argon (LArTPC) represent a new neutrino detection technology and a number of large scale detection systems have been proposed world wide. In the US such a detector would be constructed as part of the US long-baseline program LBNF at the Sanford Underground Research Facility SURF at Homestake Mine in South Dakota. LArTPC detectors offer exceptional sensitivity to electron neutrinos through the $^{40}Ar(\nu_e, e^-)^{40}K$ reaction. The detection of a supernova neutrino signal with a LArTPC detector would therefore provide important complementary information to the detection with a water Cherenkov detector. Combining information from water Cherenkov detectors and LArTPC detectors should enable the decomposition of the supernova neutrino signal into its electron neutrino, electron anti-neutrino, and heavy flavor neutrino components. Predicted event rates for the proposed LBNF detector for a 10 kPc supernovae are substantial, of the order of several thousand.

### 3.5.3 Organic Scintillators

Organic scintillator detectors offer lower energy thresholds and are therefore of prime importance for future solar neutrino measurements. This has been demonstrated impressively by BOREXINO, a joint European US effort, located at the Gran Sasso underground laboratory in Italy. BOREXINO provided the first direct measurement of pp-chain neutrinos from the sun, and the best upper limits on neutrinos from the CNO cycle. The main challenge for such measurements is to identify, characterize accurately, and reduce all backgrounds. On the North American continent, a SNO+ (Canada) liquid scintillator phase with its larger size and greater depth compared to BOREXINO has the potential to further improve solar neutrino measurements, provided backgrounds can be reduced sufficiently. LENA is a proposed next generation large scale organic scintillator detector in Europe that would provide unprecedented opportunities to detect low energy neutrinos from the Sun.

### 3.5.4 Inorganic Scintillators

Inorganic scintillator detectors are also used for dark matter searches and offer the potential of ultra low thresholds and reduced backgrounds as they do not suffer from $^{14}C$ background that tends to dominate organic scintillators. This may open the door to future high precision measurements of the pp-chain solar neutrino flux. Due to their low thresholds and their sensitivity to all neutrino flavors such detectors would also provide complementary information energy spectra and flavor composition of the neutrino flux from a Galactic supernova event. Proposed large scale detectors include CLEAN, a liquid neon detector proposed by a joint US-Canadian collaboration, and other projects based on liquid helium, neon, or xenon.

### 3.5.5  Metal Loaded Scintillators

Scintillation detectors can be loaded with isotopes that act as targets for charged current interactions with neutrinos. The outgoing charged particle is then detected in the scintillator. This approach offers opportunities for low thresholds and high precision spectral measurements of solar neutrinos, including pp-chain and CNO neutrinos. Proposed large scale detectors are LENS, a $^{115}$In loaded organic scintillator based detector at the Kimballton Underground Research Facility KURF, and ASDC (Advanced Scintillation Detector Concept), a water-based liquid scintillator detector proposed by a US-German collaboration that could for example be loaded with $^7$Li for precision measurements of pp- and CNO solar neutrinos.

### 3.5.6  Lead-based detectors

Lead has a large cross section for antineutrinos and can be used as the basis for a supernova detector with capabilities that complement those of other schemes. The analog states populated in both charged-current and neutral-current neutrino interactions are neutron unstable, and the detection of the neutrons forms the basis of an efficient supernova detector. Furthermore, neutrinos with sufficient energy can populate states that decay with the emission of two neutrons, giving a measure of the neutrino temperature. A detector based on this principle, HALO, is now in operation in SNOLAB. Although the target mass of 79 metric tonnes is relatively low, the detector is nevertheless sensitive to a supernova well beyond the galactic center, and could be readily scaled up given more lead and neutron detectors. Consideration has been given to providing the capability of detecting an initiating charged-current interaction by Cherenkov radiation, using a transparent liquid with dissolved or combined Pb instead of solid Pb metal.

### 3.6  Nuclear Theory

Nuclear theory plays a critical role in nuclear astrophysics (1) to predict nuclear quantities (such as nuclear properties, reaction rates, nuclear matter properties, or transport properties) (2) to calculate corrections to nuclear quantities due to the extreme astrophysical environments and (3) to extract astrophysically-relevant nuclear quantities from indirect experimental approaches.

Nuclear theorists already work on a wide range of topics that are important for astrophysics. In this section we summarize the most important areas.

It is worth noting that for processes that involve largely nuclei out of reach of experiments (such as the r-process) the uncertainties in nuclear theory predictions strongly hamper constraints on astrophysical models that can be obtained from observations. Progress therefore requires a reduction in these uncertainties. A first step, significant in its own right, is accurately quantifying the uncertainty in existing calculations. In the next few years, the effort to quantify the error in nuclear theory predictions should allow us to make more meaningful statements about processes like the r-process.

### 3.6.1  Nuclear Theory for Neutron Stars

Nuclear theory plays a critical role in the study of neutron stars (see section 2.6) because the extreme conditions realized in these ultra-dense compact objects cannot be directly accessed in terrestrial experiments. Recent developments in theory, observations and astrophysical simulations have begun to unravel how the properties of neutron-rich nuclei and dense nuclear matter shape astrophysical phenomena such as supernovae (see section 2.3) , gamma-ray bursts (see section 2.4) , x-ray bursts (see section 2.6.5) , neutron star mergers (see section 2.4) and related phenomena involving neutron stars. The equation of state (EOS) of dense matter, its thermal and transport properties, and the neutrino and related weak interaction rates in extreme nuclei and nuclear matter influence key observables: x-ray and gamma rays from neutron star bursts and transients, and neutrinos, gravitational waves, and nucleosynthetic yields from supernova and neutron star mergers (see section 2.6.6) .

**Ab Initio approaches:** A quantitative understanding of the interplay between many-body forces and non-perturbative effects in nuclei and dense matter is emerging though advances in nuclear theory and computational nuclear physics. Modern potentials based on effective field theory (EFT) naturally include an estimate of theoretical errors and provide a consistent framework to include many-body forces and couplings to external fields and currents. This, combined with advances in nuclear many-body theory, Quantum Monte Carlo methods, and access to high-performance computing have led to important breakthroughs: The EOS of uniform dense neutron matter at zero temperature, and the many-body response functions relevant to neutrino interactions rates have been calculated using ab initio theory. These calculations motivate and define future work to better pin down the role of non-perturbative effects and many-body forces in neutron-rich nuclei and dense matter. The challenge now is to extend these theoretical and computational methods to calculate the properties of asymmetric matter containing a small admixture of protons, heterogenous phases of dense matter encountered in the neutron star crust, and include the effects due to superfluidity, superconductivity and finite temperature. This will require developments in EFT to extend the current potentials to describe matter at higher density and will rely on a better understanding of power counting in the momentum expansion, three-body forces, and the role of the $\Delta$ and on advances in QMC, self-consistent Green's function methods, and many-body perturbation theory.

**Mean Field approaches:** Simulations of supernovae and neutron star mergers rely on an input EOS, thermal and transport properties, and weak interaction and neutrino interaction rates that span a vast range of density, temperature and composition. Here, models based on mean field theory with phenomenological interactions, and density functional theory will continue to play a central role. They remain the only practical methods at hand to cover the vast range of ambient conditions encountered in astrophysics. Theoretical developments to improve the nuclear energy density functional using insights from ab-initio theory and experiment are necessary. These improved models are also critical to interpret and motivate terrestrial experiments in intermediate and heavy nuclei, and to explore correlations between nuclear structure measurements and specific neutron star properties.

**Phases of nuclear matter in the neutron star crust:** Models to interpret bursts and transient phenomena observed in accreting neutron stars and magnetars rely on the properties of the neutron star crust. Matter in the denser regions of the crust where $10^{11} < \rho < 10^{14}$ g/cm$^3$ is characterized by a rich phase structure in where exotic spherical and non-spherical (sometimes called pasta) neutron-rich nuclei coexist with neutron matter in a phase that is both solid and superfluid at low temperature. To describe the ground state properties and transport properties of such heterogeneous matter, computational many-body methods such as QMC, (time-dependent) density functional theory, and molecular dynamics need to be extend to handle large numbers of particles, band structure effects in the neutron wave function and long distance correlations induced by the Coulomb interactions. Calculations of the thermal, mechanical and transport properties, and the non-equilibrium nuclear reactions and neutrino cooling rates which remain rudimentary but have identified the key questions to address and directions for future work. By combining large scale microscopic computations and effective field theory at low temperature, future work can calculate more reliably the superfluid gap, the low energy spectrum of collective excitations, correlation functions needed to determine the transport properties. The extension of these methods to larger temperatures is necessary to describe matter in the density range $10^{10}$ g/cn$^3$ ‹ϱ‹$10^{14}$ g/cm$^3$ encountered in supernova and neutron star merger is also within reach and is an important direction for future research. Interestingly, it has been realized recently that such hot heterogeneous matter can also be realized in low energy heavy-ion collisions in the laboratory, and these theoretical methods can be applied to interpret such experiments.

**Transport Theory for heavy ion collision experiments:** Advances in transport theory and intermediate energy heavy-ion phenomenology are critical to extract useful information about the high-density EOS from heavy-ion collision experiments. These experiments are the only means of creating nuclear matter at higher than normal nuclear densities in the laboratory and can provide information on the nuclear equation of state, both at supra-nuclear and sub-nuclear densities. Careful modeling of reactions within transport theory is required that takes into account the different stages of the collision from compression to decompression and to fragmentation, and addresses the need for the extrapolation of thermodynamic quantities from elevated to low temperatures. There are benefits in employing the same framework consistently for the simulations and the extrapolation, but that requires a compromise between the level of realism and feasibility of reaction simulations. The modeling of fragmentation has close links with the modeling of pasta in neutron-star crusts. In the near future, an improved correlation between physical inputs and observables of heavy ion collisions is needed that can be achieved, for example, through extensive cross-comparisons of models that use the same physics inputs. This is particularly important for mapping out of the symmetry energy at high density. To progress, nuclear transport theory must advance in terms of a realistic description of cluster production by further exploiting the time reversibility of fragment production and break-up processes. Currently nuclear transport theory is predominantly semiclassical in nature and should gradually transition to a full quantum mechanical approach in order to achieve a consistent description of energetic central reactions, peripheral and low-energy central reactions, and nuclear structure. The limits beyond which the current semiclassical transport breaks down are not very well defined either. Different options exist for advancing to practical quantal transport, ranging from the use of non-equilibrium Green's functions, through describing a diffusion process in an over complete basis of states for the collisions, to using a set of Schrödinger equations for the systems, with stochastic potentials.

**Phases of nuclear matter in the neutron star core:** The recent discovery of a 2 solar mass neutron star implies that the high density EOS is stiffer (has a larger pressure at a specified density) than expected from simple models of high density matter that predict a phase transition to matter containing hyperons, meson condensates and de-confined quark matter. To ascertain if a stiff EOS rules out such phase transitions it is important to refine these models by accounting for effects of correlations in the ground state. The competition between the attractive two-body and repulsive three-body hyperon-nucleon forces in high density matter is an area of active research where we can anticipate important progress. Here, ab-initio methods, phenomenology of hyper-nuclei and lattice QCD methods to directly extract the hyperon-nucleon and hyperon-hyperon phase shifts, will improve the state of the art. These advances will yield new insights on the maximum mass of neutron stars. Models of strongly interacting dense quark matter and field theoretic methods to directly calculate the EOS at finite chemical potential from QCD using re-summation techniques can also yield useful insights about the high density EOS at the neutron star maximum mass. Complimentary information is contained in the transport properties of novel high density phases, that have been shown to be distinct from those expected in nuclear matter. Further work in this area enables the use of transport phenomena in neutron stars to either discover or disfavor the existence of phase transitions in the neutron star core.

### 3.6.2  Nuclear Theory for Nucleosynthesis

Exploring nucleosynthesis scenarios necessitates simulating both the astrophysical sites considered and the possible nucleosynthesis mechanisms. The most important nuclear properties required in this context are nuclear masses, weak interaction rates such as $\beta$-decay, electron capture, and neutrino interactions, as well as nuclear reaction rates, primarily for the capture of neutrons, protons, alpha particles on a variety of targets, and nuclear fission barriers and fragment distributions. Masses and reaction rates are needed for modeling potential r-process sites, for determining neutron fluxes in the s process, for assessing the importance of additional processes (p-, rp-, $\nu$p-, $\nu$-, $\alpha$-, i-process, etc), as well as for describing Big Bang nucleosynthesis and predicting solar neutrino fluxes (see sections 2.1.4, 2.2.7, and 2.7). Experimental data are limited and need to be complemented by nuclear theory. In addition, reaction theory, for example for transfer reactions, is needed to extract relevant information from nuclear physics experiments. Existing nuclear structure descriptions and nuclear reaction theories will have to be refined and extended to meet the needs of nuclear astrophysics. In some instances, new theories will have to be developed to address the challenges posed by isotopes away from stability, both neutron-rich and neutron-poor species.

Nuclear theories range from microscopic ab initio approaches to macroscopic phenomenological descriptions. A variety of approaches are required to connect to experimental results, to identify systematic trends, and to guide and test theories across the nuclear chart. Nuclear-structure formalisms, such as the various shell models (ab initio No-Core Shell Model, Traditional Shell Model, Monte-Carlo Shell Model, Shell-Model Monte Carlo approach, Symmmetry-Adapted Shell Model, Continuum Shell Model, etc.), the Greens-Function Monte-Carlo approach, Coupled-Cluster Model, Density-Functional Theories (DFT) and their extensions (RPA, QRPA, continuum RPA) find application in different, but overlapping areas of the nuclear chart and need to be enhanced and extended in order to connect to each other, cover the isotopes of interest, and provide the structure input required by nuclear reaction theories. Similarly, nuclear reaction theories (Faddeev, Faddeev-Yakubovsky, Alt-Grassberger-Sandhas formulations, R-matrix theory, single-channel and coupled-channels direct reaction theories, multi-step direct and multi-step compound descriptions, statistical Hauser-Feshbach theory, etc.) need to be revised and extended in order to make use of newly-available experimental and theoretical structure information and vastly-improved computational capabilities if they are to address the wide range of nuclear physics issues relevant to astrophysics. In addition, structure and reaction theories need to be more closely integrated with each other.

Nuclear theory challenges that are particularly relevant to astrophysics include the extension of current theoretical descriptions i) from laboratory to stellar energies, and ii) from stable nuclei to exotic, short-lived isotopes. Expanding the reach of theory to these novel regimes requires a comprehensive understanding of the reaction mechanisms involved (including direct, semi-direct, pre-equilibrium, compound processes), as well as detailed knowledge of nuclear structure (including particle thresholds, single-particle and cluster structure, collective phenomena) and mechanisms for incorporating feedback from experiment. Many nuclei of interest have to be treated as open many-body systems and special consideration of stellar environments need to be included such as nuclear excitation, the influence of high density environments, and electron screening,

**Nuclear structure theory** is needed to predict properties of isotopes that play a role in astrophysical environments. Ground states as well as excited states of stable and exotic nuclei are of interest. Ideally, continuum effects should be included in the description.

*Ab initio approaches* describe the nucleus using a well-defined microscopic Hamiltonian with nucleon degrees of freedom, treat the internal relative motion correctly, and obtain the relevant observables by solving the quantum many-body equations without uncontrolled approximations. The development of such approaches has made considerable progress over the past decade in describing light nuclei. Growing computational resources and improved numerical algorithms, a move from realistic phenomenological and meson-theoretical nucleon-nucleon potential models to nuclear interactions derived from chiral perturbation theory, the inclusion of three-nucleon (and even more-nucleon) forces, and the inclusion of the continuum are showing their effects in nuclear astrophysics. Ab-initio approaches can now treat reactions such as $^3$H(d,n)$^4$He, $^3$He(d,p)$^4$He, $^7$Be(p,$\gamma$)$^8$B, etc. An accurate description of the $^{12}$C Hoyle state, which has been identified as a major problem in nuclear astrophysics, is a near-term goal, and we can expect that modern methods will solve another longstanding problem, namely a theoretical description of the $^{12}$C($\alpha,\gamma$)$^{16}$O reaction (see 2.2).

*Shell model:* While extensions of the ab-initio approaches push towards the treatment of systems with A>16, predictions of nuclear energies, spectroscopic factors, level densities, electromagnetic multipole transitions, Gamow-Teller strength functions, and other properties of medium-mass nuclei are typically made using the large-basis traditional shell model, the shell-model Monte-Carlo approach, or Density-Functional approaches and their extensions. Here research is focused on improving effective Hamiltonians and pushing the computational boundaries, for example to include cross shell excitations that are important in the rp-process (see section 2.6.5). Complementary developments are underway to extend the reach of ab-initio approaches by using symmetry-adapted bases to overcome model-space limitations. These developments are important for obtaining improved predictions for $\gamma$ and $\beta$ strength functions, which are needed for nucleon capture calculations and predicting $\beta$-decay rates, respectively.

*DFT and QRPA:* Nuclear properties of heavier nuclei with mass numbers above 100 are typically obtained from density-functional theories and approaches that include correlations beyond DFT, such as the Random-Phase Approximation (RPA) and the Quasi-Particle RPA. Recent work in the context of DFT has focused on improving the energy-density functional itself and on quantifying uncertainties; a long-term effort to derive a functional from the underlying forces has been initiated. Such efforts are important for obtaining mass predictions with associated uncertainties.

Correlations beyond those included in DFT approaches are critical for calculating strength functions, such as those needed as input for statistical reaction calculations and for predicting $\beta$-decay rates. (Q)RPA calculations are technically challenging, but have benefited from increasing computational capabilities and new numerical techniques, making it possible to remove some of the approximations used in the past. For instance QRPA calculations for deformed nuclei, using a finite-range interaction, are now possible. Work that remains to be done includes further computational improvements to make a wider range of calculations feasible, and the inclusion of couplings to more complicated excitations and the continuum to properly predict spreading of unbound states and decays. Such developments will be valuable for reliably calculating properties of excited states and for providing microscopic input for reaction calculations.

*Fission:* As the r-process reaches nuclei in the mass region of A=200-250, fission will occur, which can significantly impact the nucleosynthesis yields through fission recycling (see section 2.1.4). Current astrophysics models use phenomenological fission models that rely on model parameters that are typically fit to existing data. These models have insufficient predictive capability. Improvements are expected to come from microscopic fission theories that make use of Density Functional Theory. Progress has been made in this area in recent years, and even fission in a plasma environment is now being considered, but both formal and computational challenges remain to be overcome to properly describe the complex processes involved.

**Nuclear Reaction Theory** has two goals that are particularly relevant to astrophysics: 1) Achieving a comprehensive description of direct, semi-direct, pre-equilibrium, and compound processes for a variety of reactions, and 2) Extracting astrophysically-relevant quantities from combinations of theory and indirect experiments.

*Direct reactions:* Reaction theory is needed to describe elastic and inelastic scattering processes, the fusion of nuclei, as well as transfers of nucleons or groups of nucleons between projectile and target. The complexity of the problem requires the elimination of possible reaction channels from explicit consideration and the introduction of optical potentials. Phenomenological optical potentials have received much attention in the past decade. The availability of a wide range of experimental data, coupled with current computational capabilities, has led to greatly improved parameterizations, in particular for nucleon-nucleus reactions. The development of optical potentials for reactions that involve deuterons or other light ions is making much-needed progress. Further improvements are needed to achieve better accuracy and to describe reactions with isotopes away from stability. This will involve including deformation and non-locality effects in the potentials and possibly revising codes that use these potentials as input. In order to reliably treat regions away from stability microscopic optical-model approaches will be needed.

Reactions involving the capture of neutrons, protons and other charged light ions, play an important role in astrophysics. Cross-section calculations require reliable optical potentials. To improve the calculations of direct-capture processes, deformation should be treated consistently and the contributions from semi-direct capture need to be considered.

*Isolated Resonances:* Low-energy binary reactions proceeding through isolated resonances are described by R-matrix theory. Typically, measured cross sections are fit by adjusting phenomenological R-matrix parameters, a procedure which allows the extraction of resonance properties and extrapolation to energy regimes for which no data is available. State-of-the-art codes have implemented multi-level and multi-channel treatments. Improvements should include the use of (calculated) structure data and extensions to treat transfer reactions that involve resonances.

*Overlapping Resonances and Statistical Approaches:* Capture proceeding via the formation of a compound nucleus is described in a statistical model, typically a Hauser-Feshbach approach. Multiple Hauser-Feshbach codes for the description of compound-nuclear reactions (including capture) are publicly available, well documented, supported, and user-friendly. These codes require nuclear structure input (discrete levels, level densities, Î³-ray strength functions, fission barriers, etc.) and optical models, as well as pre-equilibrium descriptions. Current codes make use of available nuclear structure databases and include a range of phenomenological models to provide the required inputs. However, recently significant deficiencies in the commonly used inputs for Hauser Feshbach calculations have been identified, in particular concerning low energy enhancements of γ-strength functions. Guided by experiments, a move towards the use of microscopic approaches for calculating the needed inputs, which has begun in the last decade, addresses this problem. Continuing these efforts is important for achieving more reliable predictions for unstable nuclei. Additional, potentially significant, improvements include an explicit treatment of deformation effects, more emphasis on fully quantum-mechanical descriptions of pre-equilibrium processes, and a better understanding of correlations between different reaction channels (width fluctuation corrections).

*Interplay Between Reaction Mechanisms:* More generally, the interplay between direct reactions, semi-direct processes, pre-equilibrium contributions, and compound reactions is not sufficiently understood. The influence of doorway states (simple configurations that couple the reaction entrance channel to more complex configurations) needs to be considered and the issue of energy averaging needs to be revisited. A better understanding of these issues will yield more reliable cross section calculations. This is particularly significant for describing low level-density regions away from stability, but it also impacts stable nuclei, e.g. in the Fe region and near closed shells.

*Charge Exchange Reactions:* Additional areas in need of attention are charge-exchange reactions (which have not received much attention recently), electron screening effects and reactions on excited states (which can be expected to play a role in astrophysical plasma environments), and nuclear reactions (and structure) in the presence of strong magnetic fields. An effort needs to be made to include theoretical uncertainties with the reaction calculations, as errors need to be propagated in the astrophysical models that use the reaction theory results.

*Connection to Experiment:* Indirect measurements will play an increasingly important role in nuclear astrophysics as the focus moves towards improving our descriptions of unstable isotopes for which properties and reaction cross sections cannot be measured directly or calculated reliably. The observables obtained from such measurements have to be related to the quantities desired, which requires reaction theory. The most important quantities one would like to extract from measurements are resonance properties (resonance energies and widths), cross sections for direct capture of charged particles, and compound-nuclear cross sections. The most likely indirect approaches used will employ radioactive-beam experiments involving inelastic scattering and transfer reactions, such as (d,p) reactions.

Current reaction theories have to be extended to reliably link the measurements to the desired information. For deuteron-induced reactions, it is important to treat both the transfer and the breakup contributions. Recent work has shown that it is important to improve the treatment of the three-body dynamics in (d,p) reactions. A reliable description of transfers to (both wide and narrow) resonance states, which would allow the extraction of resonance parameters, does not exist. Further, the interplay of direct and compound processes is not sufficiently understood for both inelastic scattering and transfer reactions. These shortcomings will affect the interpretation of radioactive-beam experiments that aim at extracting nuclear structure information as well as indirect measurement that aim at determining compound cross sections.

### 3.6.3  Nuclear Theory for Neutrinos

The nuclear physics community has played a leading role in neutrino astrophysics, particularly in the areas of the sun (see section 2.2.7), supernovae (see section 2.3), and compact object mergers (see section 2.4).

We are at the point now where we can use solar neutrinos to test the assumptions of the standard solar model with great precision, and can better understand neutrino oscillations inside supernovae, where neutrino-neutrino scattering has recently been shown to induce flavor oscillations. One particular need in this context are neutrino interactions with nuclei in highly excited states that are populated in hot environments. These rates are currently being modeled on sd-shell configuration interaction (CI) calculations, but the necessary extension to the pf shell will require much expanded computational efforts.

Neutrino signals from future supernovae provide us with the best way of understanding the supernova mechanism (see section 2.3.7). We hope to use these data to obtain information about the symmetry energy of nuclear matter, the neutron star maximum mass, and the elements produced. However, we have much to do theoretically, in order to be able to interpret any future supernova neutrino signal.

Neutrinos scatter often within the proto-neutron star, which forms the core of the supernova. What are the characteristics of the neutrinos that emerge from this core? What are their energies, flavor distribution and time profile? In order to make the most of the next supernova neutrino observation, we require an improved understanding on the nuclear response for neutrinos with energies of 10s of MeVs on target nuclei. Transition Operators and Transition Strength Functions are needed. For astrophysics, the most important operator at low energy is Gamow-Teller, but at higher energy one needs forbidden operators as well. The community needs systematically-derived EFT operators, at the same level of understanding as interactions, that are implemented systematically.

To disentangle the detected supernova signal we also need to understand the way neutrinos oscillate. There are two primary open questions in this regard. One is the occurrence of oscillations due to the neutrino self-interaction, which is a complicated non-linear problem. Secondly we need to understand how neutrinos flavor transform in the presence of a turbulent medium. These oscillations are due to the interaction with electrons, and include both MSW-type transitions and parametric resonances.

Neutrinos carry away a large fraction of the energy in a compact object merger as well (see section 2.4). Our theoretical understanding of the dynamics of these objects is undergoing a rapid change with the inclusion of improved neutrino transport. Much of the dynamics is determined by the neutrino physics. In addition, the neutrinos oscillate in new ways, not seen in supernovae, the sun or terrestrial experiments. This newly discovered Matter-Neutrino Resonance occurs close to the surface of the object, where dynamics and nucleosynthesis are likely to be impacted. Again, there is much to do theoretically in order to understand the evolution of these objects, and the elements that they produce.

These investigations will change not only our understanding of how supernovae and compact object mergers evolve, but also our understanding of the nucleosynthesis. Any change in the dynamics of these events changes the temperature and density conditions, which are the defining characteristics for the element synthesis. Neutrinos affect element formation directly as well. The conversion of neutrons into protons and vice-versa through the weak interactions with neutrinos determines the neutrino neutron to proton ratio and therefore the outcome of the nucleosynthesis.

### 3.6.4  Nuclear theory needs

Nuclear astrophysics theory is on the cusp of an new era. Unique opportunities are opening up with new experimental capabilities at FRIB, ATLAS, and university-based laboratories, as well as astronomical observations from radio, optical, gamma-ray telescopes, as well as neutrino and gravitational wave detectors. It is essential that nuclear theorists are closely connected to both experimental nuclear physics and observational astrophysics. These connections are essential to finding the site of the r-process, determining the nature of dense matter, and elucidating the role of nuclear reactions in astrophysics.

There is a particularly important connection between nuclear astrophysics theory and FRIB. FRIB opens many new and exciting opportunities but it also presents great challenges for the theory community. In order for FRIB to realize its full scientific potential, the nuclear theory effort needs to be strengthened in a structured manner and the connection between nuclear theory and astrophysics enhanced.

Collaborative efforts between nuclear theory and experiment are crucial to plan and interpret experiments at nuclear facilities, as well as to guide, test and apply new theory developments. Similarly, a close relationship between nuclear theorists, astrophysicists, and astronomers is needed to properly interpret observations and identify gaps in current nuclear theory knowledge. Also needed are collaborations with computer scientists, in order to tackle the computational challenges involved in nuclear theory, both in describing the complicated quantum-mechanical many-body problem involved, and for handling the vast number of reactions that need to be considered for a comprehensive description of the various astrophysical phenomena. Furthermore, connections with the nuclear data community as well as astrophysical data compilations should be strengthened, as this will bring in needed expertise in creating evaluations, disseminating theoretical data. An important aspect for astrophysical applications are estimates of the uncertainty of theoretical data and their incorporation in data bases and models that predict astrophysical observables.

It is critical to realize the scientific potential of nuclear facilities such as FRIB and neutrino detectors, and to connect them to the astronomical observations. The nuclear physics ingredients needed for the astrophysics context provide non-trivial challenges to nuclear theory. To address these challenges, a well-trained workforce, a range of state-of-the-art nuclear facilities, and strong connections between the different sub-disciplines are essential. There are some pre-existing structures from which the nuclear astrophysics community can continue to benefit. JINA continues to have a large impact on the experimental nuclear astrophysics community by creating connections between theory and experimental nuclear astrophysics activities at a broad range of nuclear accelerator facilities in the US and abroad, between nuclear physics and astrophysics, and between theory and observations. The INT plays an important role in stimulating the exchange of ideas and new developments in nuclear theory across the various subfields of nuclear physics, and many programs have connections to theoretical astrophysics. New capabilities in high performance computing offer unique opportunities for nuclear theory that are of importance in astrophysical applications. Support for creating the necessary connections between theory and computational science, such as the current SciDAC NUCLEI effort, will be important to take advantage of these opportunities. A new planned FRIB theory center, as well as focused multi-institutional collaborations in theory, will be important to stimulate the theoretical developments needed to take full advantage of the unprecedented capabilities of FRIB.

Nevertheless, a critical need remains in the nuclear astrophysics community. The nuclear theory workforce is far too small to address the challenges that will be presented by the enormous influx of new data that will be obtained by FRIB, neutrino facilities and astrophysical observation. Above all, what is needed is an increased and timely investment for creating a focused decade-long collaborative effort that can induct into the workforce on the order of 10 junior scientists to develop nuclear theory and simulations to model extreme astrophysical phenomena.

## 3.7 Astrophysics Theory

Astrophysics theory is essential in nuclear astrophysics as it connects nuclear physics with related observables. There are a number of theoretical challenges that have been mentioned in the science section. A major thrust in the next 10 years will be multi-dimensional models of a range of astrophysical sites of interest to nuclear astrophysics: X-ray bursts (see section 2.6.5), novae (see section 2.5.4) , fast rotating stars (see section 2.2.5), core collapse supernovae (see section 2.3.4), type Ia supernovae (see section 2.5.4), neutron star mergers (see section 2.4.4), and low metallicity stars (see section 2.2.5). In many cases full 3D models of the entire evolution of the stellar object will not be possible. What will be possible is the modeling of key aspects of all these scenarios in 3D. This is important to get a handle on scenarios were 3D effects have a drastic impact on observables and nucleosynthesis. Examples include the supernova explosion mechanism, the rise time of X-ray bursts, mixing of white dwarf matter in classical novae, or hydrogen entrainment in low metallicity stars. Such 3D studies can then be used to inform and adapt 1D models that are still needed as workhorses to predict nucleosynthesis for a wide range of elements and for a wide range of stellar parameters. Another important theoretical challenge are atmosphere models of neutron stars and predictions of mass loss through stellar winds.

While it is important to develop further the broad suite of astrophysical models needed to address the open questions in nuclear astrophysics, it is equally important to carry out the work needed to make the connection between nuclear physics and astrophysics. This is essential if one wants to apply astrophysical models to nuclear astrophysics problems such as nucleosynthesis. Such connections require major efforts in implementing large nuclear reaction networks, in finding ways to overcome the computational challenges that come with a full treatment of the nuclear physics, and thorough analysis of the resulting nuclear processes, in particular related to the sensitivity of observables to nuclear physics uncertainties. In the past, it has often been difficult to carry out such interdisciplinary activities - they may not fall into traditional funding categories, or traditional areas of work in the respective subfield (and neither in nuclear physics nor in astrophysics is the full expertise for such work available). Some individual collaborations and, at a larger scale, JINA has helped to overcome this issue to some extent (see section 3.11). However, cultural differences between nuclear physics and astrophysics remain a significant hurdle that must be overcome in the future to ensure nuclear astrophysics programs at nuclear facilities and observatories are properly guided and results are used to address the open questions in the field.

## 3.8 Computational Astrophysics

The dramatic impact of computation on astronomy and astrophysics is manifested in many ways. Modern numerical codes are now being used to simulate and understand the evolution, explosion, and nucleosynthesis of stars (see section 2.1.5), how the elements are injected into the interstellar medium, molecular clouds, and extant planetary systems, and the cosmic evolution of the abundances (see section 2.8.5). They are also essential to processing astronomical spectral databases whose sizes now exceed one terabyte into abundance data that are usable by the nuclear astrophysics community (see section 2.1.7). The largest codes may have in excess of a million lines and run on supercomputers that have more than 100,000 cores, generating datasets that occupy one to a hundred terabytes of storage. The most widespread codes may be driven by communities with hundreds of users and run on desktop class machines that generate a significant fraction of the published literature. Such codes are now an indispensable part of the nuclear astrophysics enterprise. They often deploy teams - astronomers, astrophysicists, computer scientists, visualization professionals, applied mathematicians, and algorithm specialists – to create, maintain, and constantly develop them.

NSF, NASA, and DOE have made substantial investments in the advanced computing and networking ecosystem over the last few decades, from national to regional to university to individual facilities. Sustained peta-scale, and soon exa-scale, computing capabilities will be available to the nuclear astrophysics community. Such capabilities will enable cutting-edge theoretical calculations and analyses that push the nuclear astrophysics frontier. One example is 3D core-collapse simulations run to 500 s to better quantify the neutron star winds, the r-process signatures, and evolution of the proto-neutron star (see section 2.3.4). Another example is routinely deploying reactions networks with 1000's of isotopes in all 2D or 3D models. Future progress in advanced computing for nuclear astrophysics will come from further parallelization, ubiquitous deployment of next-generation 100 GB/s internet connectivity in tandem with Globus Online, distributed cloud storage systems, extracting actionable knowledge from big data, and, potentially, social computing.

Similarly, Spectroscopic stellar surveys that are scheduled over the next 5-10 years, or are recently completed (SEGUE, LAMOST, APOGEE, HERMES, GAIA, a variety of LSST followups, and the proposed 10 meter spectrographic survey telescopes), as well as all-sky surveys that will provide significant new information on supernovae and other transient events (LSST, SkyMapper) (see section 3.9). Nuclear astrophysics will have terabytes of spectroscopic data and petabytes of photometric all-sky data that need to be analyzed and compared to simulations (see sections 2.1.7 and 2.8.4)

These new technological capabilities and data driven science will enable qualitatively new physical modeling in topics relevant to nuclear astrophysics. Exploiting these new capabilities for nuclear astrophysics will require new software instruments (e.g., run-time visualization for 100TB of 1PB data sets) and sustained funding support for focused multi-institutional research collaborations.

## 3.8.1 Recommendations to address needs

1) Make long-term investments in appropriately focused research collaborations and codes that can make be uniquely effective in tackling some of the most difficult problems in modern nuclear astrophysics. The collaborations would be devoted to a specific nuclear astrophysics problem or topic that is believed to be ripe for a breakthrough within five years. One example, would be deploying 2D or 3D hydrodynamic simulations of the $^{13}$C pocket in low- and intermediate mass stars to produce a breakthrough in quantifying the main s-process.

2) Encourage the astro computation community to maintain common and clearly-documented data formats and standards. This will facilitate direct collaborations as well as encouraging the data to be publicly available, and will allow analysis codes to accommodate multiple data sources. For example, current challenges exist in reaction rate databases, nucleosynthetic yield data from stellar evolution calculations, and supernova simulations as input into interstellar or molecular cloud mixing calculations or large-scale chemical evolution simulations (see section 3.10).

3) Motivate the creation of "data libraries" for simulation inputs and outputs. This maximizes the reach and impact of the simulations, since the data can be useful for more science than the authors originally intended, thus increasing the science-per-dollar (or cpu-hour) of the simulations (see section 3.10).

4) Inspire multiple collaborations to maintain their simulation and analysis codes as open source projects. Having open source access to a variety of codes that do the same type of simulation or analysis is a necessary part of the scientific process. This improves transparency, cross-portability, and trust in the simulation results. For example, in a few subfields of astrophysics some results from non-open-source codes are starting to be dismissed as unreliable, and this early trend may propagate to other sub-fields of astrophysics. In addition, open source instruments lower the barrier to entry for the next-generation of researchers and can significantly increase the amount of science produced by a community (e.g., the MESA or GRID projects).

5) Encourage the development of an open source, community driven and supported radiation transfer code. This is a critical, but missing, piece of infrastructure that connects stellar models with the light curves and spectra obtained from observations.

6) Persuade collaborations to engage in vigorous code comparisons and verifications as a mechanism to improve the fidelity of the codes and build trust in the results. Comparison to observations requires trustworthy simulation results. For example, the puzzling variation in the nucleosynthetic yields from AGB stars hampers galactic chemical evolution studies. This can often be traced back to assumptions about the treatment of convection and mixing, but in some cases can be attributed to a discovered numerical instability.

7) Encourage policies that balance capacity computing with capability computing on the largest supercomputers. Capability computing uses all or a large fraction of the supercomputer to solve a few large problems (i.e., hero calculations). In contrast, capacity computing uses the supercomputer to solve a large number of smaller problems (e.g., surveying a parameter space). Many, but not all, members of the working group thought the balance was skewed too much towards capability computing.

8) Develop and implement an annual Summer School on numerical algorithms, parallel techniques, Big Data, and advanced computing in nuclear astrophysics.

## 3.8.2 Big Data

The phrase "Big Data" refers to large, diverse, complex, longitudinal, and/or distributed data sets generated from instruments, sensors, computational models, images and/or all other digital sources.

Big Data in nuclear astrophysics aims to advance the core scientific and technological means of managing, analyzing, visualizing, and extracting useful information from large, diverse, distributed and heterogeneous data sets so as to:

- Accelerate the progress of scientific discovery and innovation in fields of broad interest to nuclear astrophysics;

- Lead to new fields of inquiry via intellectual fusion of fields of interst to nuclear astrophysics that would not otherwise be possible;

- Encourage the development of new data analytic tools and algorithms;

- Facilitate scalable, accessible, and sustainable data infrastructure across the nuclear astrophysics community's efforts;

Today, the federal funding agencies and private enterprise recognize that the scientific and engineering research communities are undergoing a profound transformation with the use of large-scale, diverse, and high-resolution data sets that allow for data-intensive decision-making, at a level never before imagined. New statistical and mathematical algorithms, prediction techniques, and modeling methods, as well as transdisciplinary approaches to data collection, data analysis and new technologies for sharing data and information are enabling a paradigm shift in scientific investigations. Advances in machine learning, data mining, and visualization are enabling new ways of extracting useful information in a timely fashion from massive data sets (e.g., 3D simulations and spectroscopic surveys), which complement and extend existing methods of hypothesis testing and statistical inference. As a result, a number of federal funding agencies are developing big data strategies to align with their missions.

Embracing a Big Data initiative will help to accelerate discovery and innovation in nuclear astrophysics. The pipeline of data to knowledge to action has tremendous potential for transforming all areas of scientific interest to nuclear astrophysics. This initiative will lay the foundations for engaging enterprise level Big Data infrastructure projects, workforce development, and progress in addressing the complex, transdisciplinary grand challenge problems in nuclear astrophysics. The field's state-of-the-art research is increasingly data-intensive and adequate sustained support for a Big Data initiative is imperative if the field is to realize its research aspirations.

## 3.9 Astronomical Observations

Nuclear astrophysics is reflected in (and often dominating) a broad range of astronomical windows and messengers. The following lists the major wavebands of interest and some of the primary science goals, measurements and sources of interest to nuclear astrophysics:

- Electromagnetic Radiation: Specific Elemental Abundances; others

- Radio: Molecular Isotopes in interstellar medium; Pulsar masses

- Sub-mm: Cold-gas Cooling Lines (CI etc.)

- Infrared: Polycyclic Aromatic Hydrocarbons (PAHs) and Dust Emission; Atomic Lines (eg. Ag)

- Optical: Metal Elemental Abundances; timing for pulsation studies

- Ultraviolet: Element Abundances (eg. Ag); Transient LCs

- X-rays: Hot-Plasma Abundances; X-ray bursts; transients; compact star structure (EOS)

- MeV Gamma-Rays: Radioactive Isotopes

- GeV Gamma-Rays: Cosmic Rays Interactions (spallation)

- TeV Gamma-Rays: Cosmic Ray Accelerators

- Meteorites and Presolar Grains: Specific Isotopic Abundances and their Ratios

- Asteroseismology, Stellar Interiors (Core Size; convection, rotation)

- Cosmic Rays: Specific Isotopes; Spallation in the Interstellar Medium

- Neutrinos: Core Collapse Supernovae; solar Nucleosynthesis

- Gravitational Waves: Binary Mergers and Core Collapse Supernova Dynamics

Comparing nucleosynthesis source model predictions to observational data is a major challenge, often beyond the capabilities of single scientists or groups. Progress therefore depends on interactions and discussions among observers across different fields and modelers/theoreticians of different sources/processes (including chemical evolution modelers). A dedicated effort to further stimulate such cross-field interaction appears to be a most-promising first step to advance nuclear astrophysics in general (see section 3.11). We need to learn which questions can be pursued, and how. We also need to learn how to best exploit the vast observational diversity in terms of validating key astrophysical models and addressing specific nuclear astrophysics questions.

Multi-messenger studies of nuclear astrophysics sites will gain in importance in the future and require coordination among different astronomy communities, astrophysicists and nuclear physicists. An example are neutron stars where merger observations with gravitational wave detectors such as LIGO, gamma-ray telescopes, and X-ray telescopes will have to be combined with X-ray observations of isolated and accreting neutron stars and, possibly, neutrino signals from supernovae (see section 2.4.5).

A lack or loss of observational facilities may incur significant setbacks in the above prospects. The nuclear-astrophysics community should speak up as appropriate in critical cases. Primary concerns at present appear to be the potential Green Bank Telescope shutdown, which could adversely impact future NS (pulsar) mass measurements using radio pulsar timing, as well as a potential reduction in Kitt Peak National Observatory availability for optical astronomy. Also, major telescope survey programs are often driven by the science issues of dark energy, dark matter, and cosmology. This incurs the risk that the also-interesting and unsolved issues of nuclear astrophysics are not addressed adequately. In particular, in the case of expensive space programs, opportunities may be thinned-out below a minimum threshold that keeps expertise for nuclear-astrophysics observations sustained.

## 3.9.1  Radio Astronomy

Current instrumentation includes the eVLA, Arecibo, GBT, and LOFAR on the ground, and the Planck space mission. Capabilities in this field include, spectroscopy resolving isotopes in molecular lines, and an imaging resolution of about milli-arcsecs.

The relevance for nuclear astrophysics lies in determinations of abundances of molecules and isotopic ratios ($^{12}C/^{13}C$, $^{16}O/^{17}O$, etc.) in the interstellar medium. Furthermore, radio pulsar observations can provide precise neutron star masses, and pulsar timing glitches can provide constraints on neutron star interior physics. Also, measurements of global supernova explosion energy and nova ejected masses can be made. Additionally, radio transients may be powered by nuclear processes.

The future instrumental perspectives are the SKA, MeerKAT, ASKAP, and ATA . Recent pulsar mass measurements have found heavy 2 solar mass neutron stars. Such high mass measurements are quite constraining on EOS models, and given the complimentary focus of nuclear laboratory measurements to constrain the EOS and symmetry energy, it would be a set-back to lose such a capability. Moreover, continued measurements of the double pulsar system (PSR J0737-3039) over the next 5-10 years could provide a direct measurement of the moment of inertia of pulsar A in this system.

### 3.9.2 Sub-mm Astronomy

Current Instruments are JCMT, CSO, SCUBA, IRAM, APEX, Mopra, the space instruments on Herschel (single-dish instruments), and BIMA, SMA, IRAM, CARMA, ATCA, and ALMA (Interferometers).

Relevant capabilities are in the mapping of dust emission (tracing star formation), and the spectroscopy of molecular lines, with imaging resolution about a few arcsec. Their relevance for nuclear astrophysics is through the provision of star formation tracers, and isotopic ratios in different molecular species also related to star formation sites. Rotational-band lines of molecules probe abundances and the gas kinematics. Moreover, mass loss around evolved objects, and the chemistry around new stars can be studied. The future instrumental perspectives are the LMT, SCUBA-2, and ALMA.

### 3.9.3 Infrared Astronomy

Current missions and telescopes include; Spitzer, Herschel, Sofia, IRAS, and instrumentation at the VLT. Capabilities include spectroscopy adequate for resolving PAH lines. Studies relevant to dust emission, and arcsec imaging.

The relevance for nuclear astrophysics includes abundances of PAHs and the chemistry of the interstellar medium. Moreover for Galactic stars photospheric abundances can be acquired in detail due to low absorption in the IR.

The future instrumental perspectives are the space missions Euclid and NASAs JWST.

### 3.9.4 Optical Astronomy

Current missions and instruments (Fig. 24) are HST, and on the ground the variety of large-aperture telescopes such as the VLT, Keck, Subaru, Gemini (N and S), and the Magellan telescopes. Capabilities include spectroscopy with resolution adequate for resolving lines of elemental species, and imaging resolution of arcsec.

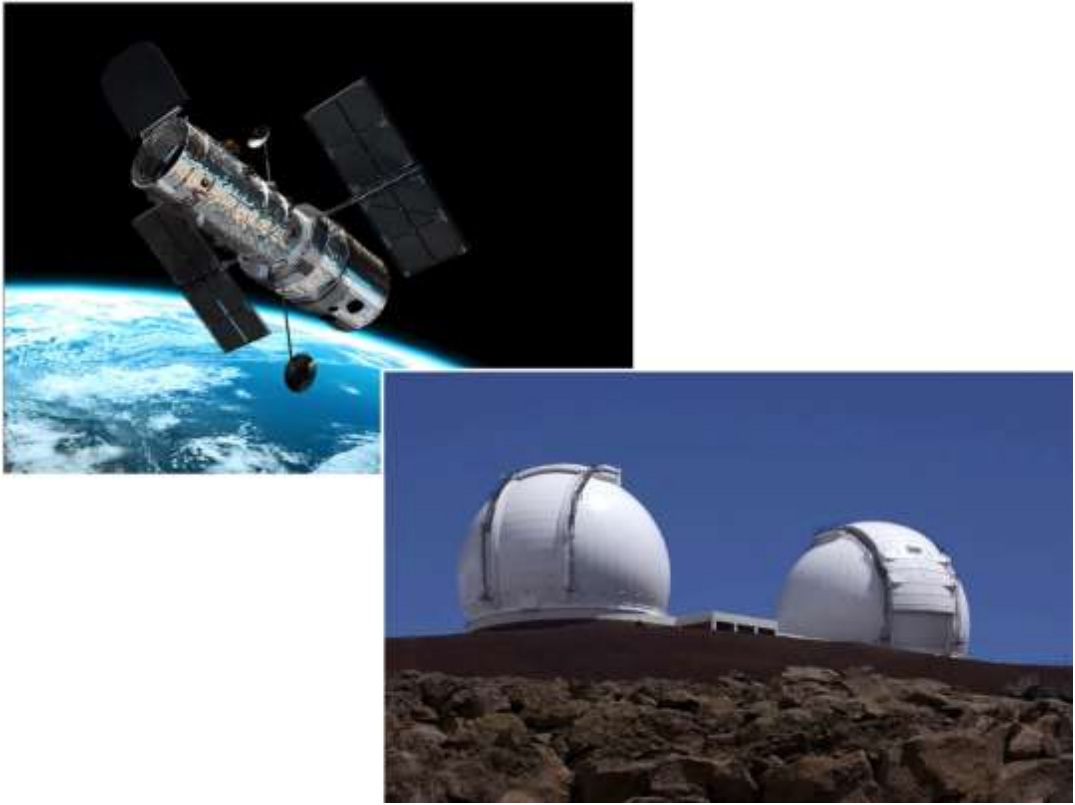

Figure 24: Ground and space based observatories enabling high resolution spectroscopy of visible and ultra-violet star light have revolutionized nuclear astrophysics with measurements of the composition of extremely metal poor stars that trace the history of chemical evolution. (Upper left: Hubble Space Telescope, Credit: European Space Agency. Lower right: Keck telescopes, Credit: NASA/Jet Propulsion Laboratory)

The relevance for nuclear astrophysics is in abundance measurements and isotopic ratios in stellar atmospheres in the Galaxy, the halo system, and nearby galaxies. Current observations are beginning to probe the abundance patterns, and in some cases, the isotopic patterns, that are thought to be produced by the very first generations of stars. Large spectroscopic surveys (SDSS/APOGEE, AEGIS, LAMOST) will be further exploited in the near future. Astrometry (and limited photometry and spectroscopy) with the Gaia mission will enable radial velocity measurements, proper motions and geometric distances for 1 billion stars, and precise luminosities for all classes of stars. Massive ground-based spectroscopic follow-up efforts to obtain radial velocities (and stellar parameters) for stars that are too faint for Gaia need to be supported vigorously. Future instrumental perspectives include the LSST, ELT, and E-ELT.

A particularly important thrust for the future are advanced capabilities in the time domain. By observing significant parts of the sky with unprecedented repetition rates new transient phenomena can be discovered and studied. This development has begun at a broad range of observatories where robotics are used to automate observations. Examples include the Las Cumbres observatory network, or the Palomar Transient Factory. Major new facilities for the future of this area will be PAN-STARRS and LSST. Automated high repetition wide field observations will pose new challenges for processing, storing, and analyzing the large amounts of data generated (see section 3.8.2).

### 3.9.5 UV Astronomy

Past missions were IUE, and FUSE. Currently, HST provides limited UV capabilities, and the GALEX mission provides 4-5 arcsec imaging and spectral resolving power of 200 and 90 in far- and near-UV bands, respectively.

The relevance for nuclear astrophysics includes measurements of abundances of light elements, and low-abundance heavy elements (for example, Ag), as well as the measurement of nova ejecta abundances. Moreover, photometry for studying the time domain behavior of transients is important.

A concern here is limited future instrumental perspectives. The group did not identify a future space capability beyond HST, for example.

### 3.9.6 X-ray Astronomy

Current missions include Chandra, XMM-Newton, Swift, Suzaku, MAXI and the recently launched NuStar.

Capabilities include spectroscopy with a range of resolving powers; as high as 1000 with the Chandra and XMM gratings, resolving power < 100 with X-ray CCDs, adequate for identifying some ion species and constraining abundances. Imaging resolution ranges from < 1 arcsec with Chandra to arcmin with Suzaku and MAXI. The recently launched NuStar mission adds $^{44}$Ti low-energy line imaging. Sensitive X-ray timing and broad-band spectroscopy provides essential constraints to understand explosive phenomena such as Type-I X-ray bursts and superbursts (see section 2.6.5). These capabilities also provide direct probes of the neutron star, its environs and the nuclear physics driving the explosions, as for example, by observing the thermal surface emission during X-ray bursts. Long term monitoring (as for example, with MAXI) can provide burst recurrence times and offer the opportunity to capture rare events such as superbursts.

The relevance for nuclear astrophysics includes abundance measurements in hot astrophysical plasmas, SNR, and WHIM abundances. In addition, X-ray light curves and spectra provide detailed probes of neutron star surfaces and nuclear physics processes relevant to thermonuclear X-ray bursts. Such measurements can, in principle, also be used to estimate global neutron star structure parameters such as the mass and radius, both crucial for EOS constraints.

Future instrumental perspectives include high spectral resolution with micro-calorimeters on Astro-H (2014); wide field surveys with eRosita (2014+); very large collecting area and fast timing (excellent for X-ray burst studies) with ESA's Large Observatory For X-ray Timing (LOFT, 2020+); fast timing and wide field monitoring with India's Astrosat (2013+); precision soft X-ray timing and medium resolution spectroscopy with NASA's Neutron Star Interior Composition EexploreR (NICER) International Space Station (ISS) payload, and possibly high throughput and high resolution spectroscopy with IXO/Athena (2024+).

There is presently a substantial effort to probe the nuclear symmetry energy with laboratory nuclear experiments (see section 2.6.6). These efforts are strongly complemented by astrophysical observations of neutron stars, which probe fundamental physics to higher densities. Recent measurements of 2 solar mass neutron stars, and the opportunity that could be provided by missions presently in development, such as LOFT and NICER, which can enable direct NS radius measurements (that directly probe the symmetry energy), suggests a strong potential for important breakthroughs in this area within the next decade.

### 3.9.7  MeV Gamma-ray Astronomy

Current Missions are INTEGRAL with SPI, Fermi-GBM, and RHESSI (Fig. 25). The Fermi-GBM provides low energy and imaging resolution but is an efficient all-sky monitor for 0.1-40 MeV. The other instruments provide high energy resolution of 3 keV, adequate for resolving isotopes, with modest imaging resolution of 3 degrees, and with sensitivities few $10^{-6}$ photons $cm^{-2}$ $s^{-1}$, they can reach Galactic sources, and supernovae to five Mpc.

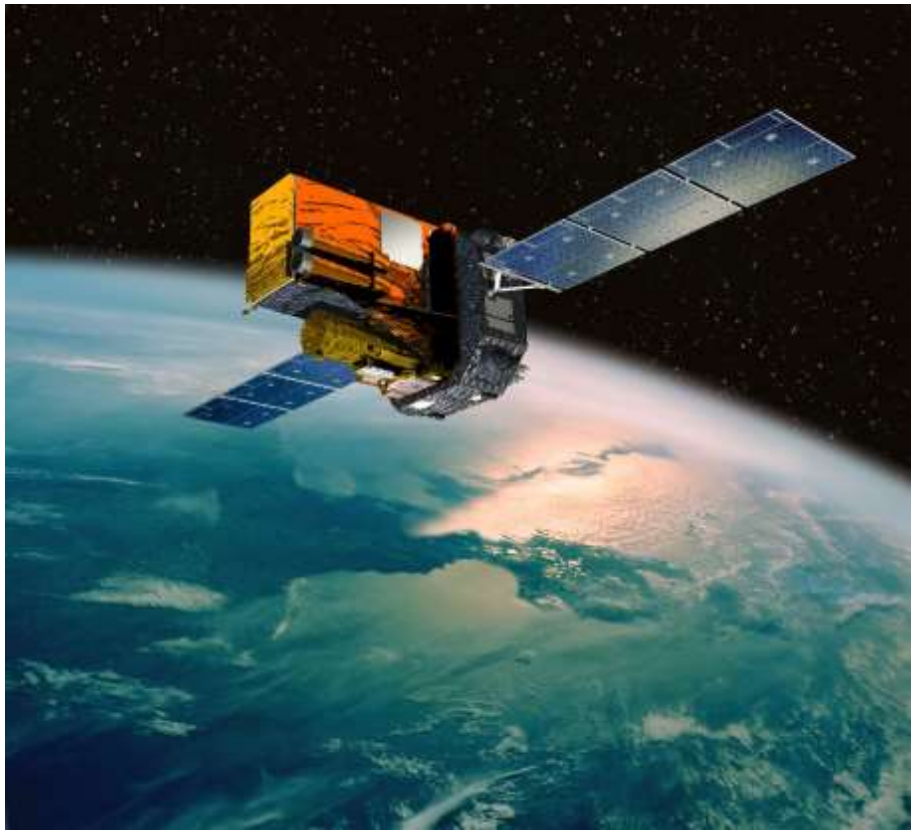

Figure 25: The direct detection of the $\gamma$-radiation from the decay of radioactive nuclei produced in stellar explosions offers a pathway to directly probe the element synthesizing nuclear processes. While existing MeV $\gamma$-ray telescopes such as INTEGRAL have provided important results, a future instrument with significantly increased sensitivity is needed to fully exploit this opportunity. (Credit: ESA - D. Ducros)

The relevance for nuclear astrophysics lies in direct measurements of radioactive isotopes from cosmic sources of nucleosynthesis. Surveys can find and measure new sources before their appearance in other observables (511 keV; $^7$Be, $^{22}$Na), or when they are embedded in dense clouds (supernovae, $^{44}$Ti). The decay of $^{56}$Ni is a diagnostic of supernova interiors both in type Ia supernovae and core collapse supernovae. $^{26}$Al from stellar groups constrains stellar-group yields (stellar and supernova), and allows the study of ISM dynamics around those. $^{60}$Fe probably originates from the same stellar groups that produce $^{26}$Al, and their isotopic ratio is an important diagnostic of multi-shell-burning structures in late evolutionary stages of massive stars. Positron annihilation relates nuclear astrophysics to the properties of cosmic rays and their propagation near sources of nucleosynthesis as well as pulsars and binaries. Much can be learned from observations of solar flares on nuclear processes in the Sun's outer layers.

In terms of instrumental perspectives there are none presently identified in the major space agencies for the time after INTEGRAL (>2020?), although Compton Telescope missions have been proposed, and technology and balloon projects are underway.

### 3.9.8  GeV Gamma-ray Astronomy

Past and current space missions include CGRO/EGRET, Agile, and Fermi.

Capabilities include spectroscopy with a resolution of 0.1 GeV, adequate for resolving nucleonic lines, and imaging with resolution of a degree.

The relevance for nuclear astrophysics is less direct than in most other fields, yet is given through the tracing of Cosmic-Rays in the Galaxy, and how this relates to nucleosynthesis sources.

Instrumental perspectives include Fermi (expected to continue for at least several years), but none identified after that in the major space agencies.

### 3.9.9  TeV Gamma-ray Astronomy

This is ground-based gamma-ray astronomy. Current experiments are H.E.S.S., MAGIC, and Veritas. Instrumental capabilities relevant for nuclear astrophysics lie in its imaging resolution of 10 arcsec. The relevance is less direct than in most other fields, but is given through constraints on CR accelerators in supernova remnants, and pulsar wind nebulae. Future instrumental perspectives are focused in HAWC and the CTA project (2014+).

### 3.9.10  Meteorites and Pre-solar Grain Studies

The Stardust mission provided samples of interplanetary particles, caught from the current medium within the solar system. Then, a rich body of meteoritic samples is available, where condensation occurred a long time ago under poorly-known conditions, but mineralogical studies help to identify origins. Of particular value are pre-solar grains identified herein, where identification is chemically through extreme resistance to acidic solvents, and observationally through extremely deviant isotopic ratios from solar-system material.

Laboratory instruments are Nano-SIMS, RIMS, Ion Microprobe. They provide capabilities of mass spectroscopy, adequate for resolving ion species., and imaging at few nm scales, i.e. down to resolving individual pre-solar grains.

The relevance for nuclear astrophysics includes measurement of abundances of specific ions in dust-producing nucleosynthesis sources; mixing in stellar atmospheres (esp. in AGB stars); solar system formation history.

Instrumental perspectives include future Nano-scale probe analysis (single grain study of isotope ratios).

### 3.9.11  Asteroseismology

Past and current missions and instruments include Kepler, Corot, MOST, and the Whole Earth Telescope. The primary relevant capability is to measure stellar oscillation modes with high precision. The relevance for nuclear astrophysics is in direct constraints on stellar interiors. Study of oscillations can determine the size of convection zones, explore mixing and measure differential rotation within stars. Oscillation frequencies can also depend on the nuclear reaction rates themselves.

No new missions are in advanced planning at present, but exo-planet search missions will likely provide such useful data. The Kepler mission is providing a wealth of new oscillation data on stars in its field of view. These observations have the capability to provide direct constraints on the stellar properties relevant to nucleosynthesis calculations. A deeper exploitation of these observations for nuclear astrophysics, however, will require further development of diagnostics to relate oscillation frequencies to nuclear reaction rates (see section 2.2.6). This is worthy of future efforts and support.

### 3.9.12  Cosmic-Ray Astronomy

Direct Cosmic Ray collectors and experiments are ACE, Pamela, ATIC, AMS-II, and indirectly AUGER. Capabilities include measurement of near-earth Cosmic Ray abundances, and Cosmic-Ray composition at VHE/UHE.

Relevance for nuclear astrophysics is in measuring abundances in cosmic rays at different energies. Future instrumental perspectives include AUGER (ground-based), Jem-EUSO (space).

### 3.9.13  Neutrino Astronomy

Current experiments include IceCube, Amanda for GeV neutrinos, Borexino for MeV neutrinos, and SK-IV. Neutrino detections (time tagged) can provide direct diagnostics of core collapse Supernovae (SN) and probe the solar nuclear energy generation process (see section 2.2.7). Excitingly, future detection of SN neutrinos could provide a resolution of the neutrino mass hierarchy. A further goal for future studies is to directly detect the p-p chain solar neutrinos and directly measure the solar luminosity in neutrinos. This can also provide constraints on the solar core metallicity.

Future instrumental perspectives include KM3Net; Laguna, SNO+, CLEAN.

### 3.9.14  Gravitational-Wave Astronomy

Current experiments include Virgo, LIGO, Geo600. The goal of these ground-based experiments is the detection of gravitational radiation in the frequency band from about 50 to 2000 Hz. Current detectors have not yet achieved source detections, and several facilities (LIGO, VIRGO) are in the process of upgrading their sensitivities. Primary source types include SNe in the Galaxy, and compact object mergers (neutron stars with neutron stars, neutron stars with black holes, or black holes with black holes) in galaxies out to 300 Mpc. Other potential sources include periodic, rotating neutron stars (pulsars), and perhaps accreting neutron stars with significant r-mode amplitudes. Space-based observatories (such as the LISA concept) are sensitive in a lower frequency band and not as directly relevant to nuclear astrophysics questions.

The relevance for nuclear astrophysics includes the potential for direct studies of binary mergers. For neutron stars the waveforms encode information about the EOS. Predictions are uncertain but 10 events per year may be seen with the upgraded detectors. Theoretical modeling of mergers has advanced substantially in recent years, and there is some indication from these simulations that such mergers may be an important site for explosive nucleosynthesis (perhaps an r-process site). Another important open question is whether or not such mergers lead to short Gamma Ray Bursts (see section 2.4).

For nearby (< 50 Mpc) events it may be possible to catch the tidal distortion phase of mergers, and this could provide constraints on the NS radius and EOS. Detections in many cases depend on accurate simulated waveforms so a rigorous theoretical program is important to facilitate and optimize the scientific return.

Future ground-based high frequency detector perspectives include Advanced LIGO 2015+; KAGRA Japan 2015/16+; Advanced Virgo 2015/16+; LIGO India 2020+. A global network of such 2nd generation detectors should enable localization of merger (and other) events and trigger multi-wavelength follow-up observations.

## 3.10 Data and Codes

Nuclear astrophysics research requires rapid and efficient exchange of astronomical, nuclear experimental, and nuclear theoretical data products and codes. There are a number of challenges that are specific to nuclear astrophysics. The field therefore cannot rely on existing data compilation, evaluation, and dissemination efforts in astronomy and nuclear physics alone. These challenges include:

- Data and codes need to be exchanged across the field boundaries of astrophysics and nuclear physics. Data therefore need to be analyzed consistently, processes need to be well documented, and data, together with uncertainties, and codes need to be selected and presented in easy to use formats and interfaces, so that they can be used by researchers in other fields who are not necessarily experts.

- Nuclear data are often obtained for nuclei under terrestrial conditions. Modifications due to the extreme astrophysical temperatures and densities are often needed to make these data applicable for nuclear astrophysics.

- Astrophysical models require specific inputs that are not always directly measured or calculated, but are derived from a variety of quantities and measurements. An example is the stellar reaction rate, which needs to be determined based on individual direct and resonant components, and each of these components may have to be derived from a combination of experimental and theoretically predicted quantities. Measurements or theoretical work related to stellar reaction rates can therefore not be used in astrophysical models unless all ingredients are reevaluated and a new stellar reaction rate is derived. Deriving a stellar reaction rate can be a very elaborate process that has to be carried out by experts in nuclear astrophysics, especially if for example, R-matrix calculations are necessary to combine the various inputs.

- Nuclear data sets for astrophysical applications need to be complete. However, experimental and theoretical data are limited to what can be obtained with current techniques. For example, even basic nuclear properties for nuclei far from stability are missing from experimental data sets, and many theoretical approaches are limited to certain mass regions, to nuclei near closed shells, or, even-even nuclei. New data therefore need to be combined with other data sets that fill in the gaps, or the missing data need to be determined in other ways, such as using simplified approaches, or interpolation or extrapolation before they can be used in astrophysical calculations. This is not straight-forward and transitions from one data set to another can lead to artifacts in the output of astrophysical models.

In addition, there are challenges related to processing large amounts of data as part of nuclear astrophysics research projects ("Big Data"). These are discussed in section 3.8.2.

### 3.10.1 Existing Data Resources

Since the last Nuclear Astrophysics Town Meeting in 1999 there have been major advances in evaluating and making publicly available astrophysical and nuclear data for nuclear astrophysics. Major efforts that make results publicly available and address specific needs in nuclear astrophysics are summarized below:

- **Big Bang Online:** http://bigbangonline.org is a Cloud computing system that provides codes and data related to Big Bang nucleosynthesis.

- **Cococubed:** http://cococubed.asu.edu/code_pages/codes.shtml provides a set of useful fortran codes for nuclear astrophysics.

- **JINA:** http://www.jinaweb.org The Joint Institute for Nuclear Astrophysics (JINA) provides a frequently updated database of currently recommended stellar reaction rates (JINA reaclib), a public R-matrix code (AZURE), and a virtual journal that identifies literature with new data for nuclear astrophysics.

- **KADoNiS** http://www.kadonis.org provides an occasionally updated and well documented data base of evaluated s- and p-process stellar reaction rates.

- **Livermore:** http://adg.llnl.gov/Research/RRSN/ provides a website with links to a broad range of nuclear and astrophysical data for use in astrophysical models.

- **MINBAR:** https://burst.sci.monash.edu/wiki/index.php?n=MINBAR.Home The Multi Instrument Burst Archive will provide data for more than 6000 X-ray bursts that are consistently analyzed.

- **MESA:** http://mesa.sourceforge.net is a modern 1D stellar evolution code that is open source, modular, takes advantage of modern computational techniques, and is well supported. It includes nuclear reaction networks.

- **NACRE:** http://pntpm.ulb.ac.be/Nacre/nacre_d.htm provides a set of evaluated rates for reactions with stable nuclei from 1999.

- **NETGEN, BRUSLIB:** http://www.astro.ulb.ac.be/pmwiki/IAA/Databases provides databases for reaction rates and tools to create tailored reaction networks.

- **NNDC** http://www.nndc.bnl.gov/astro/ The National Nuclear Data Center (NNDC) provides stellar neutron capture rates evaluated based on the NNDC evaluated nuclear data.

- **NuGrid** http://www.nugridstars.org makes available a number of tools and data products for nucleosynthesis calculations, including a virtual box based nova model.

- **nucastro.org** http://nucastro.org provides various nuclear astrophysics data sets based on theoretical calculations with the Hauser-Feshbach approach.

- **nucastrodata.org** http://nucastrodata.org provides a computational infrastructure for nuclear astrophysics, including tools to evaluate and calculate reaction rates and to carry out reaction network calculations.

- **SAGA:** http://saga.sci.hokudai.ac.jp/wiki/doku.php provides a database for stellar abundances.

- **STARLIB:** http://starlib.physics.unc.edu provides stellar reaction rates derived with a novel Monte Carlo method for estimating experimentally-based reaction rates, and associated uncertainties, in a statistically meaningful manner.

- **Webnucleo:** http://nucleo.ces.clemson.edu provides a variety of public codes for nuclear astrophysics, including the reaction network code libnucnet. XML is used as data format, and tools are available for converting and manipulating data.

### 3.10.2  Future Data Developments

**Continuity:** Continuous support for existing database efforts in nuclear astrophysics is critical. Without long term continuity databases become quickly outdated, and new data that are often obtained using significant resources cannot be used in astrophysical calculations. Continuous support is therefore essential for rapid progress in the field, for taking advantage of nuclear and observational data, and to ensure researchers are not reaching the wrong conclusions because of the use of outdated nuclear data.

**Evaluation:** a community-wide effort is needed to identify and evaluate important stellar reactions. Researchers should reach out to the astrophysics modeling community to provide them with needed input as well and to the nuclear data community for their expertise in evaluations. The community must develop a set of best practices for rate evaluations, and communicate these widely together with the necessary data, tools, and codes. This must include the evaluation and proper determination of uncertainties. A series of workshops may be a good approach to achieve this goal.

**Transparency:** While already a number of open source nuclear astrophysics codes exist, it will be important in the future to expand the number of open source codes. Open source has many advantages, including broader use (advancing the science more rapidly) and community input and contributions on code improvements. On the other hand, concerns about return of investment for funds and efforts by individual institutions and researchers, and concerns about ongoing code support have to be addressed.

**Ease of Use:** Ease of use could be improved for many public nuclear astrophysics codes. Possibilities include cloud computing or virtual boxes to avoid compatibility, update, version, backup, or cyber security issues. GUIs could be customized for different users. No single data format will work for all the diverse phenomena in nuclear astrophysics but robust database storage with custom graphical user interfaces are an excellent solution for many cases. The use of XML as a standardized but flexible format is being explored. By choosing a standard format like XML, it would be possible to take advantage of freely available tools (such as XSLT in the case of XML) for converting the data to other formats.

**Distribution and Hubs:** Multiple distribution sites are currently quite effective in satisfying diverse user needs. However, there are also drawbacks from such an approach. It would therefore be interesting to explore a unifying HUBZero-based approach for nuclear astrophysics. Such an approach has been successful for other communities, for example nano technology (nanoHub.org). A hub could be created together with the broader nuclear physics community. HUBZero is an open source system that enables a wide variety of content (lectures, animations, codes, databases, tutorials, and so forth) to be put online. A set of well-developed and easy-to-use protocols allows users to install executable versions of their own research codes (in Fortran, C, C++, etc.) on the HUB without having to write their own web software. HUBZero middleware then allows these codes to be executed on the full complement of computer resources available on the the Open Science Grid, and standardized graphical interfaces allow the results to be visualized with a variety of plotting routines. By streamlining the installation of codes on the web, a HUB allows a community to leverage the efforts of a much larger fraction of its membership to develop shared resources than if those members developed and maintained their own separate web sites. Furthermore, by adopting a HUBZero-based approach, the nuclear astrophysics community could also benefit from the considerable experience of other communities that have their own HUBs.

**Cloud Computing Opportunities:** "cloud computing" services may also be a transformative vision for the future of the field. This opens up many possibilities including: having a digital assistant who automatically collects relevant masses, level schemes, references; a way for experts to easily upload supplemental information for your evaluations; having all major databases just one mouse click away; having an evaluation template automatically filled out for you; running analysis and application codes without compatibility, updates, backups, or cyber security issues; designing custom views of datasets from a variety of visualization tools; having a "virtual expert" online 24/7 to consult with questions; sharing your large data sets easily with colleagues; easily uploading your evaluation and visually tracking its progress for reviews, revisions, and acceptance; using a pipeline to process your evaluated data for use in simulations codes; running and visualizing these simulations, then sharing the results with colleagues

## 3.11 Centers

Nuclear astrophysics requires coherent research efforts and rapid exchange of information and results across the field boundaries of astrophysics and nuclear physics, and between theory, experiment, and observations. There are many centrifugal forces at work that prevent such coherence - including the lack of common language and common training in nuclear physics and astrophysics, growing specialization, boundaries and different priorities between funding agencies, lack of opportunities for exchange, and other cultural differences. The formation of centers that can overcome these divergent forces is therefore essential for nuclear astrophysics. In the early stages of nuclear astrophysics (mid 20th century) such centers formed around the eminent personalities of the field. However, with growing specialization, broadening of the field, the increased complexity of the questions to be addressed, and changes of funding patterns larger scale government or privately supported centers are essential in todays research environment. Such centers serve as intellectual focal points for the field, they stimulate the development of common goals, they provide resources needed to connect research efforts in various subfields, they facilitate the rapid exchange of ideas and data across field boundaries, they provide interdisciplinary education and development opportunities for young researchers in the field, and they stimulate leading edge and innovative research.

In the US the Joint Institute for Nuclear Astrophysics (JINA) serves as such a center for nuclear astrophysics. JINA has had an enormous impact on the field and brought together nuclear experimental, nuclear theory, theoretical astrophysics, observational astronomy, and computational astrophysics communities. JINA has stimulated the formation of similar centers across the world, as other countries recognized the importance of centers for nuclear astrophysics. Examples include the VISTARS, NAVI, and EMMI centers in Germany, the Munich Universe Cluster also in Germany, an initiative at the Institute for Advanced Studies in Sao Paulo Brazil, and the Shanghai Center for Nuclear Astrophysics in China. It will be important for the US community to continue its leadership in this area, especially in light of new large scale experimental (FRIB) and observational (LSST, LIGO) facilities (together with many new opportunities at smaller university laboratories) that will require interdisciplinary research networks to exploit their potential for addressing the forefront questions in nuclear astrophysics.

# 4 ACKNOWLEDGEMENTS

Numerous members of the nuclear astrophysics community have contributed to this White Paper through writing, editing, presentations, organization, and other synergistic activities. The following individuals have made significant text contributions:

Almudena Arcones

Dan Bardayan

Lee Bernstein

Timothy Beers

Alex Brown

Edward Brown

Carl Brune

Jeff Blackmon

Art Champagne

Alessando Chieffi

Aaron Couture

Pawel Danielewicz

Roland Diehl

Mounib El Eid

Jutta Escher

Brian Fields

Carla Fröhlich

Falk Herwig

Raph Hix

Christian Iliadis

Bill Lynch

Brad Meyer

Gail McLaughlin

Bronson Messer

Tony Mezzacappa

Filomena Nunes

Brian O'Shea

Madappa Prakash

Boris Pritychenko

Ernst Rehm

Sanjay Reddy

Grischa Rogachev

Bob Rutledge

Hendrik Schatz (Chair 2012 and Co-Convener 2014)

Michael Smith

Tod Strohmayer

Ingrid Stairs

Andrew Steiner

Frank Timmes

Dean Townsley

Michael Wiescher (Co-Convener 2014)

Remco Zegers

Vladimir Zelevinsky

Michael Zingale

We would also like to thank Sherry Yennello for leading the local organization of Nuclear Astrophysics Town Meeting in 2014, and the APS Division of Nuclear Physics for sponsoring the meeting. We also thank the organizers of the Town Meeting on Neutrinos and Fundamental Symmetries for contributions. We thank the Joint Institute for Nuclear Astrophysics, a NSF Physics Frontiers Center, for its support and organization of the 2010 Nuclear Astrophysics Town Meeting that initiated the writing of this White Paper.

## APPENDIX A

Please note in the original document submitted to the LRP Writing Group in January 2015 the following 'Joint Summary of the NAP and LENP Town Meetings ' was placed before the NAP Executive Summary emphasizing the synergy and common goals of the two communities. This same joint summary was repeated in the LENP WP.

## JOINT EXECUTIVE SUMMARY FROM THE NUCLEAR ASTROPHYSICS AND LOW-ENERGY NUCLEAR PHYSICS TOWN MEETINGS

In preparation for the 2015 NSAC Long Range Plan (LRP), the DNP town meetings on Nuclear Astrophysics and Low-Energy Nuclear Physics were held at the Mitchell Center on the campus of Texas A&M University August 21-23, 2014. Participants met in a number of topic-oriented working groups to discuss progress since the 2007 LRP, compelling science opportunities, and the resources needed to advance them. These considerations were used to determine priorities for the next five to ten years. Approximately 270 participants attended the meetings, coming from US national laboratories, a wide range of US universities and other research institutions and universities abroad.

The low-energy nuclear physics and the nuclear astrophysics communities unanimously endorsed a set of joint resolutions. The full text of the resolutions adopted at the Town Meetings is included at the end of this document. The joint resolutions were condensed from the individual recommendations of the two town meetings in order to recognize the highest priorities of the two fields. The resulting joint resolutions reflect hard choices, as not all priority items from the individual meetings are included. The communities strongly hope that the broader needs can be realized in time.

The resolutions represent a plan to address the compelling scientific opportunities in the study of atomic nuclei and their role in the cosmos. The intellectual challenges of our subfields were captured well in overarching questions from the 2012 NRC Decadal Survey of Nuclear Physics:

- How did visible matter come into being and how does it evolve?

- How does subatomic matter organize itself and what phenomena emerge?

- Are the fundamental interactions that are basic to the structure of matter fully understood?

- How can the knowledge and technological progress provided by nuclear science best be used to benefit society?

Answers to these questions require a deeper understanding of atomic nuclei, both theoretically and experimentally, than we have now. The last decade already saw considerable progress in understanding of the nucleus and its role in the universe. New ideas, combined with new experimental techniques, major leaps in computing power and impressive improvements in experimental capabilities resulted in discoveries leading to quantitative and qualitative changes in our understanding of nuclear and astrophysical phenomena.

A comprehensive discussion of the broad scientific opportunities and challenges will be provided in the individual town meeting reports. Low-Energy nuclear physics research addresses the existence of atomic nuclei, their limits, and their underlying nature. It also aims to describe interactions between nuclei and dynamical processes such as fission. The ultimate goal is to develop a predictive understanding of nuclei and their interactions grounded in fundamental QCD and electroweak theory. Nuclear astrophysics addresses important scientific questions at the intersection between nuclear physics and astrophysics: the chemical history of the Universe, the evolution of stars, stellar explosions, the nature of dense matter, and the ultimate fate of visible matter.

Both fields are poised for breakthroughs. New astronomical data, such as high resolution abundance distributions from very early stars; the neutrino fluxes from our sun, nearby supernovae, or possibly even the Big Bang; and anticipated gravity waves detected from merging neutron stars, will only be interpreted successfully if the relevant nuclear processes are understood. Exploration of elements and isotopes at the extremes of N and Z will provide the insights needed for a comprehensive understanding of nuclei. This exploration may reveal novel quantum many-body features and lead us toward a deeper understanding of complex quantum systems and of the mechanisms responsible for the emergent features found in atomic nuclei.

More broadly, science and society rely on our understanding of the atomic nucleus. Its relevance spans the dimensions of distance from 10-15 m (the proton radius) to 12 km (neutron star radius) and the evolutionary history of the universe from fractions of a second after the Big Bang to today; i.e., 13.8 billion years later. As reaffirmed by the 2012 National Academies of Sciences' decadal study *"Nuclear Physics: Exploring the Heart of Matter,"* the path to understanding the nucleus requires the completion of the Facility for Rare Isotope Beams (FRIB) and its effective operation. Unprecedented access to a vast new terrain of nuclei will result in scientific breakthroughs and major advances in our understanding of nuclei and their role in the cosmos, and will open new avenues in cross-discipline contributions in basic sciences, national security, and other societal applications.

Based on this first joint resolution was: **1. The highest priority in low-energy nuclear physics and nuclear astrophysics is the timely completion of the Facility for Rare Isotope Beams and the initiation of its full science program.**

While FRIB is the top priority of both subfields, there are other capabilities needed to reach the scientific goals of the field. In arriving at a joint set of resolutions, the Town Meeting participants addressed priorities for the field as a whole. What emerged is a coherent plan that pursues key scientific opportunities by leveraging existing and future facilities. The plan involves continuation of forefront research activities, development of needed theory, and initiation of a focused set of new equipment initiatives. While many specific ideas were discussed, it was decided to approve wording that made clear that the community is asking for the base set of needs while recognizing that some initiatives may have to be delayed. The consensus minimum need is outlined in the following.

The science goals of Low-Energy Nuclear Physics and Nuclear Astrophysics require studies with probes ranging from stable and radioactive nuclei, to photons, neutrons, and electrons. While FRIB will explore uncharted regions of the nuclear chart and produce rare isotopes important for study of explosive environments, there is a key component of the program that links this exploration to studies of near-stable isotopes. The ATLAS stable beam facility has world-unique capabilities that will enable necessary precision studies near stability and at the limits of atomic number. Many aspects of stellar evolution, in particular the long quiescent phases of stellar evolution defining the conditions and providing the seed for subsequent cataclysmic developments, are driven by reactions of stable isotopes that will be studied at stable-beam accelerators. The electron beam at JLab provides a unique capability for probing the short-range part of the nuclear force in nuclei and the modification of nucleons in the nuclear medium. The photon beam at HIÎ³S and the neutron beams at Los Alamos provide unique, important information. The university accelerator labs have a special role. They contribute cutting edge science, targeted research programs of longer duration, critical developments of techniques and equipment, combined with hands-on education.

For success in the long term, including success of FRIB, we must adequately operate current low-energy facilities and support the research groups that utilize these facilities to advance science and to foster an experienced and engaged research community. Adequate operations require that planned upgrades, already in baseline budgets, proceed, and current levels of research activities be enhanced. Significant facility reductions have occurred since 2007, including the closure of HRIBF and WNSL. These closures, coupled with limited availability of research hours at facilities worldwide, has resulted in a loss of research opportunities that can be restored in part by the development of a multi-user capability at ATLAS.

The Town Meetings recognized that the NSF has played a leading role in the establishment and operation of one of the nation's flagship nuclear science facilities, NSCL, as well as in supporting university laboratories that carry out forefront nuclear physics and nuclear astrophysics research. There is the opportunity for continued leadership by support of cutting edge nuclear physics and transformative research extending well into the era of FRIB operations. This LRP can make the case for continued strong nuclear physics funding by NSF. Continued NSF leadership can be accomplished by an exciting plan that includes the effective operations at NSCL and NSF university laboratories, proposed upgrades to ReA3 at NSCL, and support for nuclear astrophysics initiatives such as an underground accelerator.

Hence, the Town Meeting participants recommended: **2a. We recommend appropriate support for operations and planned upgrades at ATLAS, NSCL, and university-based laboratories, as well as for the utilization of these and other facilities, for continued scientific leadership. Strong support for research groups is essential.**

Significant progress in the theory of nuclei and the astrophysical environments in which they are embedded is a key component of the scientific goals of the field. A strong theory effort needs to go hand in hand with future experimental programs. The proposed FRIB Theory Center, a modest-scale national effort comprising a broad theory community, will therefore be important for the success of FRIB. (A brief overview of this effort will be summarized in a separate document.) The FRIB Theory Center will complement the NSF funded Joint Institute for Nuclear Astrophysics that facilitates the communications within the nuclear astrophysics community. JINA's efforts are essential for the interpretation of nuclear physics experiments and stellar observations in the framework of new theoretical developments. Its continuous operation is a main goal for the nuclear astrophysics community. This need for broad theory support resulted in part two of the second recommendation: **2b. We recommend enhanced support for theory in low-energy nuclear science and nuclear astrophysics, which is critical to realize the full scientific promise of our fields.**

In addition to the needs articulated above, a targeted set of new instrumentation and accelerator investments are necessary. FRIB will be a world-leading accelerator and will yield the best science when coupled with world-leading equipment. The community has identified and prioritized key detector systems and re-accelerator upgrades that will enable effective utilization of FRIB with the highest science potential. In the 2007 LRP the major new detector endorsed was GRETA and, at that time, it was recognized that an astrophysics separator would be needed (which is now named SECAR). These state-of-the-art systems are still top priority and highly anticipated by the community. While existing equipment can and will be used at FRIB, major advances in other detector and separator technology continue to take place. To this end, the community has developed exciting ideas for critical new equipment. Not all can be realized immediately, but a targeted suite to address the highest-priority research programs is needed.

In the area of accelerator investments, there is an opportunity for NSF to play a leading role by upgrading ReA3 and by establishing an underground accelerator for nuclear astrophysics. The community has long advocated for these developments.

In order to prepare the future and advance the science goals of the community, the third part of the second resolution states: **2c. We recommend targeted major instrumentation and accelerator investments to realize the discovery potential of our fields.**

Two additional joint resolutions were adopted to support the main conclusions of the Computational Town Meeting and the Education and Innovation Town Meeting.

## Full Text of The Joint Resolutions

Science and society rely on our understanding of the atomic nucleus. Its relevance spans the dimensions of distance from $10^{-15}$ m (the proton's radius) to 12 km (the neutron star radius) and timescales from fractions of a second after the Big Bang to today, i.e. 13.8 billion years later. As reaffirmed by the 2012 National Academies of Sciences' decadal study *"Nuclear Physics: Exploring the Heart of Matter"*, the path to understanding the nucleus requires the completion of the Facility for Rare Isotope Beams (FRIB) and its effective operation. Unprecedented access to a vast new terrain of nuclei will result in scientific breakthroughs and major advances in our understanding of nuclei and their role in the cosmos, and will open new avenues in cross-discipline contributions in basic sciences, national security, and other societal applications.

- **The highest priority in low-energy nuclear physics and nuclear astrophysics is the timely completion of the Facility for Rare Isotope Beams and the initiation of its full science program.**

In support of our science goals we must continue forefront research, exploiting existing facilities, develop new capabilities and equipment, and enable major advances in nuclear theory.

- **We recommend appropriate support for operations and planned upgrades at ATLAS, NSCL, and university-based laboratories, as well as for the utilization of these and other facilities, for continued scientific leadership. Strong support for research groups is essential.**

- **We recommend enhanced support for theory in low-energy nuclear science and nuclear astrophysics, which is critical to realize the full scientific promise of our fields.**

- **We recommend targeted major instrumentation and accelerator investments to realize the discovery potential of our fields.**

Realizing the scientific potential of Low-Energy Nuclear Science and Nuclear Astrophysics demands large-scale computations in nuclear theory that exploit the US leadership in high-performance computing.

- **We endorse the recommendations of the 2014 Computational Nuclear Physics Meeting: "*Capitalizing on the pre-exascale systems of 2017 and beyond requires significant new investments in people, advanced software, and complementary capacity computing directed toward nuclear theory.*"**

Education, outreach, and innovation are key components of any vision of the future of the field of nuclear science. Our fields play a leading role in education and training of the nation's nuclear science workforce. They are ideally positioned to advance applications in medicine, energy, national security, and material science. The health of this field is required to train the talented national workforce needed to assure continuing societal benefits in these critical areas. Continuation of this role is a major goal of our fields.

- **We endorse the recommendation of the DNP Town Meeting on Education and Innovation.**

## APPENDIX B

## Nuclear Astrophysics Town Meeting 2012 – Program

### Monday October 8, 2012

20:00 - 22:00  Cash Bar, Room: Vespucci-Byrd

### Tuesday October 9, 2012

07:00 - 08:00  Light Refreshments, Lindbergh AB Hall

**Session I** Chair: Ingo Wiedenhoever, Florida State U., Room Lindbergh AB

| 8:15 - 8:30 | Introduction | Hendrik Schatz | Michigan State U. |
|---|---|---|---|
| 8:30 - 8:50 | Stars | Art Champagne | U. of North Carolina |
| 9:00 - 9:20 | Stellar Observations with KEPLER Constraining Nuclear Processes | Sarbani Basu | Yale U. |
| 9:30 - 9:50 | Open Questions in Nucleosynthesis up to Zn | Alex Heger | Monash U. |

10:00 - 10:30  Coffee Break, Lindbergh AB Hall

**Session II** Chair: Carla Froehlich, North Carolina State U., Room Lindbergh AB

| 10:30 - 10:50 | Open Questions in s- and p-processes | Kerstin Sonnabend | Goethe U. Frankfurt |
|---|---|---|---|
| 11:00 - 11:20 | Open Questions in r- and vp-processes | Rebecca Surman | Union College/U. of Notre Dame |
| 11:30 - 11:50 | Explosive Hydrogen Burning | Jeff Blackmon | Louisiana State U. |

12:00 - 13:30  Lunch, Hotel Lobby-Atrium and Flight Deck Lounge (overflow)

**Session III** Chair: Reiner Kruecken, TRIUMF, Room Lindbergh AB

| 13:30 - 13:50 | Near Field Cosmology and First Star Signatures | Tim Beers | NOAO / MSU |
|---|---|---|---|
| 14:00 - 14:20 | Galactic Chemical Evolution | Chiaki Kobayashi | U. of Hertfordshire |
| 14:30 - 14:50 | Neutrino Physics and Weak Interactions | George Fuller | U. of California, San Diego |
| 15:00 - 15:20 | Neutrino Experiments | Nikolai Tolich | U. of Washington |

15:30 - 16:00  Coffee Break, Lindbergh AB Hall

**Session IV** Chair: Thomas Rauscher, University of Basel, Room Lindbergh AB

| 16:00 - 16:20 | Neutron Stars and Dense Matter | Sanjay Reddy | U. of Washington |
|---|---|---|---|
| 16:30 - 16:50 | Prospects with LIGO | Christian Ott | Caltech |
| 17:00 - 17:20 | Core Collapse Supernovae | Chris Fryer | LANL |
| 17:30 - 17:50 | Prospects with NUSTAR and $\gamma$ rays | Steve Boggs | U. of California, Berkeley |

18:00 - 19:00  Dinner at Working Group Location, Earhart Foyer and Lower Level Foyer

19:00 - 21:00  Working Groups I

| Working Groups I | Room |
|---|---|
| Stars and Stellar Evolution (Champagne, Chieffi, Herwig) | Earhart |
| Neutrino Astrophysics (Fuller, Qian) | Balboa |
| Neutron Stars and Dense Matter (Lynch, Prakash, Ruthledge, Stairs) | Ericsson |
| Core collapse Supernovae, Neutron Star Mergers and GRBs (Hix, Ott) | Cortez |
| Thermonuclear Explosions: Type Ias, Novae and X-ray Bursts (Bardayan, Zingale) | LindberghAB |
| First Stars, Chemical Evolution, Cosmology (Fields, O'Shea) | Coronado |

## Wednesday October 10, 2012

07:00 - 08:00  Light Refreshments, Lindbergh AB Hall

**Session V** Chair: Kate Jones, University of Tennessee, Room Lindbergh AB

| 8:15 – 8:35 | Novae and Type Ia Supernovae | Dean Townsley | U. of Alabama |
|---|---|---|---|
| 8:45 – 9:05 | X-ray Bursts and Neutron Star Crusts | Andrew Cumming | McGill U. |
| 9:15 – 9:35 | Opportunities with Grains | Andrew Davis | U. of Chicago |

09:45 - 10:00  Coffee Break at Working Group Location, Earhart Foyer and Lower Level Foyer

10:00 - 12:00  Working Groups II

| Working Groups II | Room |
|---|---|
| Equipment and Facilities: Radioactive Beams, Neutrons (Couture, Rehm, Zegers) | LindberghAB |
| Equipment and Facilities: Stable Beams and Gamma Beams (Brune, Iliadis, Wiescher) | Lindbergh C |
| Astronomy (Diehl, Strohmayer) | Balboa |
| Theory (Arcones, E. Brown, McLaughlin, Nunes) | Earhart |
| Astro Computation (Mezzacappa, Timmes) | Cortez |
| Data and Codes (Meyer, Prytichenko, Smith) | Coronado |

12:00 - 13:30  Lunch, Hotel Lobby-Atrium and Flight Deck Lounge (overflow)

**Session VI** Chair: Robert Tribble, Texas A&M University, Room Lindbergh AB

| 13:30 - 13:50 | Intersections with Planets and Astro Biology | Frank Timmes | Arizona State U. |
|---|---|---|---|
| 14:00 - 14:20 | Opportunities at NIF | Dennis McNabb | LLNL |
| 14:30 - 14:50 | The Role of Centers | Michael Wiescher | U. Notre Dame |
| 15:00 - 15:20 | Data | Richard Cyburt | Michigan State U. |

15:30 - 16:00  Coffee Break, Lindbergh AB Hall

16:00 - 18:00  Reports of working groups and wrap up - 5 min per working group

## Twitter

Submit questions and constructive comments via Twitter at any time during the meeting:
• Use hashtag #NucATown anywhere in your tweet
• Search for #NucATown to see tweets, or follow @NucAstroTown12, we will retweet there
• It's not too late to join – sign up for a twitter account at twitter.com (or download the app),
  login, press the tweet button in the upper right and type your question

# Joint DNP Town Meetings on Nuclear Structure and Nuclear Astrophysics

August 21-23, 2014

**Mitchell Institute, Texas A&M University**

Home

Nuclear Astro Meeting

Low Energy Meeting

Venue

Accommodations

Program

Organizing Committee

Working Groups

Sponsors

Registration

Webinar

Past Meetings

White Papers

## Program

**Thursday, August 21st: Mitchell Physics Building**

**8:15 AM - 10:35 AM Thursday Morning Joint Astro/LENP Plenary Session**
Room: 203/204/205 Chair: M. Riley

8:15 AM - 8:20 AM Welcome
8:20 AM - 8:35 AM Charge and LRP process (J. Hardy)
8:35 AM - 9:05 AM Theory of Nuclei and Their Reactions (W. Nazarewicz)
9:05 AM - 9:35 AM Experimental Studies of Nuclei (R.V.F. Janssens)
9:35 AM - 10:05 AM Nuclear Astrophysics Overview: Report from 2012 Town Meeting (H. Schatz)
10:05 AM - 10:35 AM Theory for Astrophysics (S. Reddy)

10:35 AM - 11:00 AM **Coffee Break**

**11:00 AM - 12:15 PM Thursday Morning Joint Astro/LENP Plenary Session**
Room: 203/204/205 Chair: C. Elster

11:00 AM - 11:25 AM Fundamental Symmetries (B. Balantekin)
11:25 AM - 11:50 AM Neutrino Physics (N. Tolich)
11:50 AM – 12:15 PM Nuclear Structure Studies at Jefferson Laboratory (J. Arrington)

12:15 PM – 1:30 PM **LUNCH**

**1:30 PM - 3:10 PM Thursday Afternoon Joint Astro/LENP Plenary Session**
Room: 203/204/205 Chair: H. Schatz

1:30 PM - 1:50 PM FRIB Status, NSCL, and the Path Forward (T. Glasmacher)
1:50 PM - 2:10 PM FRIB Major Equipment (M. Smith)
2:10 PM - 2:30 PM ATLAS White paper Summary (G. Savard)
2:30 PM - 2:50 PM ARUNA White Paper Summary (I. Wiedenhoever)
2:50 PM - 3:10 PM Experimental Initiatives in Nuclear Astrophysics (C. Brune)

3:10 PM - 3:35 PM **Coffee Break**

**3:35 PM – 4:35 PM Thursday Afternoon Joint Astro/LENP Plenary Session**
Room: 203/204/205 Chair: M. Wiescher

3:35 PM - 3:55 PM Report from the Workshop on Computational Needs for Nuclear Physics (J. Carlson)
3:55 PM - 4:15 PM Report from the Town Meeting on Education and Innovation (M. Thoennessen)
4:15 PM - 4:35 PM Applications Overview (S. Wender)

**Short break**

**4:45 PM - 5:45 PM Discussion Session Education and Outreach** Room: 203/204/205 Chair: M. Thoennessen

**6:00 - 7:30 PM Reception: Cyclotron Institute**

**Friday, August 22nd: Mitchell Institute/Mitchell Physics Building**

**8:30 AM - 12:00 PM LENP WG (Working Group Organizers in Parenthesis)**

10:15 AM - 10:35 AM **Coffee Break**

- Theory for Low-Energy Nuclear Science (Carlson, Hagen, Elster, Nazarewicz, Reddy) Room 205
- Nuclear Structure and Reactions Experiment (Carpenter, Fallon, Gade, Higinbotham, Jones, Wuosmaa) Room Hawking Auditorium - Presentations Set 1 Set 2
- Nature of Dilute and Dense Nuclear Matter and the Equation of State (Horowitz, Natowitz, Sobotka, Tsang) Room 204
- Applications and Nuclear Data (Stoyer, Lapi, Vetter, Kondev, McCutchen) Room 203

12:00 – 1:00 PM **Lunch**

**1:00 PM – 2:30 PM Astro WG 1 + LENP WG 2 (Working Group Organizers in Parenthesis)**

- [Supernova models, mergers models and chemical evolution](#) (Hix, Froehlich) MI102
- [RIB Beams, explosive nucleosynthesis](#) (Rehm, Zegers, Bardayan, Blackmon) Room 204
- [Stable and gamma beams, stars, ....](#) (Champagne, Brune, Rogachev) Room 213
- [Nuclear theory for astrophysics](#) (McLaughlin, Reddy, Escher) Room 205
- [Applications and Nuclear Data](#) (Stoyer, Lapi, Vetter, Kondev, McCutchen) Room 203
- [Nuclear Structure and Reactions Experiment](#) (Carpenter, Fallon, Gade, Higinbotham, Jones, Wuosmaa) Room Hawking Auditorium - Presentations [Set 3](#)

2:30 PM - 3:00 PM **Coffee Break**

**3:00 PM– 4:30 PM Astro WG 2 (Working Group Organizers in Parenthesis)**

- Astrophysics theory and computing (Timmes, Messer) Room 205
- [Neutron beams, s-process, NIF](#) (Couture, Bernstein) Room 213
- [Data for Astrophysics](#) (Meyer, Smith) Room 203
- [Neutron Stars Science, experiments, theory](#) (Lynch, Steiner) Room 204

4:30 PM - 4:45 PM **Break**

**4:45 – 6:45 PM Joint Astro/LENP Session (Working Group Organizers in Parenthesis)**

- [Experimental Facilities and Instrumentation](#) (Gross, Howell, Macchiavelli, Savard, Sherrill, Wiedenhoever) Room Hawking Auditorium
- Future Computational Needs in Nuclear Physics (Carlson) Room 203 - Chaired by M. Savage

**Saturday, August 23rd: Mitchell Institute**

**8:00 AM – 11:30 AM Joint Astro/LENP Session**
Room: 203/204/205

8:00 AM - 9:30 AM [Working Group Reports](#)

9:30 AM - 9:50 AM **Coffee Break**

9:50 AM - 11:45 PM [Resolutions and Outcomes](Resolutions and Outcomes)

11:45 **Adjourn**

[Webmaster](Webmaster) Last updated October 28, 2014